\documentclass[useAMS,usenatbib]{mn2e}

\usepackage{graphicx}
\usepackage{names}
\usepackage{times}
\begin{document}

\title{Spiral structure in nearby galaxies \\ I. Sample, data analysis, and overview of results.  }
\author[S. Kendall, R. C. Kennicutt, Jr., C. Clarke]{S. Kendall$^1$\thanks{E-mail: sak39@ast.cam.ac.uk}, R. C. Kennicutt$^1$, C. Clarke$^1$\\
$^1$ Institute of Astronomy, University of Cambridge, Madingley Road, Cambridge CB3 0HA}
\date{submitted, accepted}

\pagerange{\pageref{firstpage}--\pageref{lastpage}}

\maketitle

\label{firstpage}

\begin{abstract}

This paper, the first of two, introduces an observational study of spiral structure in galaxies chosen from the SINGS survey. Near infrared (NIR) and optical data are used to produce mass surface density maps, and from these the morphology of the disc is examined. The aim of this work is to characterise the prevalence of spiral structure in this sample and, in the cases where a clear spiral pattern is found, include the findings in a comparative study (reported in paper II). A two-armed (`grand design') spiral pattern is found in approximately half the galaxies studied,
including all those that are designated as grand design in the optical, but
also  including some, but not all, optically flocculent galaxies. 
It is found that the level of non-axisymmetric structure in the galaxies'
mass distributions is only modestly higher in those galaxies that are
classified as `grand design' compared with  those that are not, implying
that non-grand design galaxies possess significant power
in higher order modes.   
There is no evidence that bars preferentially trigger the spirals, but they do appear to stir up non-axisymmetric structure in the disc. In contrast, there is evidence that strong/close
tidal interactions with companion galaxies are associated with  strong
two-armed spiral structure in the infrared, though there are a number
of galaxies with relatively weak 
infrared spiral structure that do not possess such companions. 

\end{abstract}

\begin{keywords}
galaxies:individual--galaxies:spiral--galaxies:structure--infrared:galaxies
\end{keywords}

\section{Introduction.}\label{intro}

In an earlier paper, \cite{2008MNRAS.387.1007K} (hereafter KKCT08), near infrared (NIR) and optical data were used to investigate the spiral structure in M81. This paper is the first of two which aim to present a similar analysis of all suitable galaxies in the \textit{Spitzer} Infrared Nearby Galaxies Survey (SINGS - \citep{2003PASP..115..928K}).

The NIR is a good choice of wavelength with which to study the mass distribution in galaxies, because the emission is dominated by red giant stars, which are good tracers of the underlying stellar mass distribution. In contrast, optical wavelengths tend to have disproportionately high contributions from massive young stars which have a much lower mass-to-light ratio (M/L) than longer lived, lower mass stars. Thus the optical is not a suitable tracer of the stellar mass unless a correction is made for these young stars. Conveniently, it is often possible to correct optical wavelengths for this effect, \citep{2001ApJ...550..212B}, and this approach is discussed further in section \ref{ch2_BVRI}.

The motivation for this work is found in the complexities, and at times controversy, surrounding the study of galactic spiral structure. Although the subject is more than fifty years old, we are still a long way from fully understanding these beautiful features. The main theoretical approaches can be divided into two main groups; firstly the quasi-stationary spiral structure (QSSS) theory proposed by \cite{1964ApNr....9..103L} and developed further by \cite{1964ApJ...140..646L, 1966PNAS...55..229L}. The counter arguments are in favour of transient spirals driven by tidal interactions or bars, such as envisaged by Toomre in his swing amplification theory \citep{1981seng.proc..111T}, or supported by solid-body rotation \citep{1979ApJ...233..539K}.

Fundamental questions which have yet to be settled fully concern the origins of galactic spirals; do  spiral arms arise spontaneously in stellar discs  or do they require some mechanism to provide a driving force (either external, for example a companion galaxy, or internal, such as a central bar or oval distortion)? 
In the latter case, does long lived spiral structure imply that the trigger
is itself longlived, in which case one would expect to see a good correlation
between grand design spiral structure and features such as
bars and companions. The effects of bars and companions have been addressed in both computational and theoretical work \citep{1981seng.proc..111T, 1992AJ....103.1089B} and observational studies \citep{1979ApJ...233..539K, 1982MNRAS.201.1021E, 1989ApJ...342..677E, 1998MNRAS.299..685S} with evidence that companions in particular may be significant (e.g \citep{1982MNRAS.201.1021E, 1998MNRAS.299..685S}).

Related to this is the question of the lifetime of these spiral patterns; are they long lived, lasting in excess of a Gyr with little or no evolution of the pattern, or are they instead transient (and if the latter, how transient)? In the QSSS picture, a rigidly rotating spiral pattern may represent a normal
mode of the system and can thus be self-sustained for many rotations. On the
other hand, if a companion is providing the necessary stimulus then a pattern could reasonably be expected to evolve as the interaction proceeds \citep{2010MNRAS.403..625D, 2008ApJ...683...94O}. Even in an isolated spiral, the pattern may evolve over little more than a dynamical time. Indeed, N-body simulations such as \citep{1984ApJ...282...61S, 2003MNRAS.344..358B} (see also \citep{2006MNRAS.371..530C, 2008MNRAS.385.1893D}) suggest that the spiral features in isolated discs may
be transient but regenerative features which are ill described as global
modes with radially constant  azimuthal wave number and pattern speed.  At
the same time, such simulations also exhibit secular evolution of spiral
structure in the sense that this `washes out' with time unless the
disc is `cooled' through the continuous creation of new stars on 
dynamically cold (near circular)
orbits. This latter effect may however be over-estimated in many numerical
studies due to resolution effects causing enhanced heating (Semelin \& Combes
2002). The effects discussed so far all relate to the evolution of structure in the
underlying mass distribution \citep{2000Ap&SS.272...31S}, but we 
note that evolving spiral features are also expected in the
opposite scenario (i.e. that of SSPSF - stochastic self propagating star formation - \citep{1976ApJ...210..670M, 1978ApJ...223..129G}) where - as an extreme - 
the spiral pattern could be largely independent of disc self-gravity but
merely be a kinematic effect associated with the shearing of chains of
star formation regions in which each star formation event triggers
its neighbour. Naturally, were it the case that spiral arms indeed
only involved the young stellar population, then one would not expect
to see significant spiral structure in the NIR.

The different spiral properties of galaxies may be loosely sub-divided into grand design and flocculent types \citep{1981ApJS...47..229E}. The discs of grand design galaxies are dominated by two well defined spiral arms - typical examples are M81 or M51.  Flocculent galaxies are much more patchy, with short wispy segments of spiral structure but no clear ordered pattern: good examples of flocculent spirals are NGC 7793 or NGC 3621. \citeauthor{1981ApJS...47..229E}  made this classification based on \textit{B} band images, which mainly traces the
response of  the young stellar populations (and underlying that, the structure
of the star forming gas). An interesting question (which also tests whether
spiral structure is indeed a self-gravitating phenomenon that involves most of the galaxy's stellar mass)  is whether the stellar mass also mirrors the \textit{B} band, or whether some optically flocculent galaxies may have grand design structure in the NIR, and vice-versa.

In order to address these questions, it is important to be able to characterise the morphology of the spirals accurately; information about the strength and shape of the pattern, particularly the amplitude of the arms relative to the disc and the pitch angle, can be compared to theoretical predictions. This information, especially when combined with simulations, is our best hope of reaching a better understanding of spiral structure. Examples of correlations that have been predicted include, for QSSS theories, a dependence of pitch angle on Hubble type \citep{1966PNAS...55..229L,1975ApJ...196..381R}. If the arms are instead expected to shear, then pitch angle should correlate with the rate of shear of the disc. Finally, if the spiral arms are caused by tidal interactions then the relationship between pitch angle and amplification factor is important \citep{1981seng.proc..111T}. The amplitude of the spiral arms is also expected to depend on Hubble type and related parameters such as mass concentration, since these are all factors which influence the Hubble classification of a galaxy. In addition, the amplitude of the spiral arm is expected to correlate with star formation, since the gas is expected to shock more strongly for larger amplitude spiral arms \citep{1969ApJ...158..123R}, and thus star formation will be enhanced due to the increase in gas concentration \citep{1998ApJ...498..541K}. In addition, information about the morphology of spiral arms is of use when considering the effects of spirals on the host galaxies; massive structures, such as the spiral arms in grand design galaxies like M81 or M51, are large enough to exert significant gravitational torques on the galaxy. Thus, the better we are able to characterise the mass distributions, the better we can model the galaxies.

To this end, the purpose of this study is to analyse the spiral structure in a range of different spiral types, ranging from grand design spirals such as NGC 5194 (M51) to optically flocculent galaxies like NGC 7793. The excellent resolution and depth of the \textit{Spitzer} data should allow us to obtain much more accurate estimates of the degree of non-axisymmetric structure in these galaxies, and as a result, provide useful constraints and tests for theories of spiral structure. This work builds on many previous surveys of spiral structure in the 
infrared - \citep{1976ApJS...31..313S, 1989ApJ...343..602E, 1998MNRAS.299..685S, 2004A&A...423..849G}, but takes advantage of the new wavelengths available from  \textit{Spitzer}, in particular the 3.6$\mu$m which is found to be an excellent tracer of stellar mass, as discussed later in this work. In addition, Spitzer offers improved resolution in the NIR, with PSFs of ~1.7 and 2 arcsec in the 3.6 and 8$\mu$m bands respectively.

This paper introduces the SINGS sample and the criteria for choosing the galaxies used in this work. The methods used to relate stellar mass to the light distribution are presented and discussed (Section 2). In Section 3, we divide
our sample into a detailed sample of 13 galaxies in which we can trace well
defined m=2 spiral arms over a reasonable radial range; for each of these
galaxies, Section 3.1 reports on 
the radial variation of arm amplitude   and also on any azimuthal offset between
the stellar spiral (as measured at $3.6 \mu$m) and the $8 \mu$m, which is assumed to be a tracer of the gas shocks. Section 3.2 lists the 13 galaxies which we deem to be
not grand design as well as a further 5 galaxies which exhibit an
evident spiral pattern but where the analysis is complicated by factors
such as a strong bar or pronounced asymmetry in the arms. 
We defer a detailed inter-comparison of the properties of our galaxies
in the detailed sample until a following paper (Paper II). For the purpose
of this paper we focus on conclusions that can be drawn from the
intercomparison of the host galaxy properties of  galaxies which either
are or are not grand design in the near infrared. Thus we examine
(Section 4) the association between grand design structure in the NIR
with optical grand design structure, with Hubble type and with the presence
of bars and of companions. We also examine whether prominent
m=2 (i.e. grand design) spiral structure is necessarily associated with
a greater overall level of non-axisymmetric structure. We summarise
our main conclusions in Section 5.   
\section{Methods.}\label{methods}

This work makes use of the infrared array camera (IRAC: \cite{2004ApJS..154...10F}) on \textit{Spitzer}. The two shortest wavelengths, 3.6 and 4.5$\mu$m, can be thought of as tracers of the stellar mass distribution. Band 4, the 8$\mu$m data, is used as a tracer of the shocks induced in the gas. The IRAC data reduction is described in \cite{2004ApJS..154..204R} and the SINGS documentation. The SINGS sample also provides complementary data in other wavelengths; of particular use to this work are the optical bands (standard \textit{B}, \textit{V} and \textit{I} filters). In a few cases, SINGS optical data were not available or incomplete, in which case Sloan Digital Sky Survey (SDSS) \textit{g},\textit{r}, and \textit{z} band data were substituted instead.

In order to select the candidate galaxies, all spirals in the SINGS sample with inclinations $\le 70^{o}$, as determined by either \cite{2006MNRAS.367..469D}, \cite{2008AJ....136.2648D}, or \cite{2008MNRAS.385..553D} (D06, D08 or dB08), were included for analysis. If multiple studies covered a galaxy and the values lie either side of 70$^{o}$, the galaxy was included in the selection. Galaxies with high inclinations are harder to analyse for spiral structure, hence the imposed cut-off. In addition, again to aid the analysis, a size constraint was applied such that only galaxies with D$_{25}$ $\ge$ 5.0 arc minutes were included. Thus, the numbers in the sample are limited; of the 75 SINGS galaxies, 44 are spirals, and only 31 satisfy the inclination and size requirements.

The method used to extract and analyse the morphology is similar to that described for M81 in KKCT08. The process of extracting the spiral structure is not expected to be as straightforward in all cases (M81 was chosen because it is a relatively simple example). A full discussion for each galaxy is reserved for Section 3.1 
but some specific examples are noted in the following section.

\begin{figure}   \centering
  \includegraphics[width=83mm]{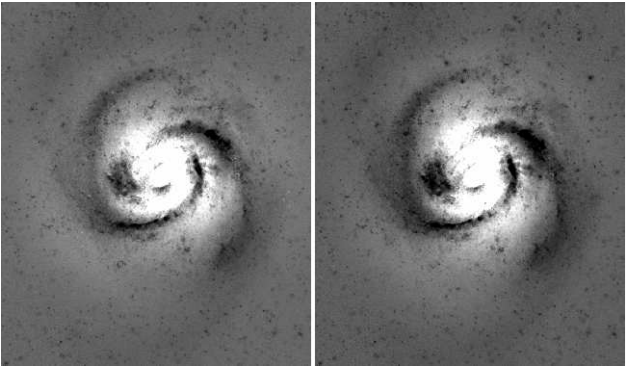}
  \caption{NGC 4321. Left; the kinematically constrained residuals from the kinematically constrained GALFIT fit (PA = --28.0 b/a = 0.89). Right; the residuals from the unconstrained fit (PA = --68.2 b/a = 0.87). This is a good demonstration that the kinematic constraints on the fit do not greatly affect the appearance of the residuals; the measured differences in amplitude and phase come from the differences in position angle and axis ratio (or ellipticity) of the sampling ellipses.}
  \label{N4321kin_phot_comp}
\end{figure}

\begin{figure}   \centering
  \includegraphics[width=60mm]{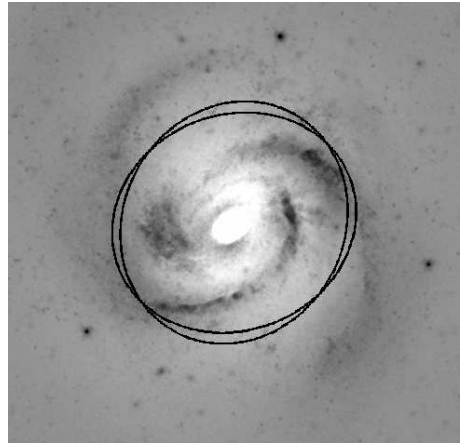}
  \caption{NGC 4321, showing the ellipses for kinematically constrained fits and the unconstrained (pure-photometric) fits produced by GALFIT, which have position angles of --28.0 and --68.2 respectively (angle increases counterclockwise from North, which is upwards in this figure). This illustrates how, even if the residuals are largely unaffected by the differences in fit, large variations in the azimuthal profiles are achieved because features in the residuals are transected at different angles and radii.} 
  \label{N4321ell_plot}
\end{figure}

The first step in extracting the non-axisymmetric structure for analysis is to fit the axisymmetric components of the galaxies using GALFIT \citep{2002AJ....124..266P}, a 2D fitting algorithm. This was achieved using a Sersic bulge, exponential disc, and tilted plane for the background. However, unlike the case of M81, initially the ellipticity (e, = 1-b/a where b/a is the axis ratio of the ellipse) and position angle (PA) of the disc were not allowed to vary freely. Instead, the values were constrained to lie within limits imposed by kinematic data from D06, D08 or dB08. The model galaxies were then subtracted from the data in order to produce residual mass surface density maps. However, on examining the results it became apparent that it is not always appropriate to use the kinematic values to constrain the fits. As a result all galaxies were also fitted with the disc parameters unconstrained. The results from the two fitting processes were then compared to decide which option is better for each galaxy. As can be seen in Figure \ref{N4321kin_phot_comp}, the differences in the appearance of the residuals between the two fitting methods are often negligible. In these cases, differences in results come largely from the shift in the sampling ellipses, as can be seen in Figure \ref{N4321ell_plot}.

In other cases, particularly for more inclined galaxies, the appearance of the residual image can be strongly affected by the different fitting parameters. For example, the kinematic ellipticity constraints worsened the fit of NGC 2841, and introduced an artificial m=2 signal into the residuals. In the majority of cases the photometric fit was preferable, as the kinematic fit to the phase was clearly unphysical.

\begin{figure}   \centering
  \includegraphics[width=83mm]{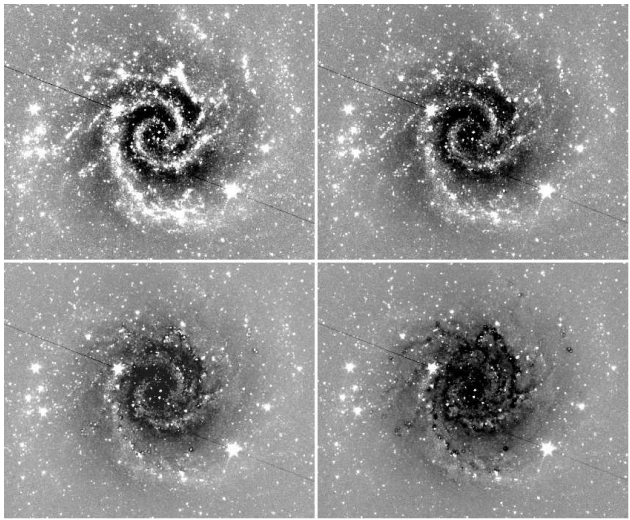}
  \caption{PAH correction; example of PAH corrections for NGC 0628, using the 3.6 $\mu$m image. Clockwise from top left are PAH corrections of 0.03, 0.05, 0.09 and 0.07. This clearly shows the effects of over-correcting for the PAH emission in the bottom two panels. For reference, 0.08 was used for M81, compared with the best-fitting correction of 0.05 for NGC 0628.}
  \label{PAH03_05_07_09}
\end{figure}

The residual images tended to have significant small-scale structure, believed to be predominantly due to polycyclic aromatic hydrocarbon (PAH) features, with some contributions from young stars. PAHs (polycyclic aromatic hydrocarbons) are large molecules which absorb starlight and re-emit in the near-to-mid IR. PAH emission is located at the boundary between HII regions and so called photo-dissociation regions \citep{2009ApJ...699.1125R} that border the HII regions. In addition to a continuum, PAHs have well defined peaks in emission. One such emission feature at 3.3$\mu$m \citep{1991ApJ...380..452T, 1981MNRAS.196..269D} falls within the IRAC 3.6$\mu$m bandwidth, and so this is the most likely cause of much of the contamination on small scales. PAH emission is also found in the 8$\mu$m waveband, and this was used to reduce the contamination in the 3.6$\mu$m data by subtracting a scaled version of the 8$\mu$m data. The exact method is as follows: the 8$\mu$m data were corrected for the stellar continuum emission by the subtraction of a scaled version of the 3.6$\mu$m model galaxy created by GALFIT. The scaling constant used was 0.232, as given by \cite{2004ApJS..154..253H}. It should be noted that, in using the model, rather than the original 3.6$\mu$m image, to subtract the stellar continuum from the 8$\mu$m data, the variation in continuum contribution over the spiral arms is not taken into account. However, the original image has been shown to contain PAH emission; by using the model the risk of affecting the PAH contribution to the 8$\mu$m emission is eliminated. The effect of this approximation (after all corrections) is that the 3.6$\mu$m flux will be slightly lower on the spiral arms than if the full 3.6$\mu$m image been used to remove the continuum, but by less than a factor of $\sim$0.05. This small change is unlikely to be noticeable above the noise in the data. The systematic effect on the phase (if noticeable) will be to reduce the offset measured between the 8$\mu$m peaks and the density maximum in the stellar spiral wave by slightly increasing the amplitude of the 8$\mu$m feature in phase with the stellar spiral. The 4.5$\mu$m data also appear to have PAH contamination, although with a smaller correction needed (for example, in KKCT08, the 4.5$\mu$m data required a factor of 0.05 of the 8$\mu$m continuum-corrected image, as compared to 0.08 for 3.6$\mu$m). This is consistent with measured ratios of 4.5/8$\mu$m flux in \cite{2006A&A...453..969F}, who found ratios in the range 0.037-0.065, with an average of 0.048. For some galaxies in this sample the small scale contamination was significant, and removing these features was not always as straightforward as reported in KKCT08 for M81. For example, on examining the galaxies during the PAH removal process it was obvious that the spatial scales and locations of fine structure features at 3.6$\mu$m (or 4.5$\mu$m) do not always exactly match the 8$\mu$m PAH features. This does not appear to be a result of resolution differences, nor can the problem of residual substructure be fixed by a larger PAH correction factor: a demonstration of the effects of increasing the PAH correction factor can be seen in Figure \ref{PAH03_05_07_09}, where four different PAH correction factors are used, and for larger values the 8$\mu$m arms begin to show up clearly as negative features, even though some regions of positive small-scale substructure remain. The conclusion that can be drawn is that some of the small scale structure is probably direct emission from young stars in star forming regions, not PAHs. This is unfortunate but not as serious as might be imagined, since these features are much smaller than the underlying spiral structure that is the goal of this analysis. 

\begin{figure}   \centering
  \includegraphics[width=83mm]{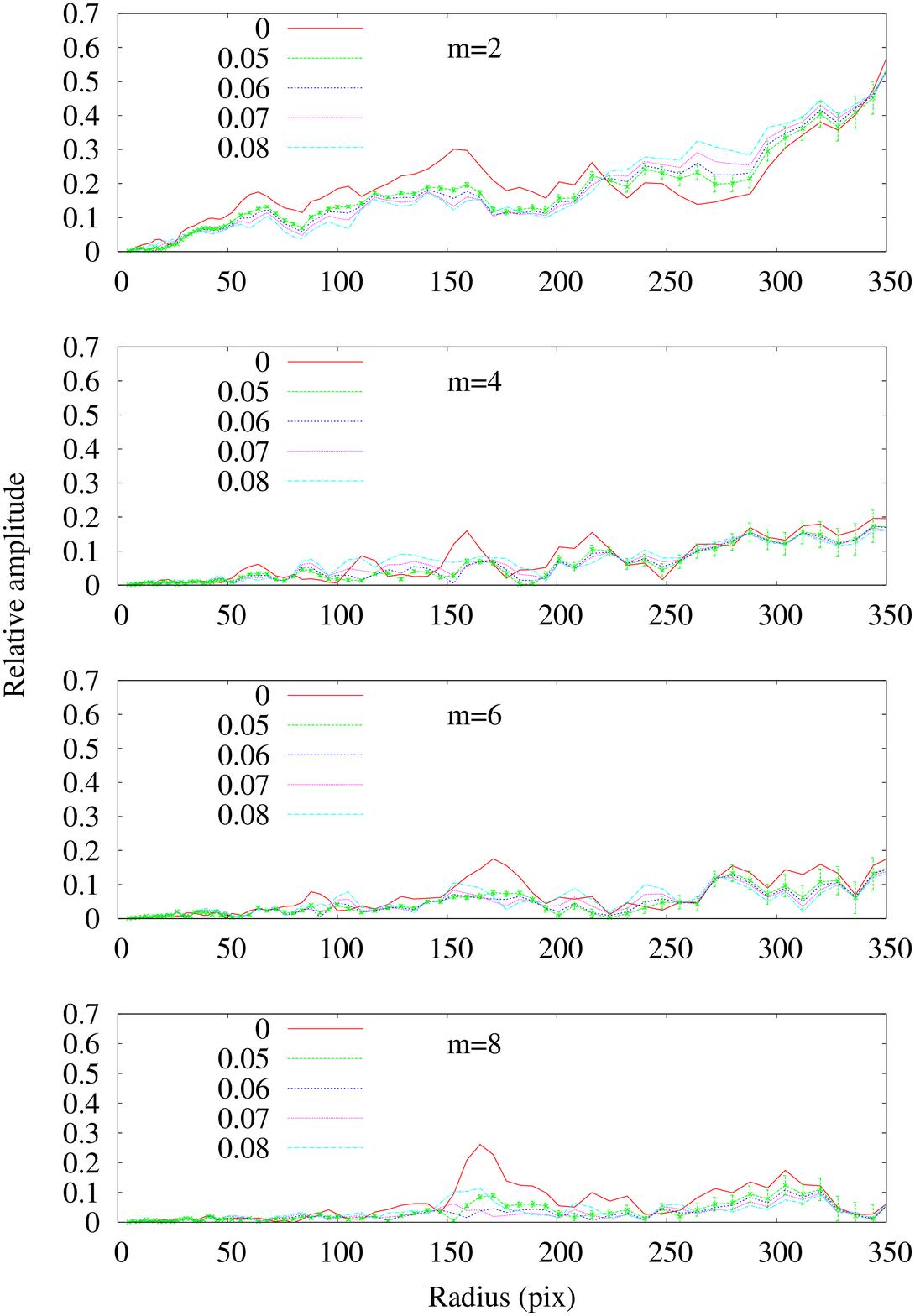}
  \caption{Relative amplitudes of different components of the CURVEFIT fit for NGC 0628 as a function of PAH correction factor (0.0, 0.05, 0.06, 0.07 and 0.08). The Fourier components plotted are m= 2, 4, 6 and 8. The error bars show 1 sigma errors due to random noise.} 
  \label{N0628_relampplot}
\end{figure}

From Figure \ref{N0628_relampplot}, it can be seen that the average fractional difference between the 0.05 and 0.07 PAH corrections for m=2 decreases from approximately 0.2 around R=50 pixels to ~0.05 around R=350 pixels. The difference between the 0.05 and 0.06 PAH corrections is about half this. Thus, although the PAH correction cannot fully remove the small scale structure in the 3.6 and 4.5 micron images (probably due to some of this emission being due to OB associations or RSGs), the measured amplitudes of the azimuthal profiles are relatively insensitive to the exact PAH correction used; a significant change in PAH correction of several tens of percent has only a moderate effect on the measured amplitude of the spiral structure (even in a late type spiral such as NGC 0628). For a direct comparison of the errors associated with a ten per cent change in the PAH correction and the random noise in the profiles (as well as the discrepancies between the amplitudes measured from different wavelength data) the figures in Section \ref{results} should be examined.

\begin{figure}   \centering
  \includegraphics[width=83mm]{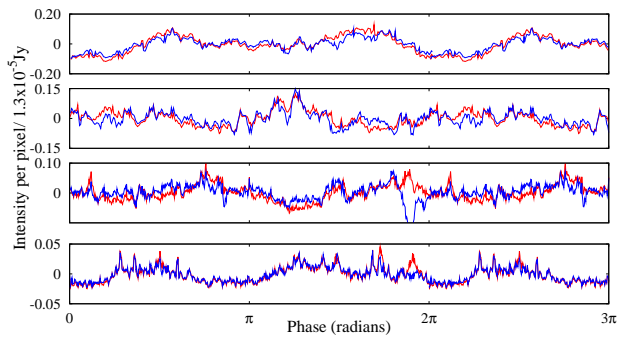}
  \caption{Azimuthal profiles corresponding to the PAH corrections of 0.05 (red) and 0.08 (blue) at radii of (top to bottom) 42, 72, 115, 222 arcsec.}
  \label{PAHcorr_compplot}
\end{figure}

\begin{figure}   \centering
  \includegraphics[width=83mm]{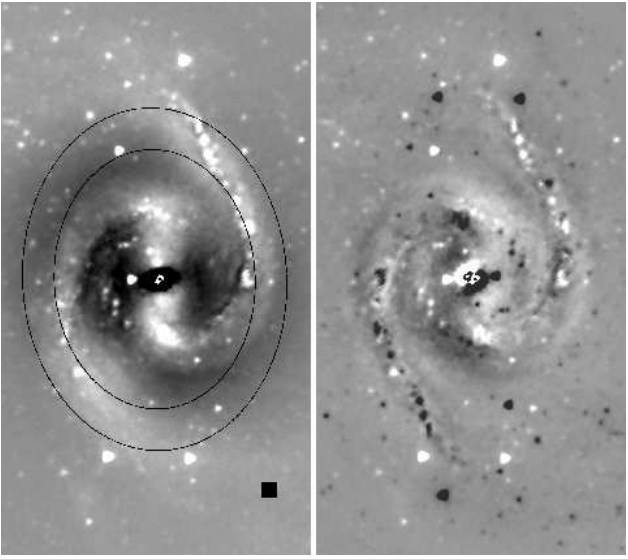}
  \caption{NGC 1566 and right, the image created by subtracting the galaxy rotated by 180$^{o}$ from itself. A square 10x10 pixels is shown in the bottom right of the left-hand frame, and two ellipses with SMA $\sim$80 and 110 pixels are marked.}
  \label{NGC1566_diff}
\end{figure}

\begin{figure}   \centering
  \includegraphics[width=83mm]{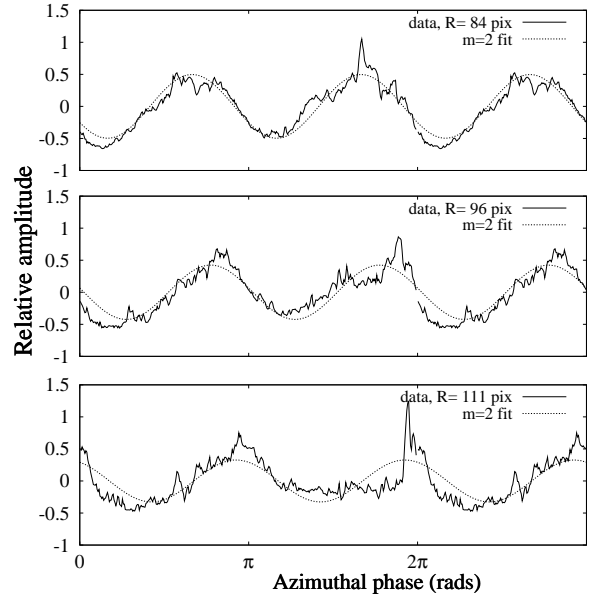}
  \caption{Azimuthal profiles for photometric fits to NGC 1566 at radii 84,96 and 111 pixels. As can be seen, the sharp features on the spiral arms, particularly the arm that peaks around $\frac{3}{2}$$\pi$-2$\pi$ ($\sim$2 o'  clock in Fig. \ref{NGC1566_diff}) do not influence the m=2 fits.}
  \label{NGC1566_prof}
\end{figure}

In the cases where the PAH correction clearly does not fully remove all small scale structure in the residual images it is possible to examine the effects of the remaining fine structure: Figure \ref{NGC1566_diff} shows NGC 1566 and an image which illustrates the degree of m=2 symmetry in the galaxy. Outside the bar region, the difference image shows the remnant PAH/young star features clearly, but almost no discernible contrast between the broader m=2 arms. Studying the small regions of contamination in the spiral arms shows that the flux can be more than three times as much in the areas of contamination than in the regions 180$^{o}$ apart in the opposite arm. However, when the size of the regions in question is taken into account, it can be seen that the effect on the profile fitting of the lowest Fourier components should be negligible; in the bottom right of the left-hand panel of Figure \ref{NGC1566_diff}, a square 10x10 pixels is shown, which is of a similar size to the features in question. At a radius of $\sim$80 pixels, a feature 10 pixels wide corresponds to less than five per cent of the m=2 component azimuthal wavelength. As can be seen in Figure \ref{NGC1566_prof}, the sharp features due to this contamination do not cause the amplitude of the m=2 Fourier component to be over-estimated.

Many of the galaxies in this sample have large foreground stars overlying their discs, which need to be removed before the data can be analysed. In the case of M81 these stars were small enough to be corrected by hand using the IRAF task IMEDIT, but for other galaxies the stars are more significant and the PSF wings are non-negligible. For this process, the DAOPHOT package within IRAF was used to remove the largest stars.

As has been illustrated, despite a non-perfect PAH removal process the residual images are still suitable for analysing the spiral morphology, albeit with a certain degree of caution in some cases. A final step used XZAP, modified as in KKCT08 to use MEDIAN rather than FMEDIAN, to remove small cosmic-ray-like regions after the PAH corrections and PSF subtractions were applied.

\subsection{Extracting the spiral structure.}\label{ch2_extract}

Following the steps described above, the residual images can be used as tracers of the non-axisymmetric mass surface density. From these non-axisymmetric mass maps, the spiral structure may be extracted and analysed. First, the IRAF task ELLIPSE was used to extract intensity as a function of azimuth for a range of radii. In order to extract the variation of intensity with azimuth for the non-axisymmetric mass surface density maps the model axisymmetric components were used to fit ellipticity and position angle as a function of radius. These input values were used to produce a series of azimuthal profiles for the residual images, and the data were combined into radial bins to give reasonably even coverage in log(R). The azimuthal profiles were analysed to obtain the strengths of all the Fourier components. To achieve this the data were re-binned to 1$^{o}$ azimuthal bins to reduce noise effects and give a uniform angular separation for Fourier analysis. The PYTHON fast Fourier transform (FFT) NUMPY.FFT was used to calculate the Fourier components of each profile in turn.

\begin{equation}
profile = \Re(\sum_{m=0}^{m=8}a_me^{i(mx + \phi_m)})
\label{ch2_eq2}
\end{equation}

The CURVEFIT routine in IDL then took the parameters determined by the FFT fitting as initial guesses for $a_m$ and $\phi_m$ in equation \ref{ch2_eq2}. A constant background noise was included in the fit, providing an error estimate for each fitted parameter. In addition, in order to examine the errors in the fits caused by the remnant PAH features, profiles were calculated for $\pm$10 per cent of the PAH correction (the PAH errors cannot be incorporated into the fit in the same way as a random noise term since the PAH errors are correlated with the peaks in the 3.6$\mu$m intensity, which would produce an unwanted weighting of the profiles with azimuth).

\begin{figure}   \centering
  \includegraphics[width=83mm]{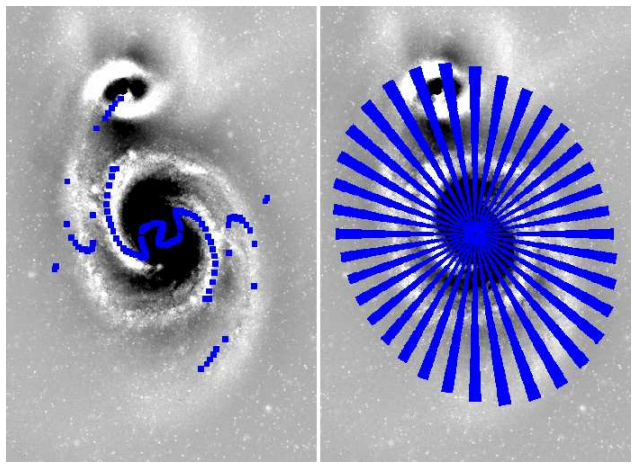}
  \caption{NGC 5194. (Left) The positions of the peaks of the m=2 Fourier component; as can be seen, the method of analysing azimuthal profiles to recover information about the spiral arms breaks down in some cases. (Right) Showing segments over which radial profiles were sampled in blue.}
  \label{NGC5194_radguide}
\end{figure}

Although the azimuthal profile method can be extremely successful in identifying and characterising spiral arms, the method fails if an arm is crossed more than once by a single isophotal ellipse (which can occur if the disc or spiral arms are warped), or if the arm runs almost parallel to the ellipse (which happens if the spiral arms are tightly wound and/or the galaxy is highly inclined). In these cases another approach is needed to determine the amplitude and phase of the arms: instead of taking azimuthal profiles the galaxy is sampled radially, producing plots of relative amplitude vs radius. The image is sampled every 10$^{o}$, in segments that are 4$^{o}$ wide, and with radial bins which are the same width as the radial steps between azimuthal profiles; an example may be seen in Figure \ref{NGC5194_radguide}. From these profiles it is possible to identify the maxima and minima in relative amplitude, and hence determine the phase and amplitude of the spiral arms.

\begin{figure}   \centering
  \includegraphics[width=83mm]{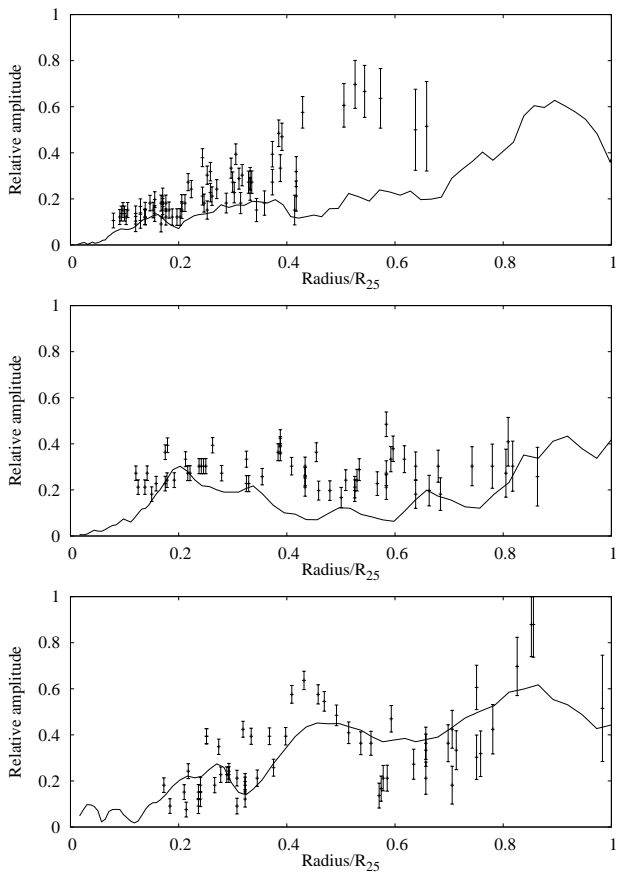}
  \caption{The relative amplitude data for NGC 0628 (top), NGC 3184 (middle), and NGC 4321 (bottom), comparing the data from radial profiles (points with error bars) and azimuthal profiles. In NGC 3184 the azimuthal profile method traces the arms accurately except between 0.5-0.7R$_{25}$, and in NGC 4321 the azimuthal profile method is in agreement only between radial limits of 0.5-0.75R$_{25}$.}
  \label{azi_rad_comp}
\end{figure}

It is worth noting that the radial profile method has some disadvantages when compared to the azimuthal profile method. In particular, it is harder to determine the amplitude of the wave because the amplitudes measured from radial profiles are more susceptible to bias from remnant PAH or young stellar emission. This is partly because the resolution of the azimuthal profiles is better, but more importantly the entire azimuthal profile is used when fitting each Fourier component, meaning that the fits are less susceptible to noise around the peak of the spiral arms. This is not the case in the radial profiles - only the regions around the maxima and minima are used; the phase should be determined from the radial mid-point of the wave (if measured trough-trough), and half the peak-trough height of the spiral gives the amplitude. These measures of spiral strength are chosen for maximum consistency with the amplitude measured from azimuthal profiles, but it should be noted that the quantities are not exactly equivalent (due to noise and radius effects). Further, because most of the noise comes from remnant PAH or young stellar emission which is located preferentially on the spiral arms, the maxima in the relative amplitude are more likely to be moved due to small-feature noise. However, difficulties in placing the minima in the relative amplitude can also affect the amplitude measurement, as well as the phase. Further, because the radial profile data are analysed by hand, the possibility of observer bias is possible; the residual images are used to guide the eye when identifying peaks in some particularly noisy profiles, and judgment calls are needed, for example to determine whether the highest point on the profile is likely to be affected by noise, or a genuine feature in the mass distribution. Finally, it is not possible to extract the Fourier components of the spiral arms from the analysis of the radial profiles. However, despite these drawbacks, the method offers the chance to obtain data that would otherwise be unavailable. Another advantage of the radial profile method, when compared to Fourier analysis, is that m=2 symmetry does not need to be assumed, unlike the analysis of Fourier m=2 components, and it is possible to analyse a non-symmetric two-armed spiral. A comparison of the data available from radial and azimuthal profile methods is presented in Figure \ref{azi_rad_comp}, in order to illustrate the differences. In the case of NGC 4321, the radial profile method simply shows more scatter, and the averages over the relevant radial range are approximately equal. But for the other two examples shown, NGC 0628 and NGC 3184, the radial profile data consistently overestimate the relative amplitude by a factor of $\sim$1.5 on average, and occasionally much more. For the two reasons outlined above - the lack of information on Fourier components and greater risk of overestimating the amplitudes - the radial profiles are only used to present data where the azimuthal profile method has failed. From Figure \ref{azi_rad_comp} it is clear though that the general trends in radius and amplitude seen in both methods are the same.

The majority of the analysis of these galaxies is carried out using the m=2 Fourier components to define the phase and amplitude of the spiral pattern. This is an important assumption, and worth discussing briefly, because spiral arms are not always (and possibly never) described by cos(2x) terms alone. The justification for this simplification is that m=2 is the dominant Fourier component in the majority of the spirals analysed here, and using the m=2 component alone reduces fine-structure contamination which increases in higher order components as finer angular scales are sampled. Finally, the use of m=2 alone allows for comparison with theoretical predictions which normally assume that spirals are two-armed, and also allows for fair inter-comparison between different galaxies. But, it is important to note that cos(2x) is not an accurate reflection of profile shapes in many cases - they often have sharper peaks, are not always perfectly symmetric about the maximum, and the cosine term imposes 180$^o$ rotational symmetry on the spiral arms. Despite these limitations, the azimuthal profiles are preferable, when the method is successful, due to the greater information available and reduced risk of contamination from fine scale structure.

\subsection{Optical data.}\label{ch2_BVRI}

Although it may initially seem surplus to requirements to include more than the SINGS NIR data in this survey, it became apparent that the data are not clean tracers of the stellar mass, and so it was decided to include other ways of measuring the stellar mass. The steps needed to obtain the stellar mass surface density and extract the spiral morphology from the optical data are described in the following section.

As with the NIR data, the colour-correction method described in KKCT08 could be applied to more galaxies. This colour correction is based in the work of \cite{2001ApJ...550..212B}, who use  stellar population synthesis models in combination with model galaxies to derive the M/L for any wavelength, and the colours of the galaxy models. From these datasets they were able to show that colour varies with M/L, and thus colour could be used to recreate the true, wavelength independent, M/L ratio for a given wavelength and colour combination.

With the exception of NGC 2403 and NGC 3198, the optical data come from the SINGS archive. The data source does not affect the steps applied; however, the calibration of some of the optical images from the SINGS archive may be unreliable (probably due to cloud cover during some observations - (Calzetti 2009, private communication)). For these purposes it is not a problem: because the images are combined in log space, all factors used to multiply images in linear space are effectively combined into a single additional constant. However, the uncertainties in calibration do mean that the zeropoints will not be physically meaningful, and so the final mass maps can only provide the relative (not absolute) mass surface density.

\begin{figure}   \centering
  \includegraphics[width=83mm]{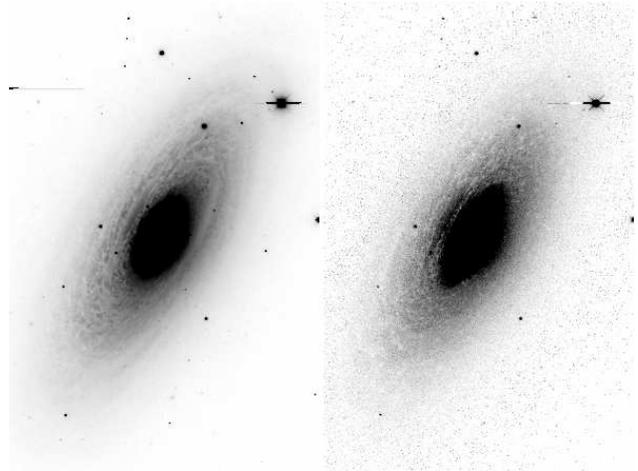}
  \caption{The I band (left) and colour corrected image (right) of NGC 2841; many dust lanes are still visible, probably due to those regions being optically thick.}
  \label{NGC2841_opt_dust}
\end{figure}

The details of the colour-correction method are largely the same as described for M81 in KKCT08, with the additional removal of foreground stars by the IRAF DAOPHOT task PSF subtraction. The conversion of background subtracted images to logarithmic scales and subsequent combination of images creates a very noisy sky. As soon as the galaxy becomes comparable to the background light levels, the signal is lost. The result is that the radial range over which the spiral structure can be traced is less than for the IRAC images (even if the spiral arms are above the noise threshold, the inter-arm regions may not be).
Another way in which the colour-corrected optical images may be flawed is due to the effects of dust. As noted by \cite{2001ApJ...550..212B} (BdJ), the effect of dust extinction is negligible to first order, since the effects of extinction from dust largely cancel out with the associated reddening. However, this is not true for optically thick regions. In a few galaxies there were significant numbers of dust lanes that appeared to be optically thick, and could not be corrected (for an example, see Figure \ref{NGC2841_opt_dust}).

For the galaxies with SDSS data, the (\textit{g-r}) colour was used in place of (\textit{B-V}) and \textit{z} was chosen as a substitute for \textit{I}. In this case, the \cite{2003ApJS..149..289B} prescription to retrieve the stellar mass is log$_{10}$(M/L$_z$) = a$_z$ + b$_z$(\textit{g-r}), with a$_z$ and b$_z$ taking values of --0.223 and 0.689 respectively. In other respects the process was identical to using \textit{B}, \textit{V} and \textit{I} band data.

After the colour-correction, the surface mass density maps were run through GALFIT, keeping the same ellipticity and position angle as for the IRAC data. The azimuthal profiles were extracted from the residual images in exactly the same way as described for the IRAC data.

\subsection{IRAC 8$\mu$m data.}\label{ch2_8mic}

\begin{figure}   \centering
  \includegraphics[width=83mm]{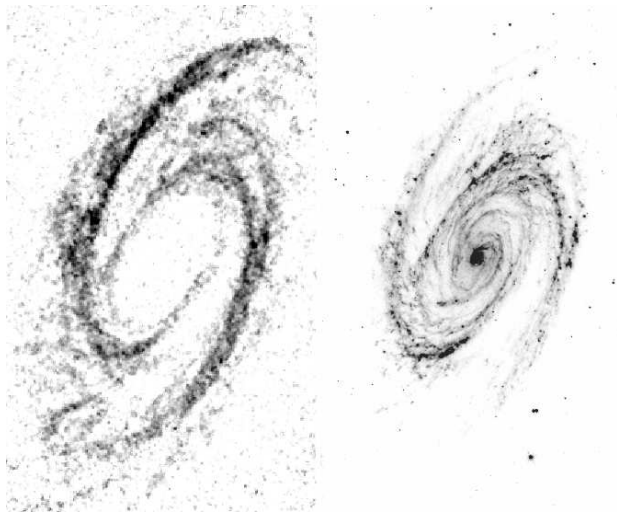}
  \caption{HI observation (known to trace shocked gas) on left, \textit{Spitzer} 8$\mu$m on right, showing the degree of agreement between the emission at the two wavelengths. Images are to the same scale. The HI observation was obtained from NED, originally published by \protect\cite{1995A&AS..114..409B}.}
  \label{figHIcomp}
\end{figure}

In addition to the stellar mass surface density, the response of the gas is of importance when considering the effects of spiral structure. The stellar density wave is expected to trigger a shock in the gas, and this can be detected through the 8$\mu$m because this wavelength largely traces dust emission. The HI 21cm line is a more conventional tracer of gas shocks, but the 8$\mu$m emission band has several advantages because any galaxy in the SINGS sample has high resolution 8$\mu$m data available. 
Emission at 8$\mu$m is dominated by dust, which tends to be concentrated in regions of high gas density at (or just behind) the shock front. To be visible in emission the dust needs to be heated, and the primary mechanism for concentrated emission is the switch-on of young stars triggered by the shock front (stars will drift downstream from the shock as they form). The link between the two wavelengths is highly plausible (see Figure \ref{figHIcomp}), although not all 8$\mu$m emission is from HI regions. The 8$\mu$m image is prepared for use by subtracting the stellar continuum emission, as described in KKCT08, but is otherwise unchanged.

\section{Results.}\label{results}

Not all of the galaxies in our sample of $31$  turned out to be suitable for further detailed analysis of their spiral structure. 
In what follows we refer to the $13$ galaxies
for which we were able to characterise the infrared grand-design spiral
structure in detail as the `detailed sample' and we provide a description
of the spiral parameters for each of these $13$ galaxies in Section 3.1.
These galaxies, together with their sizes, axis ratios and position
angles  are listed in
Table \ref{ch4_tab2}. 
There are a further $5$ galaxies which we judged by eye to be describable
as grand design in the near infrared but where we were unable to 
characterise their structure in detail. We briefly describe these
$5$ galaxies in Section 3.2 and refer to these galaxies collectively
as the `additional sample' . When we refer to the (NIR) grand design sample,
we refer to the combined sample of $18$ galaxies, comprising both the
detailed sample and the additional sample.  

  We classify the remaining $13/31$ galaxies in our sample as being
non-grand design (i.e. without a predominating m=2 mode) 
in the 
near infrared. These galaxies (together with the $5$ grand design galaxies
that we could not characterise in detail) are listed in Table \ref{ch4_tab4}, together
with their sizes and inclinations. We note that non-axisymmetric structure
is discernible in almost all the galaxies listed in Table \ref{ch4_tab4}: the only
possible exception is NGC 1291 and even then the lens-shaped ring (visible most clearly in the 8$\mu$m image ) may in fact be two very tightly wound spiral arms
(see Fig. 12). The galaxies in Table \ref{ch4_tab4} are a mixture of ringed structures,
and those where spiral structure is either of low amplitude or dominated
by modes other than m=2 
(see for example the archetypal
short flocculent
spiral structure in NGC 7793 in  Fig. 12).  This means that
we also exclude from grand design status 
such galaxies as NGC 4254 
which has 
clear three armed structure as well 
as those with a  strong $m=1 $ mode such as  NGC 1512 and NGC 4725 
which display predominantly
one-armed structure (although with evidence for a weak second arm
in both cases).   

  The galaxies in our total sample cover the entire range of Elmegreen
classes in terms of the optical classification of their spiral
structure (i.e. they include all Elmegreen classes apart from
10 and 11, whose use was discontinued by \cite{1987ApJ...314....3E}).
Their breakdown by Elmegreen class is as follows: 10 are in arm class 1-4, 14 in arm class 5-9, and 7 in arm class 12. Figure 12 illustrates a selection
of $3.6 \mu$m  infrared images of galaxies in our sample together with their
Elmegreen arm class in the caption. 

 On the whole there is a general
correlation between a galaxy being classified as grand design in the
optical and in the near infrared. This is demonstrated in Table \ref{armclass_tab}, where 
the optical structure is classified by Elmegreen class: 1-4 corresponds to galaxies with weak or chaotic spiral structure, class 12 to well defined grand design spiral structure and classes 5-9 to intermediate cases.
We find no galaxies that are optically grand design which do not
show grand design
structure in the NIR. This confirms that such structure is not merely manifest
in the pattern of recent star formation (which tends to dominate the optical
light) but represents  structure also in the underlying stellar mass
distribution. We do however find (in line with the findings of \cite{1991Natur.353...48B, 1996ApJ...469L..45T, 1999AJ....118.2618E}) that some optically flocculent galaxies in our sample show grand design structure in the near infrared.

Among the galaxies that are grand design in the NIR, we find a variety
of morphologies (although all, by definition, have a strong $m=2$ disturbance).
Some grand design galaxies are beautifully regular and symmetric, for example NGC 1566 or NGC 3031. However, this regularity is not seen across the whole 
sample: galaxies which appear to have distortions or asymmetries in their arms
 include NGC 5194 and  NGC 4321.  These  
spirals are largely symmetric under a 180$^o$ rotation, but the the arms do not
 always follow a smoothly varying shape; clear kinks are evident at one or more
 radii. In other cases, for example NGC 3627 or NGC 4254, the symmetry between the arms is broken (see discussion in Section 3.2) or else a  predominantly 
two armed spiral bifurcates and splits into sub-branches as in
NGC 1566  or NGC 3938.

\begin{figure*}  \centering  
 \vbox to220mm{\vfil
   \includegraphics[width=45mm]{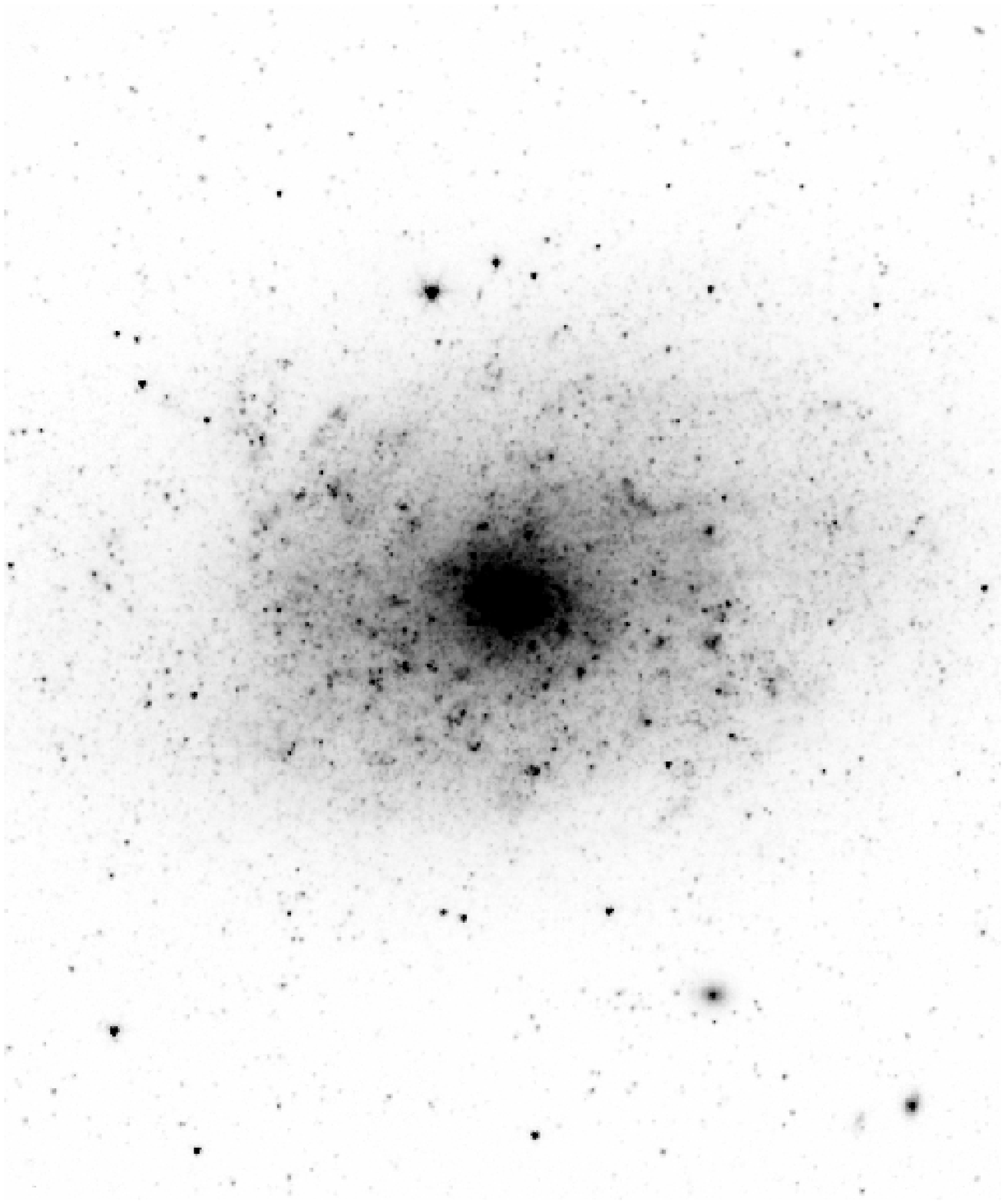}
   \includegraphics[width=45mm]{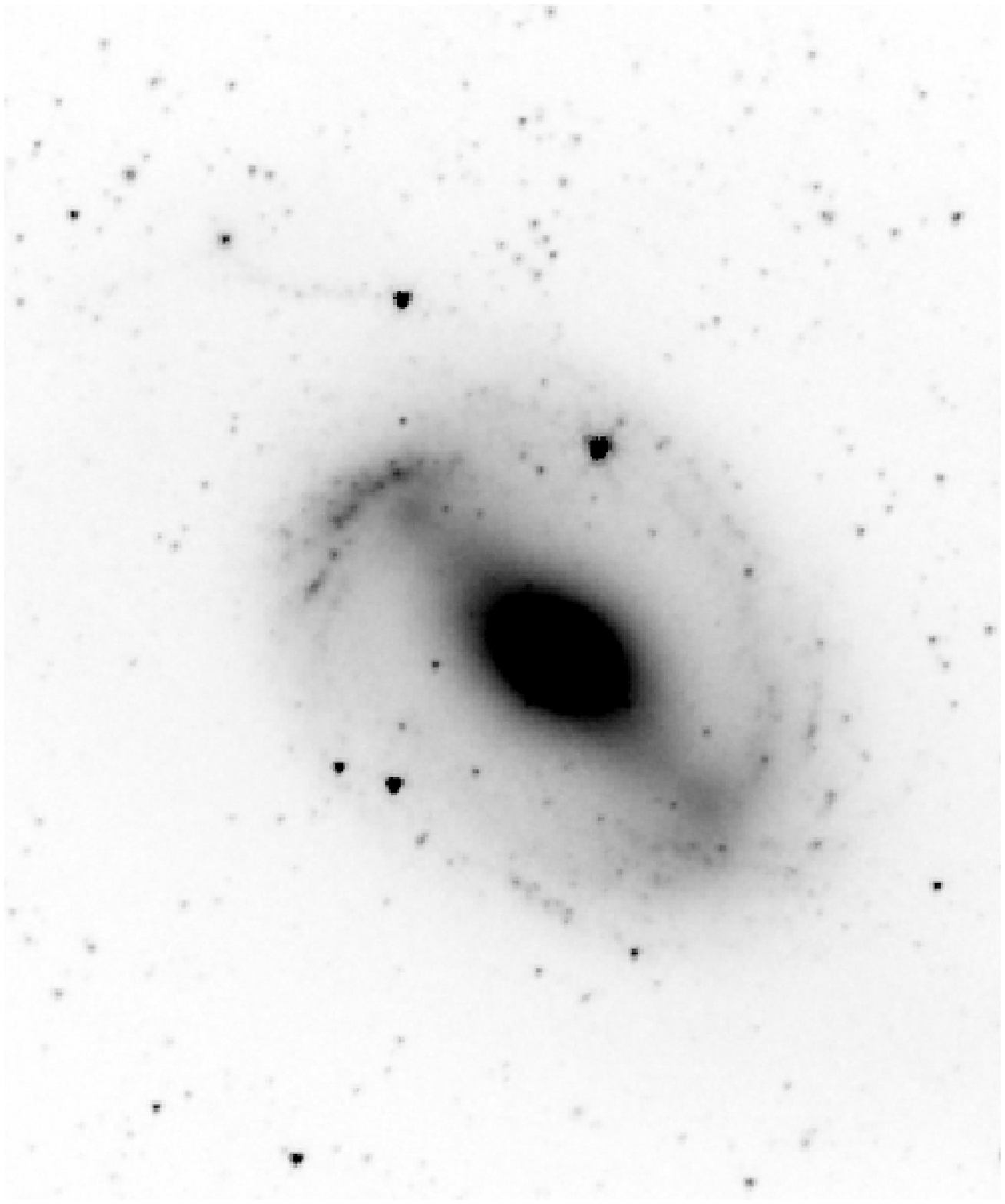}
   \includegraphics[width=45mm]{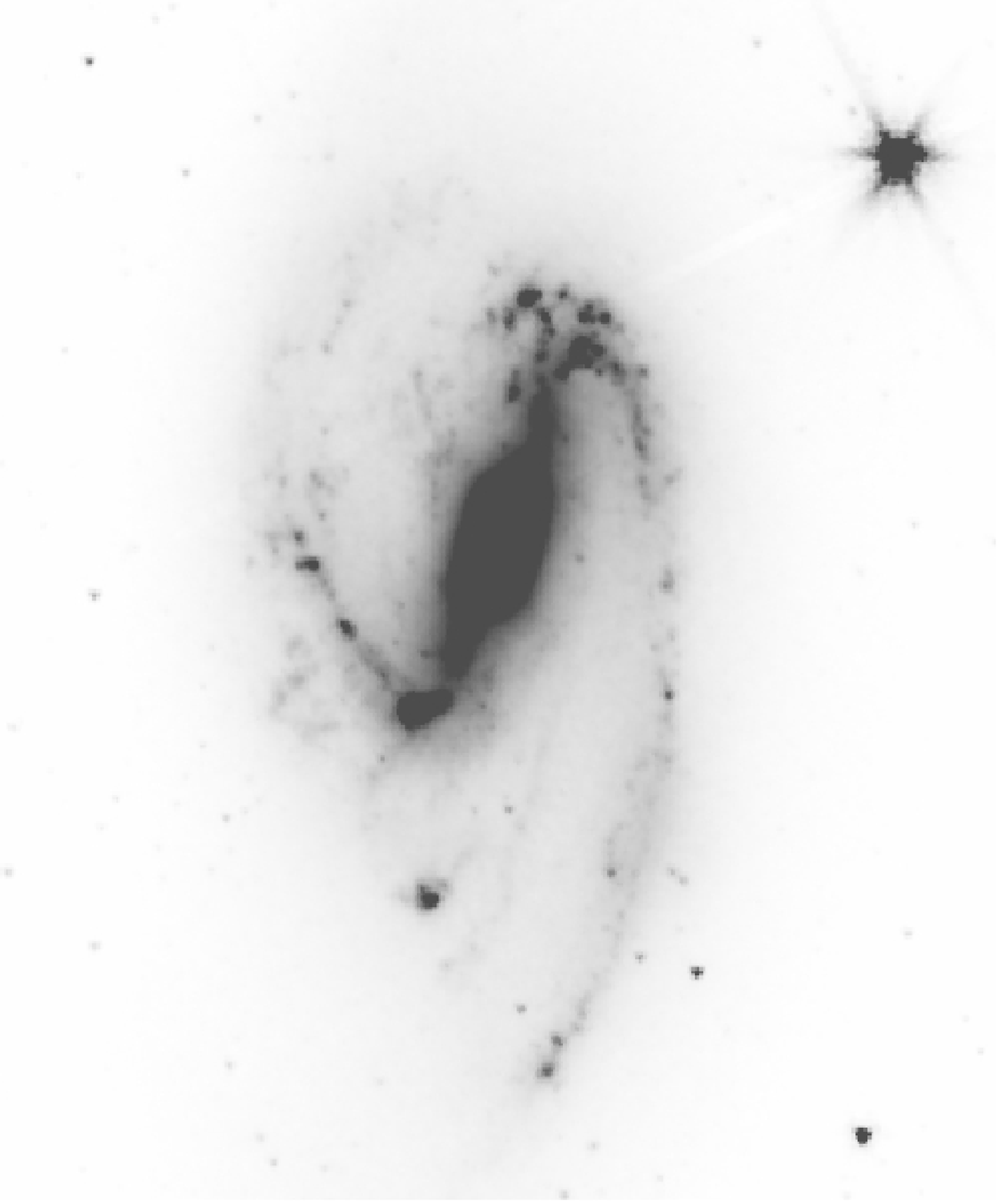}\\
   \includegraphics[width=45mm]{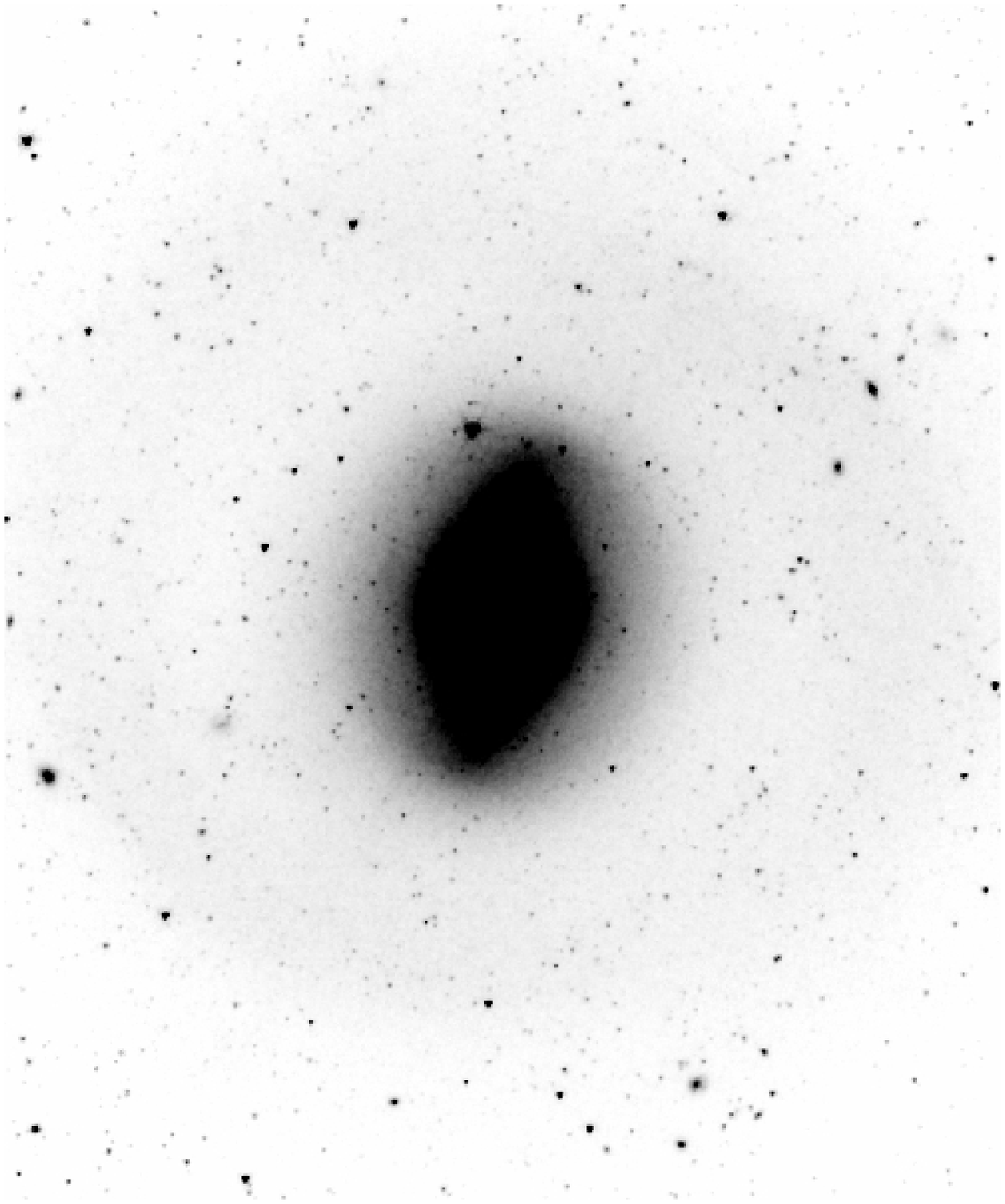}
   \includegraphics[width=45mm]{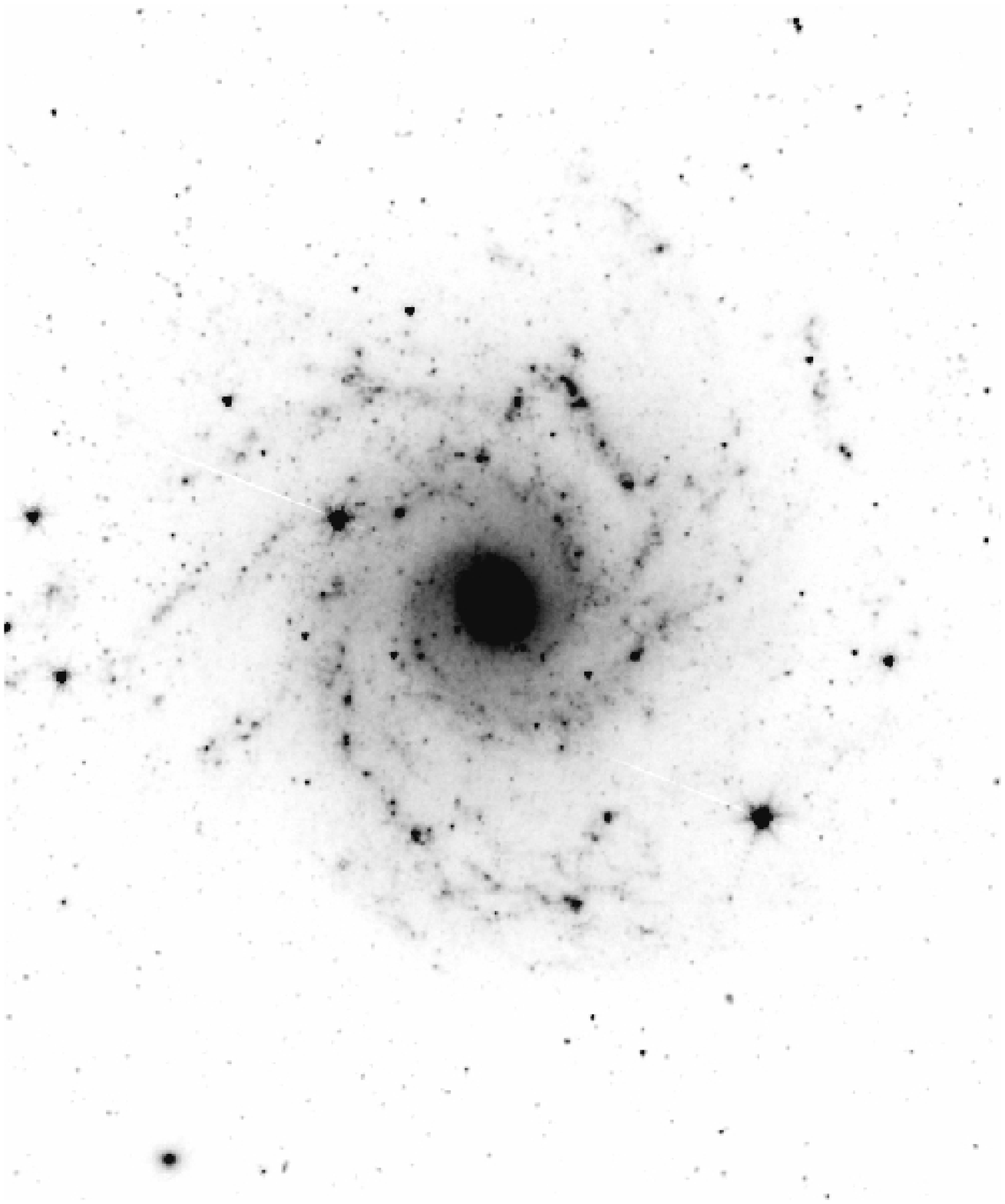}
   \includegraphics[width=45mm]{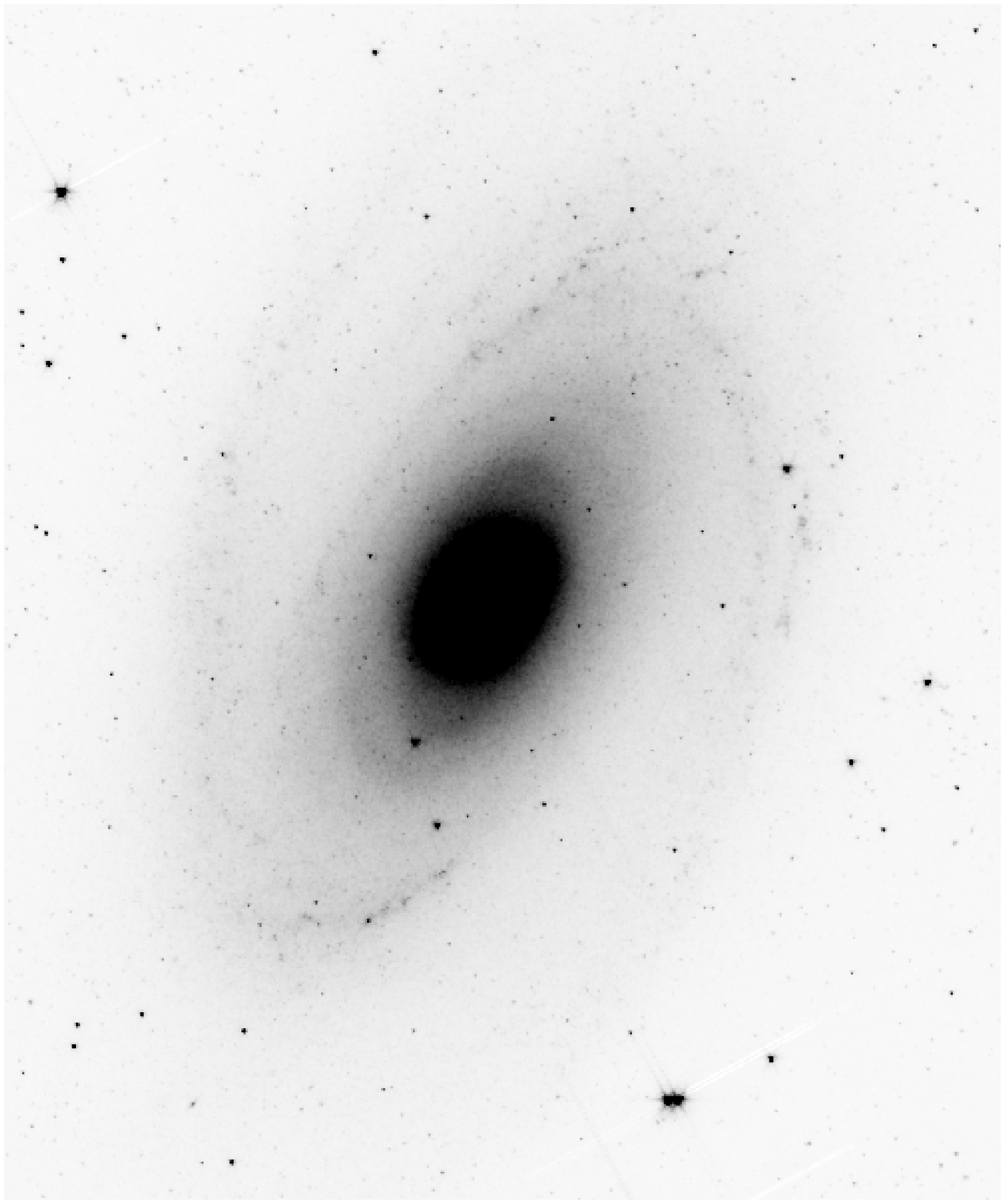}\\
   \includegraphics[width=45mm]{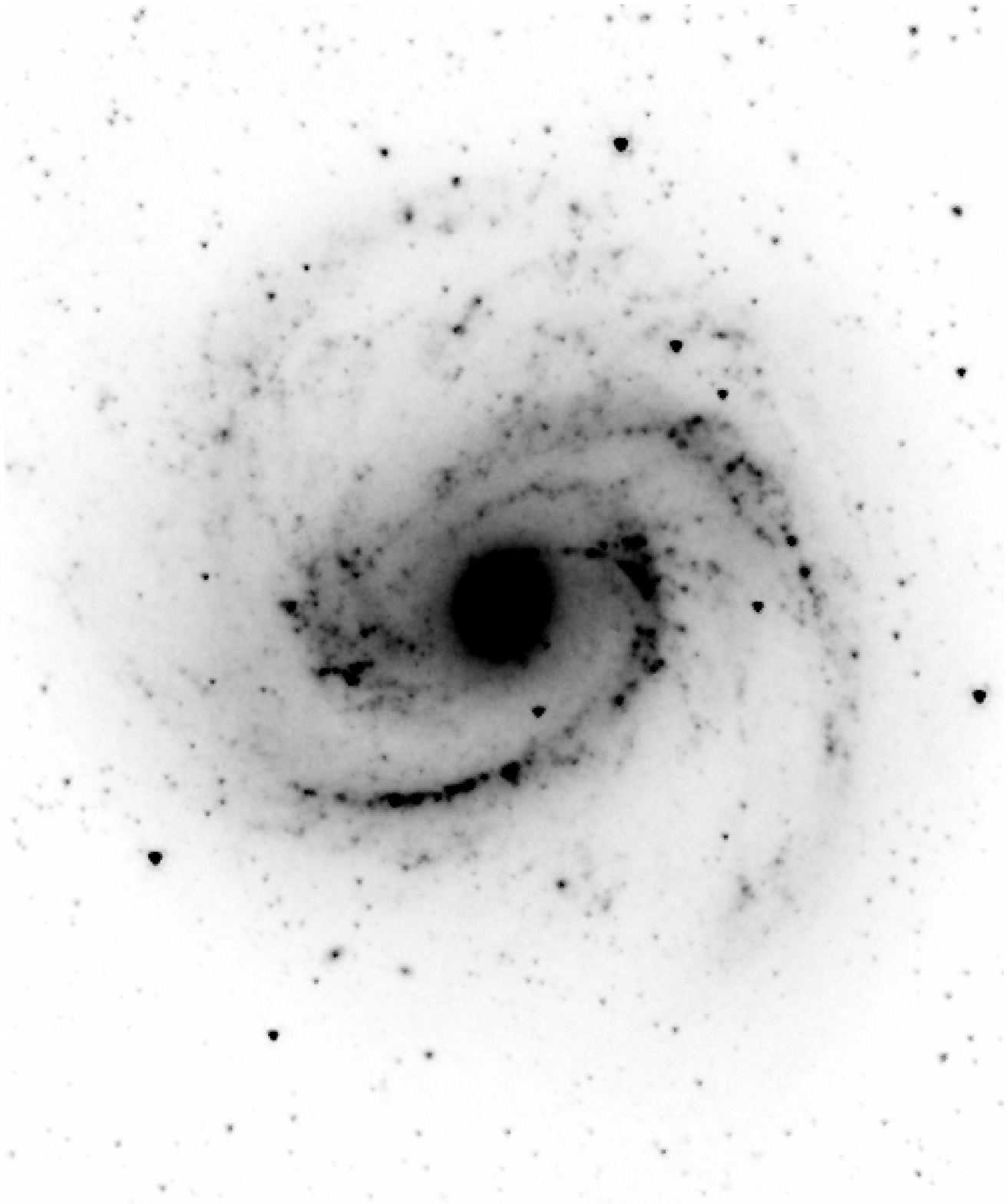}
   \includegraphics[width=45mm]{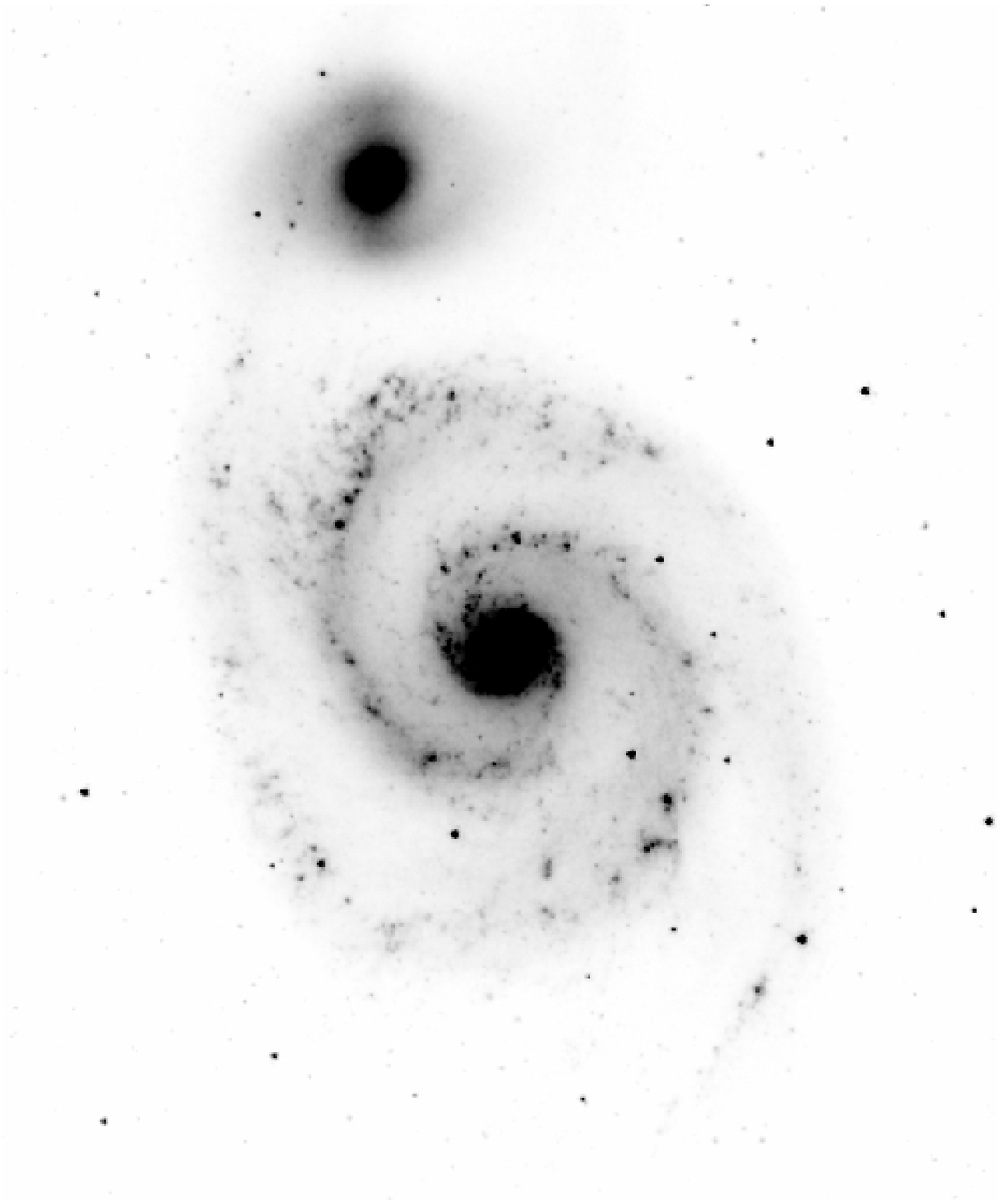}
   \includegraphics[width=45mm]{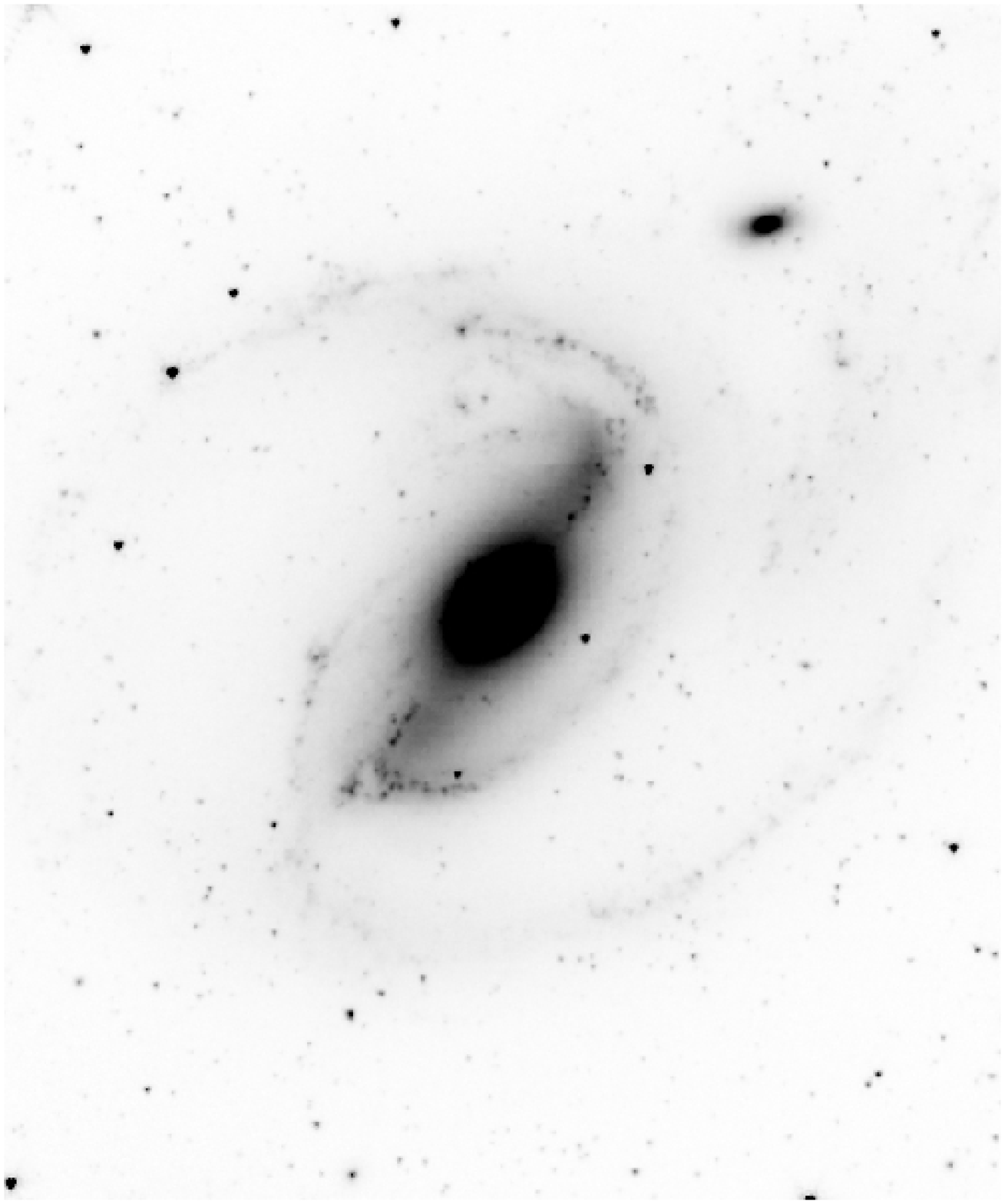}\\
   \includegraphics[width=45mm]{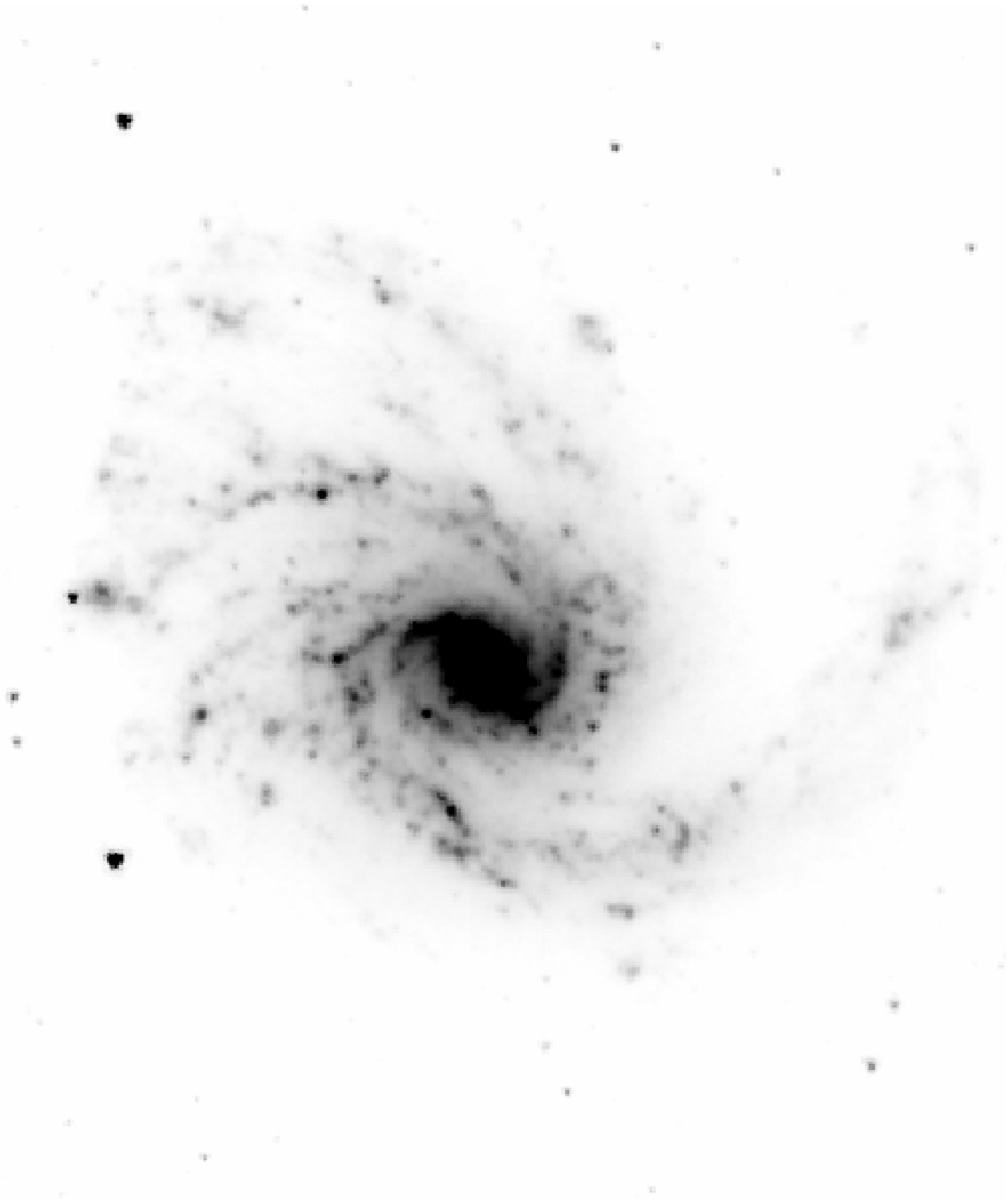}
   \includegraphics[width=45mm]{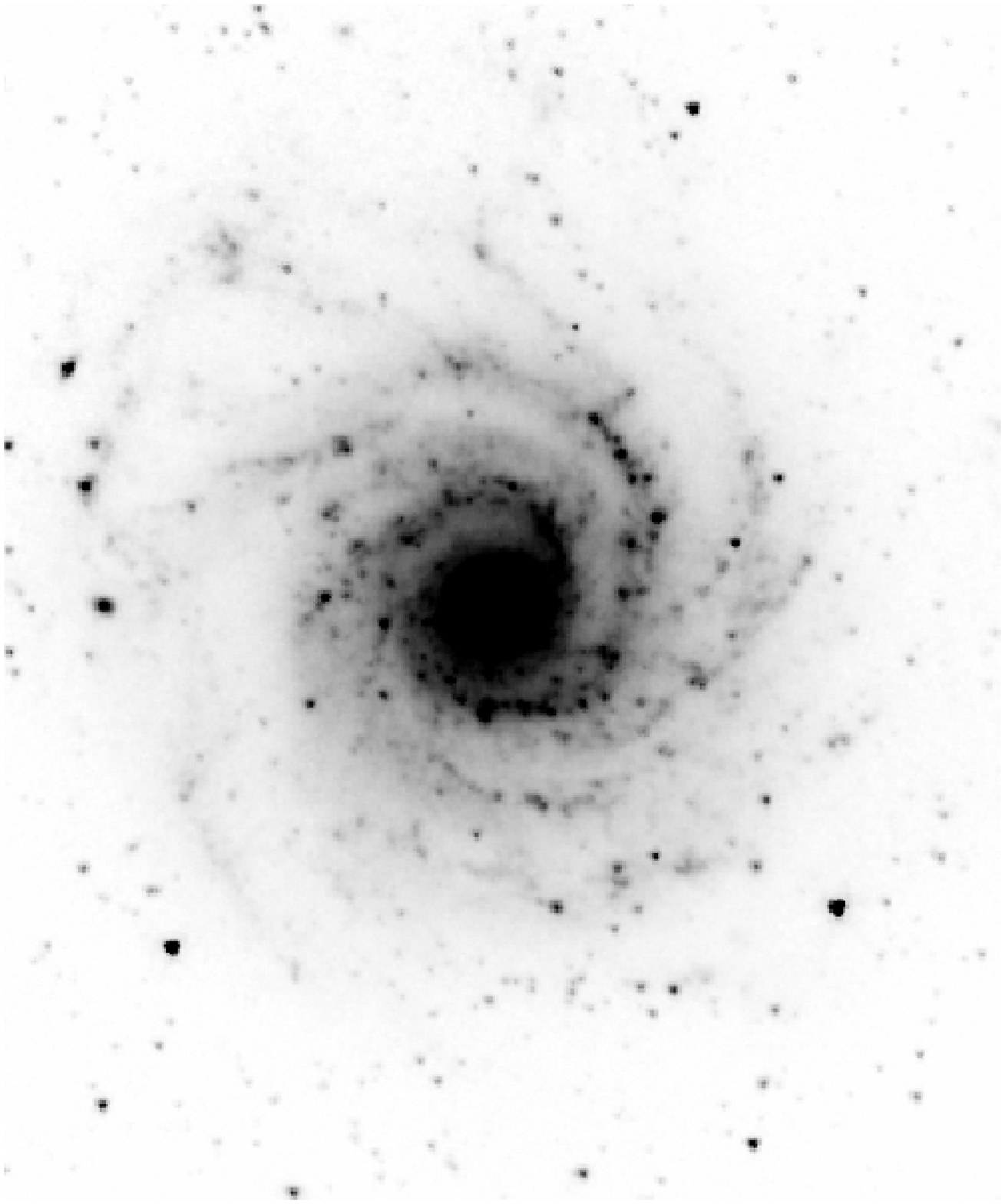}
   \includegraphics[width=45mm]{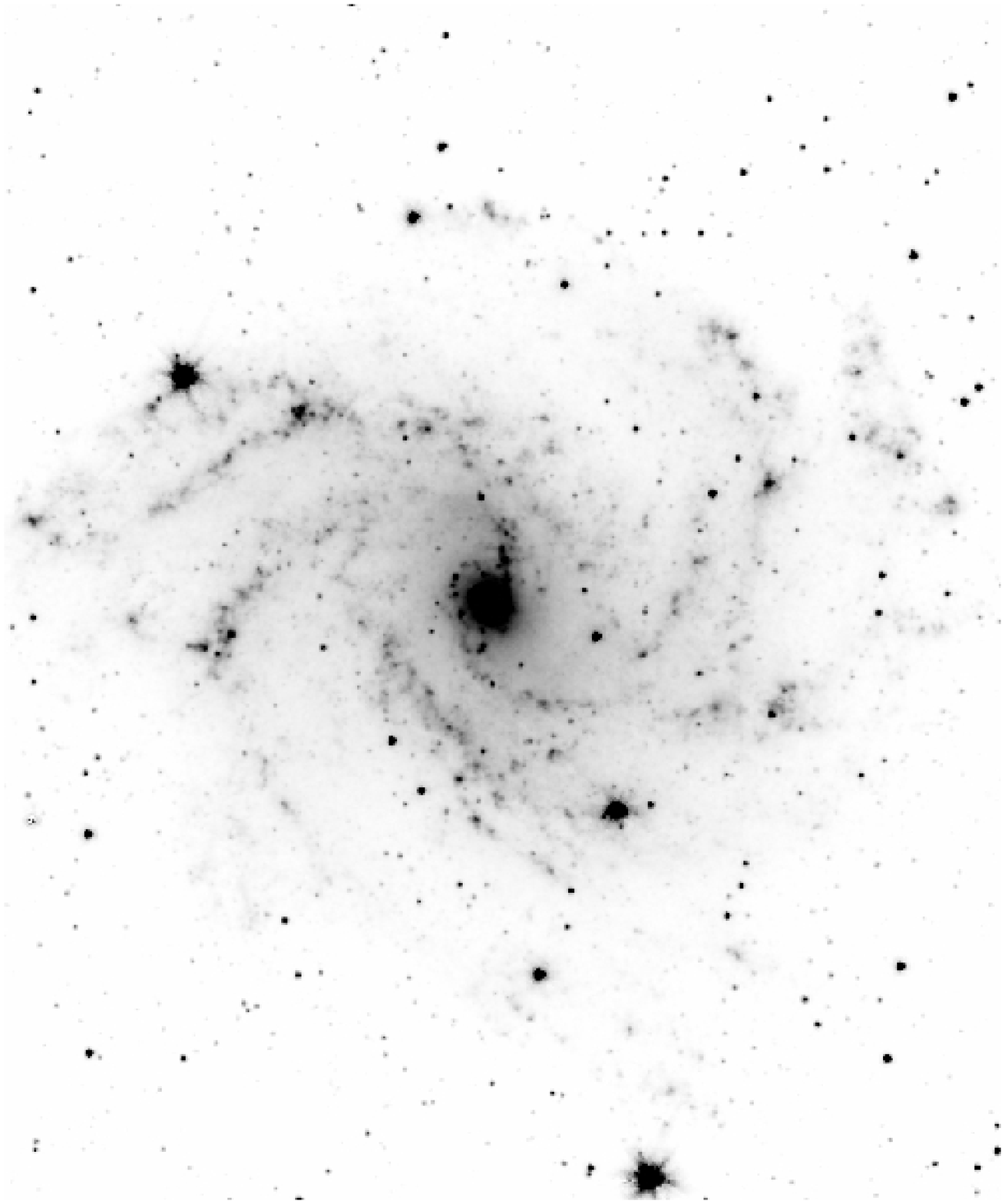}
   \caption{An illustration of the variety of galaxy morphologies  at 3.6 $\mu$m(Elmegreen arm class indicated in brackets). Top row, (l-r); NGC 7793 (2), NGC 1512 (6), NGC 3627 (7). Second row; NGC 1291 (8),  NGC 0628 (9), NGC 3031 (12). Third row; NGC 4321 (12), NGC 5194 (12), NGC 1097 (12). Bottom row; NGC 4254 (9), NGC 3938 (9), NGC 6946 (9).}
  \label{morph_overview}
 \vfil}
\end{figure*}

As expected given the range of morphology, the angular and radial extent of the stellar spiral arms varies hugely across this sample. The galaxies that have the most extended and continuous spiral arms in the stellar mass are NGC 0628 and NGC 5194, where the arms wind over angles in excess of 360$^o$ (nearly 540$^o$ in the case of NGC 0628). Even in the less extreme cases, many of the grand design galaxies have spiral arms which can be followed for 180-360$^o$. In contrast, the galaxies which are not found to have underlying continuous spirals in their stellar mass have patchy arm segments that are continuous for no more than a few tens of degrees.

 The 8$\mu$m images allow us also to study the  distribution of warm dust in
these 
galaxies, and hence the morphology of gas associated with regions of star
formation. We find that in general the gas is more filamentary
than the stellar distribution on  small scales and that clumpy gas is
associated with spurs and feathering (beautiful examples exist in NGC 3031
and NGC 5194).  There is a general tendency for galaxies that show
clear grand design structure in their stellar component to also
show well organised structures in warm dust, although in some cases
the inherent clumpiness of the $8 \mu$m maps makes the definition
of clear spirals problematical. Even in cases where both the
stellar distribution and the gas show clear grand design
structure, the features are not necessarily spatially coincident.
In particular we go on to analyse below whether there is
evidence for any systematic azimuthal offset between the crests
of the stellar distribution and the peaks in dust emission.

\begin{table}
  \centering
  \begin{tabular}{|l|l|l|l|}
    \hline
    &Class 1-4  &  Class 5-9 &  Class 12\\
    \hline
    NIR Grand Design (detailed) & 3 & 6 & 4 \\
    NIR Grand design (additional) & - & 2 & 3 \\
    NIR Non-grand design & 7 & 6&   -  \\
    \hline
  \end {tabular}
  \caption{ Breakdown of galaxies by optical (Elmegreen class) and NIR
classification (this paper). Increasing Elmegreen class denotes an
increasing degree of grand design structure in the optical. The additional
sample comprises galaxies that are deemed to be grand design  in the NIR but where
there are issues preventing their inclusion in the detailed sample 
(see section \ref{ch4_results_other} for further details).}
  \label{armclass_tab}
\end{table}

The final morphological factor that has not yet been addressed is the effect of bars. From examining this sample, the data are consistent with the usual link between bars and rings; rings are much more common in barred galaxies, but not ubiquitous (for example, NGC 1097 is strongly barred but not ringed). NGC 1097 is 
an example of a galaxy exhibiting both a bar and grand design spiral structure:
in fact it is the bar that prevents our being able to characterise
its spiral structure over a large radial range and causes us to exclude
this galaxy from our detailed sample 
- see further discussion of this galaxy in Section 3.2.
We discuss the correlation between bar status and grand design spiral
structure in Section 4. 
\begin{figure}   \centering
  \includegraphics[width=60mm]{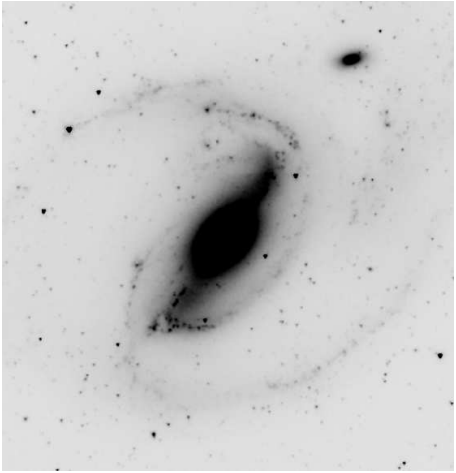}
  \caption{NGC 1097, showing the strong bar and tightly wrapped spiral arms.}
  \label{NGC1097}
\end{figure}

\subsection{Grand design spirals (detailed sample).}\label{ch4_results_granddesign}

Table \ref{ch4_tab2} shows the sub-sample of galaxies chosen for further detailed analysis of their spiral structure (hereafter the `detailed' sample). Here `grand design' is interpreted purely as those galaxies that have reasonably coherent and extensive spiral arms in the stellar mass distribution (rather than the more precise definition introduced by \cite{1982MNRAS.201.1021E}). In the majority of cases   this means  that m=2 spiral arms  dominate over other Fourier
components, although there are a few examples of galaxies in the detailed sample
for which power is more equitably spread among modes of different m.
A comparison of the results from each of the galaxies in this section will be carried out in Paper II.

\begin{table}
  \centering
  \begin{tabular}{|l|l|l|l|}
    \hline
    Galaxy & R$_{25}$ (arc min) & axis ratio (b/a) & PA (degrees)\\
    \hline
    NGC 0628 (M74) & 5.25 & 0.99 & 20 \\
    NGC 1566 & 4.15 & 0.77& 3 \\
    NGC 2403 & 10.95 & 0.41 & --58 \\
    NGC 2841 & 4.05 & 0.44 & --30 \\
    NGC 3031 (M81) & 13.45 & 0.52 & --28 \\
    NGC 3184 & 3.7 & 0.96 & --8 \\
    NGC 3198 & 4.25 & 0.35 & 39 \\
    NGC 3938 & 2.7 & 0.95 & 15 \\
    NGC 4321 (M100) & 3.7 & 0.86 & --28 \\
    NGC 4579 (M58) & 2.95 & 0.77 & 85 \\
    NGC 5194 (M51) & 5.6 & 0.76 & 22 \\
    NGC 6946 & 5.75 & 0.81 & 71 \\
    NGC 7793 & 4.65 & 0.66 & --81 \\
    \hline
  \end {tabular}
  \caption{Galaxies that have traceable spiral structure. R$_{25}$ from \protect\cite{2003PASP..115..928K}, the axis ratios and disc position angles are
the values from the GALFIT disc fits, whether constrained by the kinematic data, or allowed to vary in a fully photometric fit.}
  \label{ch4_tab2}
\end{table}

\subsubsection{NGC 0628}\label{ch4_0628}
NGC 0628 is classified as a SAc type galaxy, and is widely regarded as being an isolated field galaxy, although it has been suggested that NGC 0660 might have been an interacting companion some time in the past (possibly as long ago as 2x10$^{9}$ years \citep{1991IAUS..146..113E}). NGC 0628 has an extended HI disc.

\begin{figure}   \centering
  \includegraphics[width=60mm]{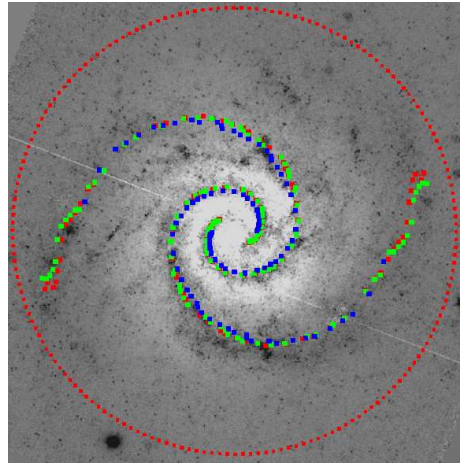}
  \includegraphics[width=60mm]{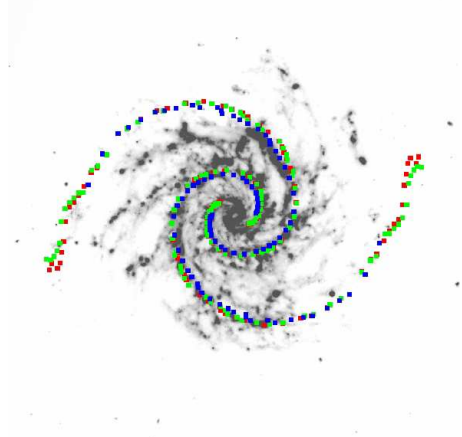}
  \caption{NGC 0628, showing the IRAC 3.6$\mu$m residual image above. The ellipse in red marks R$_{25}$ (5.25 arc minutes). The points in red, green and blue show the positions of the maxima of the m=2 Fourier components for the IRAC 3.6, 4.5 $\mu$m and \textit{I} band colour corrected phases respectively. The data are only plotted over the radial range for which a logarithmic spiral can be traced (the same range is used to calculate the pitch angle). Below is the 8 $\mu$m image, again showing the positions of the maxima of the m=2 Fourier components
of the 3.6 $\mu$m image. All images presented have standard alignment, with North up and East to the left of the page.}
  \label{n0628_resids}
\end{figure}

\begin{figure}   \centering
  \includegraphics[width=60mm]{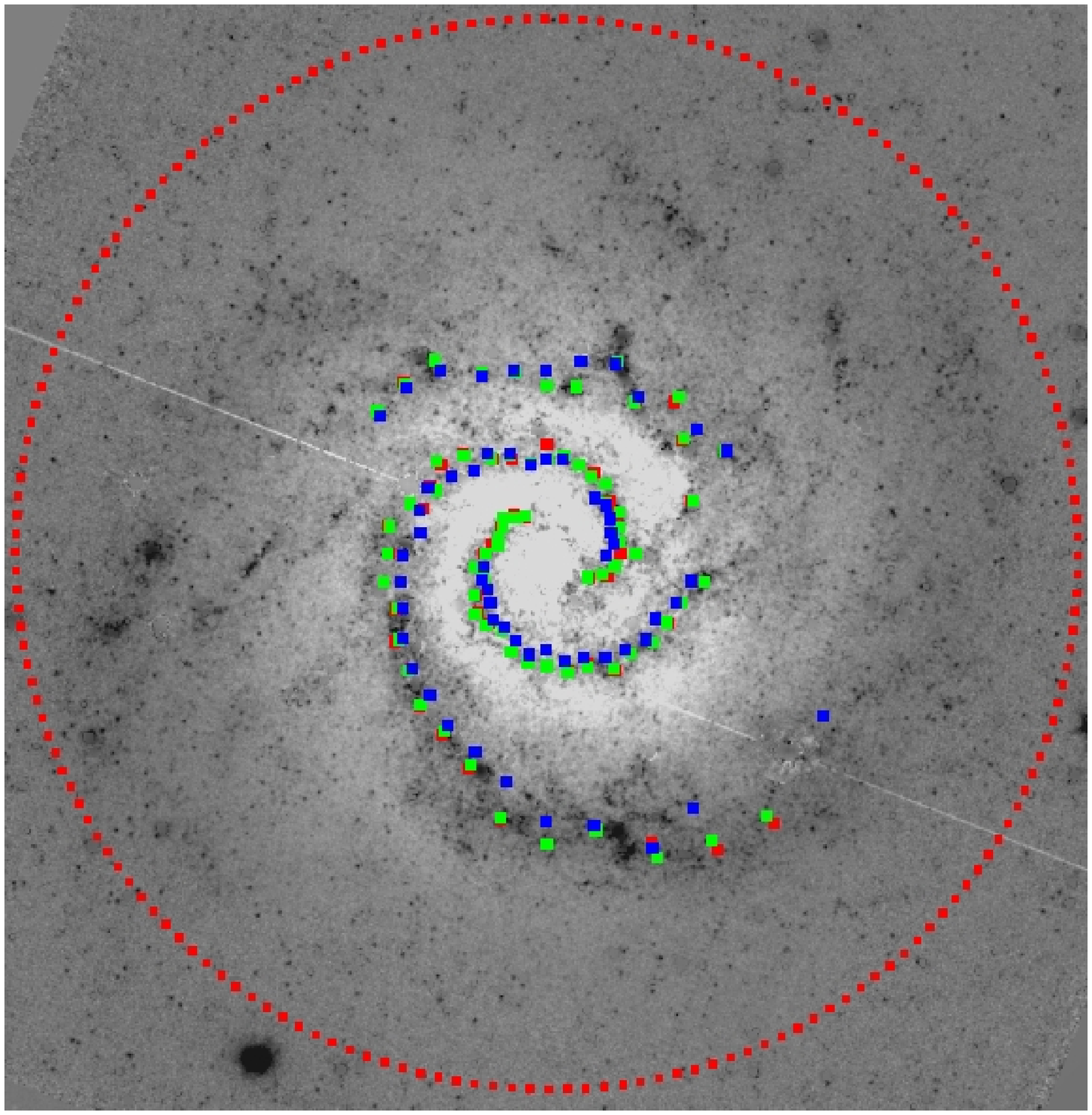}
  \includegraphics[width=60mm]{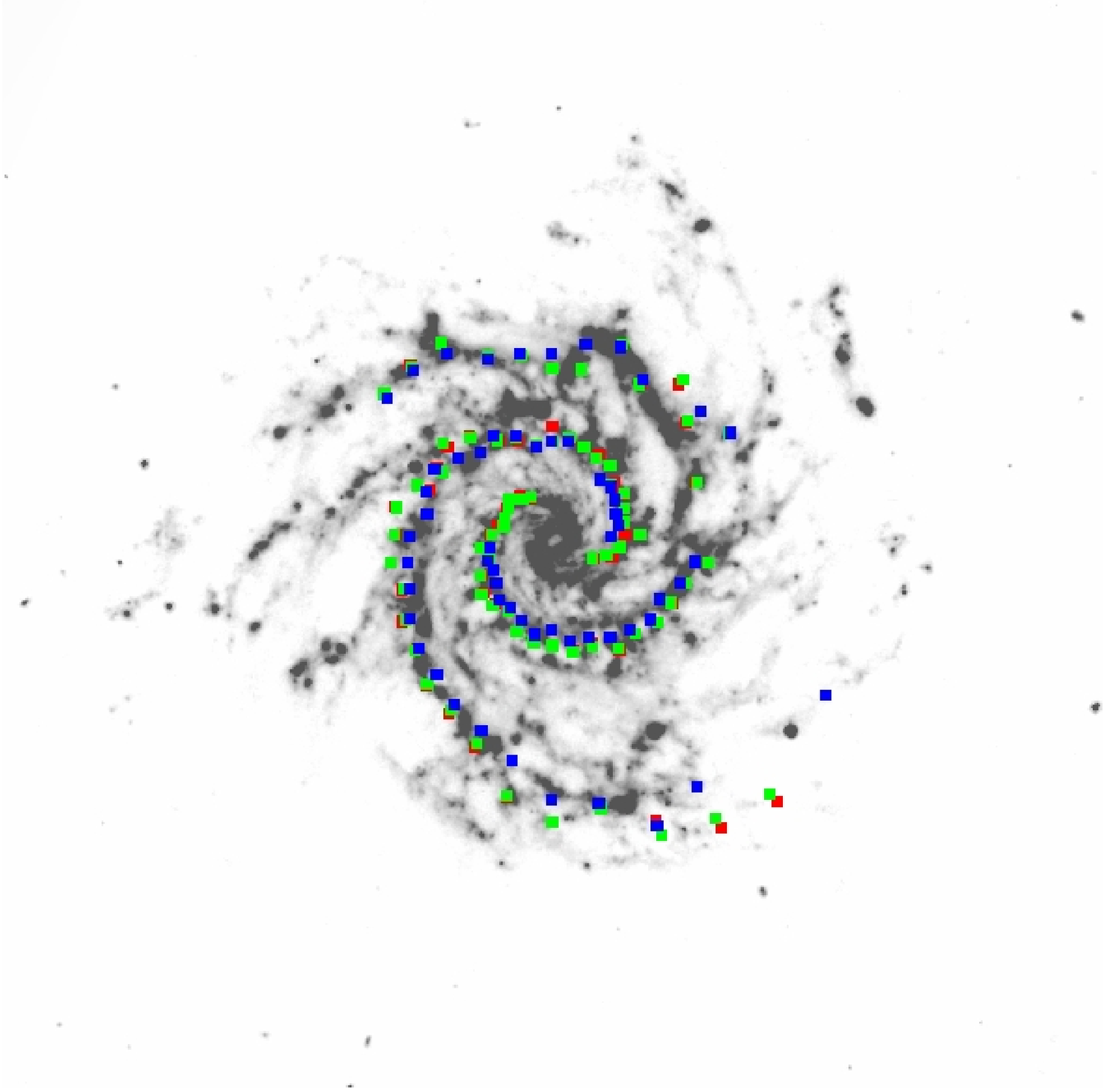}
  \caption{NGC 0628. As in  Figure \ref{n0628_resids} but with the maxima of
the spiral arms being determined from the radial profiles method.}
  \label{n0628_resids2}
\end{figure}

\begin{figure}   \centering
  \includegraphics[width=83mm]{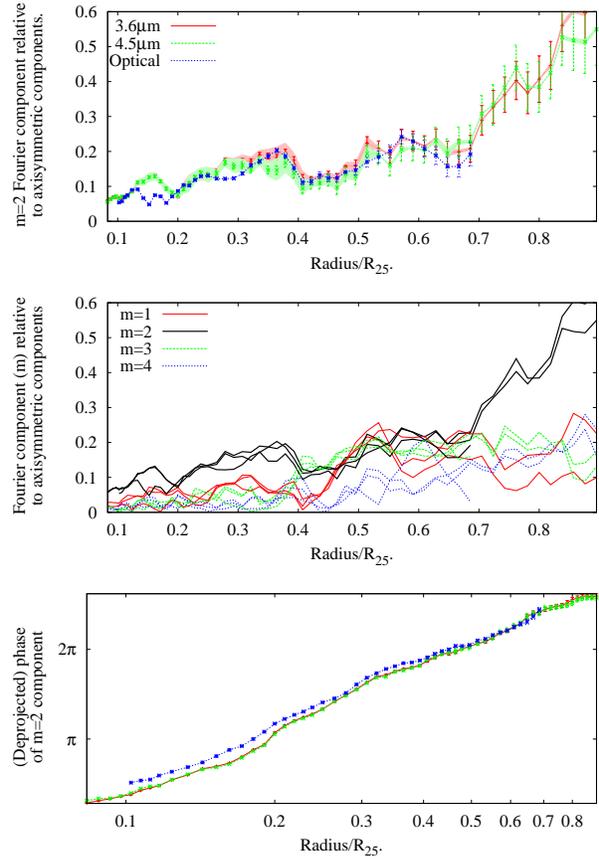}
  \caption{Azimuthal profile data for NGC 0628. The results are shown for the radial range over which the spiral is approximately logarithmic. Top figure; the m=2 components only. Shaded regions show the errors due to a 10 per cent change in the PAH correction factor, and the error bars show the errors associated with the random noise in the data. The optical data have a smaller radial range than the IRAC data because the signal to noise is much worse at larger radii in the colour-corrected data. Middle figure; this compares the strengths of the first four Fourier components. Multiple lines for a given value of m correspond to analyses
based on $3.6,4.5 \mu$m and colour-corrected I band images. Bottom figure; the phase of the m=2 Fourier component is plotted as a function of radius (this data is used to calculate the pitch angle). Colours are as for the top figure.}
  \label{n0628_prof_dat}
\end{figure}

\begin{figure}   \centering
  \includegraphics[width=83mm]{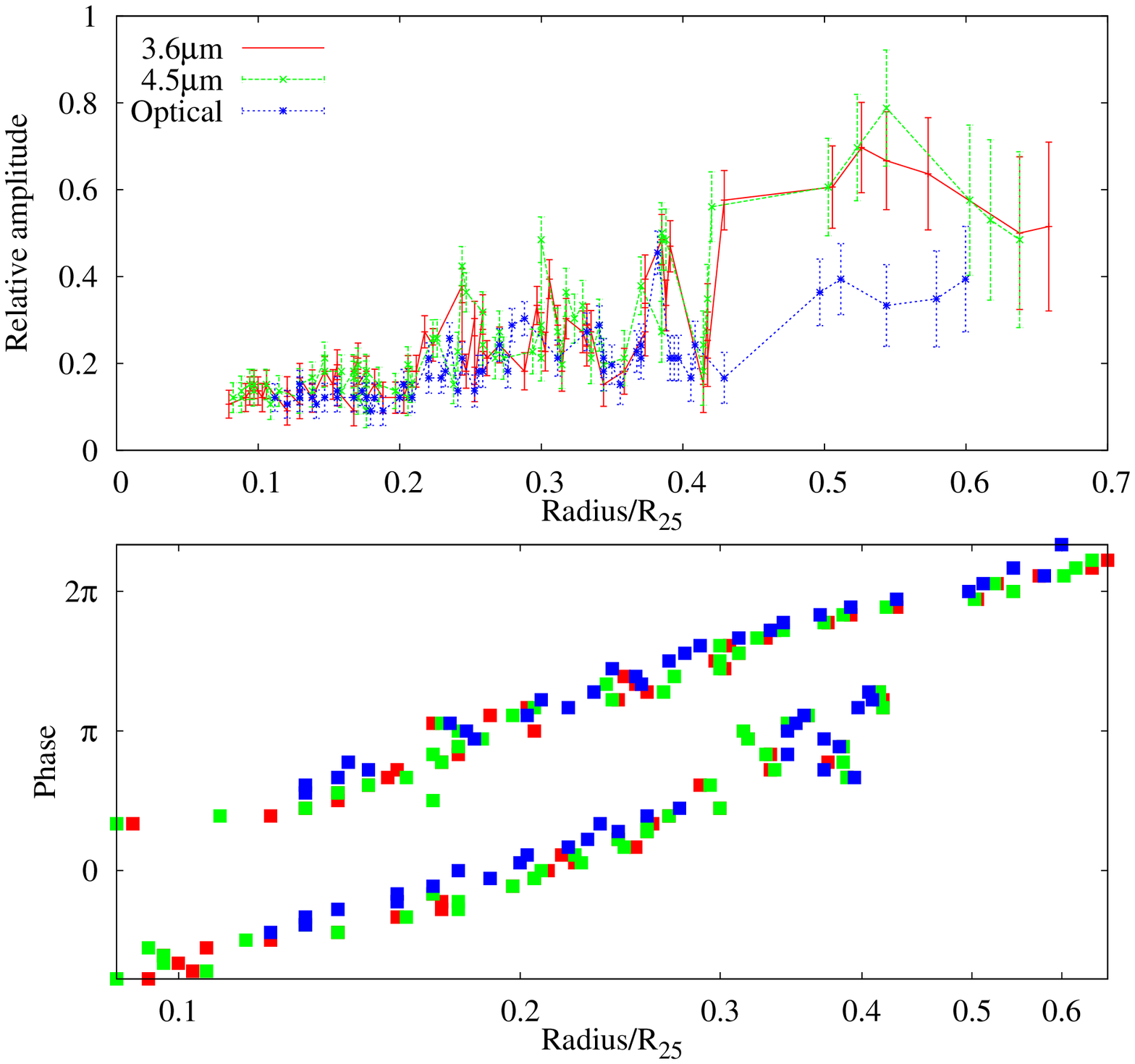}
  \caption{Radial profile data for NGC 0628. Top figure; shows the relative amplitude (half peak-to-trough) of the spiral. This measure of spiral strength is discussed further in section \ref{ch2_extract}. Bottom figure; shows the phase of the peaks (this data is used to calculate the pitch angle). The same data may be seen overplotted on the residual image and 8$\mu$m image in Figure \ref{n0628_resids2}. Colours are as for the top figure.}
  \label{n0628_raddat}
\end{figure}

It is worth noting early on that the m=2 Fourier component imposes 180$^o$ rotational symmetry on the arms. In the case of NGC 0628 it can be seen that the arms are not perfectly symmetric, especially in the gas response, and only one arm appears to be fully traced by the m=2 phase over the entirety of its length. However, it should also be noted that the residual contamination from the PAHs/young stars can be misleading to the eye. The underlying mass distribution is not necessarily traced exactly by the gas response, so the underlying arms (which are hard to discern in NGC 0628 at larger radii) may in fact be more symmetric than appears, and the symmetry masked by contamination and low amplitude. In contrast, 180$^o$ rotational symmetry is not imposed on the phases 
determined by the radial profile method, but the method is more susceptible to contamination from localized regions such as star forming regions. Thus, by comparing the plots of phase vs radius for the two methods  - Figures \ref{n0628_prof_dat} and \ref{n0628_raddat} - and the pitch angles calculated from these data, it can be seen that the difference is not large. The average pitch angle measured by the azimuthal profile method is 17$^o$, whereas from the radial profiles the pitch angle is 16$^o$. The errors in these quantities are $\sim$1$^o$, suggesting that the difference between the two methods is not greater than the uncertainties. For a further discussion of general trends in the pitch angles and associated errors, see Paper II.

From Figure \ref{n0628_prof_dat} it can be seen that the relative amplitude of the m=2 component in NGC 0628 shows a similar trend to that observed for M81, although the relative amplitude reached is larger by around a third. When compared to Figure \ref{n0628_raddat}, the contrast in relative amplitudes determined for the two methods is striking. Although the same general trends are seen in both sets of data, there is much more noise in the relative amplitudes determined from the radial profiles (to be expected given that this method is more susceptible to contamination), and the relative amplitude is higher on average when measured by the radial profile method. The systematic differences between the two methods are discussed further in Paper II. 

\begin{figure}   \centering
  \includegraphics[width=83mm]{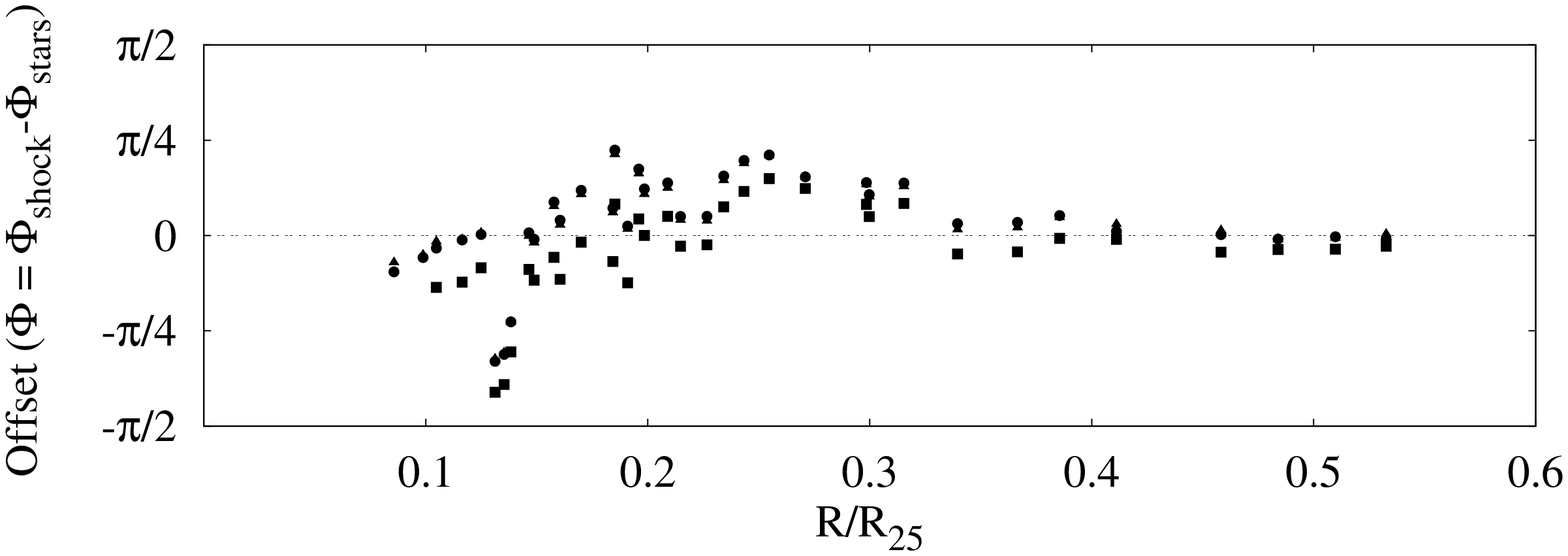}
  \caption{The offset between the gas shock (assumed to be traced by the 8$\mu$m emission) and stellar mass for NGC 0628; the triangles, circles and squares represent the 3.6$\mu$, 4.5$\mu$m and \textit{I} band colour corrected data respectively.. The offset function is twice the offset (offset function  =  2($\phi_{8{\mu}m}$ - $\phi_{stellar-arm})$). In the case of NGC 0628 an upstream gas shock will give a positive offset.}
  \label{offset_n0628}
\end{figure}

Figure \ref{offset_n0628} shows the offset between the maximum of the stellar spiral m=2 Fourier component and the shocks in the gas, identified from the 8$\mu$m image. As is noted in the figure, the offset function used by \cite{2004MNRAS.349..909G} for a two armed spiral is twice the physical offset. The expected sign of the offset for an upstream shock is noted in the figure caption. \footnote{In all cases, we deduce the direction of galactic rotation by assuming a trailing spiral pattern and we describe shocks as `upstream' if they occur ahead of the spiral arm in the direction of galactic rotation. Such structures are however only `upstream' as far as the gas is concerned at radii within corotation since it is only at such radii that the gas flow through the spiral arms is in the same direction as galactic rotation. Our nomenclature thus implicitly assumes that the radius of corotation lies at or beyond the outer extent of the spiral pattern.} The shock in NGC 0628 appears to start to move upstream but then returns to the spiral arm. As can be seen when comparing the radial range of the offset data with the phase data from the stellar mass, the gas shock can not be traced as far in radius as the stellar spiral because it becomes difficult to follow, but there is a hint from the western arm that the gas starts to shock more upstream of the spiral arms again at larger radii. The errors in the offsets are dependent on two measurements; the phase of the m=2 and the phase of the shock. The errors in the phase of the m=2 mode of the stellar mass component, calculated by CURVEFIT, are indicated in Figure \ref{n0628_prof_dat} (although a larger systematic error is possible if the assumption of measuring the offset of the shocks from the m=2 component maximum is incorrect). As well as the error in the phase of the stellar spiral, there is an uncertainty in measuring the position of the shock, which is estimated to be no larger than $\sim$0.1 radian (and decreasing with increasing radius). Again however, there is a possible systematic error since the switch-on of young stars is required to trigger PAH emission: thus, if anything, the shocks will be slightly upstream of the measured positions within corotation. The combined errors from the non-systematic effects are in general smaller than the scatter already apparent between the different wavelengths (due to the differences in m=2 component phase measured from the different wavelengths).

\subsubsection{NGC 1566}\label{ch4_1566}
NGC 1566, of type SABbc, is the brightest member of the Dorado Group. An HI study of the group carried out by \cite{2005MNRAS.356...77K} lists it as having two small companions, although it is not clear if the three galaxies are interacting (indeed, \cite{2000ApJ...541..565K} lists NGC 1566 as being apparently isolated).

\begin{figure}   \centering
  \includegraphics[width=60mm]{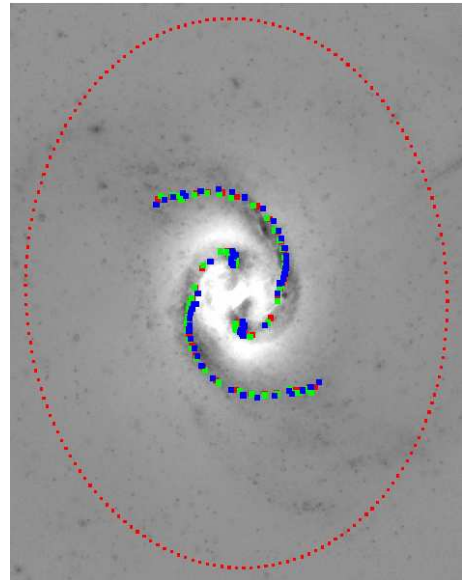}
  \includegraphics[width=60mm]{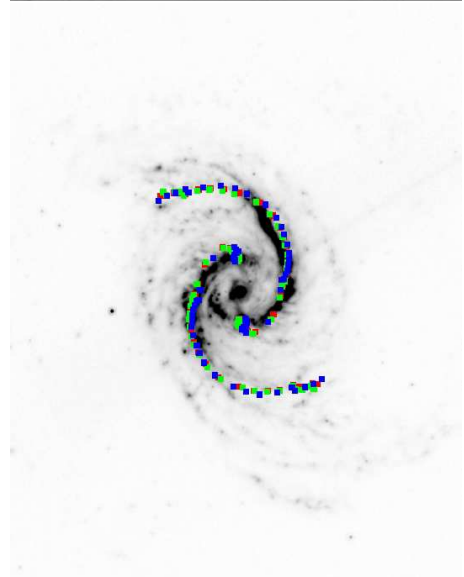}
  \caption{NGC 1566 (as Figure \ref{n0628_resids}). R$_{25}$ = 4.15 arc minutes.}
  \label{n1566_resids}
\end{figure}

\begin{figure}   \centering
  \includegraphics[width=83mm]{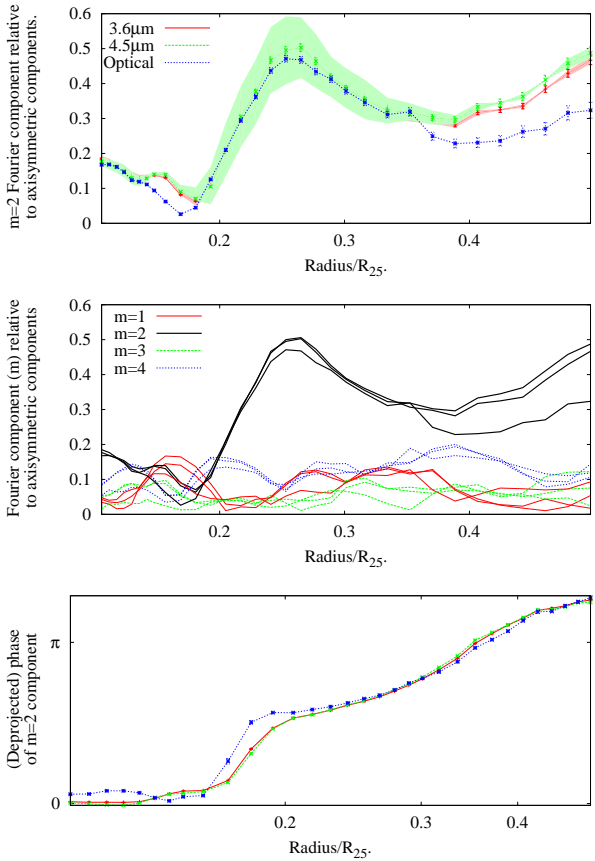}
  \caption{Azimuthal profile data for NGC 1566 (as Figure \ref{n0628_prof_dat}).}
  \label{n1566_prof_dat}
\end{figure}

Figure \ref{n1566_prof_dat} shows that NGC 1566 displays an unusual trend in relative amplitude with radius when compared to the majority of spirals in this sample. The peak at small radii in relative amplitude may be caused by a resonance with the bar; \cite{2009AJ....137.4487B} study the Fourier amplitudes in spiral galaxies with bars, and see a wide range of behaviours of the m=2 component in the region outside the bar. NGC 1566 is not the only galaxy to show this type of behaviour in the spiral arms (other examples from their sample being NGC 6384, NGC 0986 and, to a slightly lesser extent, NGC 6221). An alternative possibility presents itself when studying the images; it can be seen that although the very brightest regions are small, the arms are still very narrow and well defined up to R $\sim$ 0.35R$_{25}$. Some of the unusual behaviour in relative amplitude may well be due to remnant contamination from underlying young stellar populations or associated PAH emission, although the fact that the \textit{I} band colour-corrected data show exactly the same trend would suggest that this is not the case. Whatever the cause, the shape of the relative amplitude data agree well with trends in relative amplitude found by \cite{1990ApJ...355...52E}, although the relative amplitude found here is less (a maximum of 0.5 compared to almost 0.7 in \cite{1990ApJ...355...52E}). The differences can be ascribed partly to the fact that \cite{1990ApJ...355...52E} use optical data (hence probably including more contamination from young stars), and partly because the relative amplitudes are not measured in precisely the same way (though since the m=2 component dominates over the radial range of interest, the difference should be small).

The spiral is lost after 0.5R$_{25}$, but this is not very surprising; although on first inspection of Figure \ref{n1566_resids} the spiral arms appear to extend to R$_{25}$ (particularly in the 8$\mu$m image), the main spiral does not extend beyond the end of the detected signal in azimuth; the structure at larger radii appear to be caused by a bifurcation in the arms. However, it is worth noting that NGC 1566 also has outer spiral arms, not traced in this work, which extend well beyond R$_{25}$, and are most clearly obvious in the optical but are present, although very faint, in the IRAC images too. The most likely reason why they are not detected with this method is that the signal is lost amongst the background noise.

From Figure \ref{n1566_prof_dat}, it can be seen that there appears to be a kink in the phase plot, but if the full range of phase vs radius data is examined the case for starting the arms at 0.1R$_{25}$ rather than 0.2R$_{25}$ is clearer. With the range used the average pitch angle is 20$^o$, and using a reduced range, with a lower cutoff at 0.2R$_{25}$, the average pitch angle is unchanged to the nearest degree.

\begin{figure}   \centering
  \includegraphics[width=83mm]{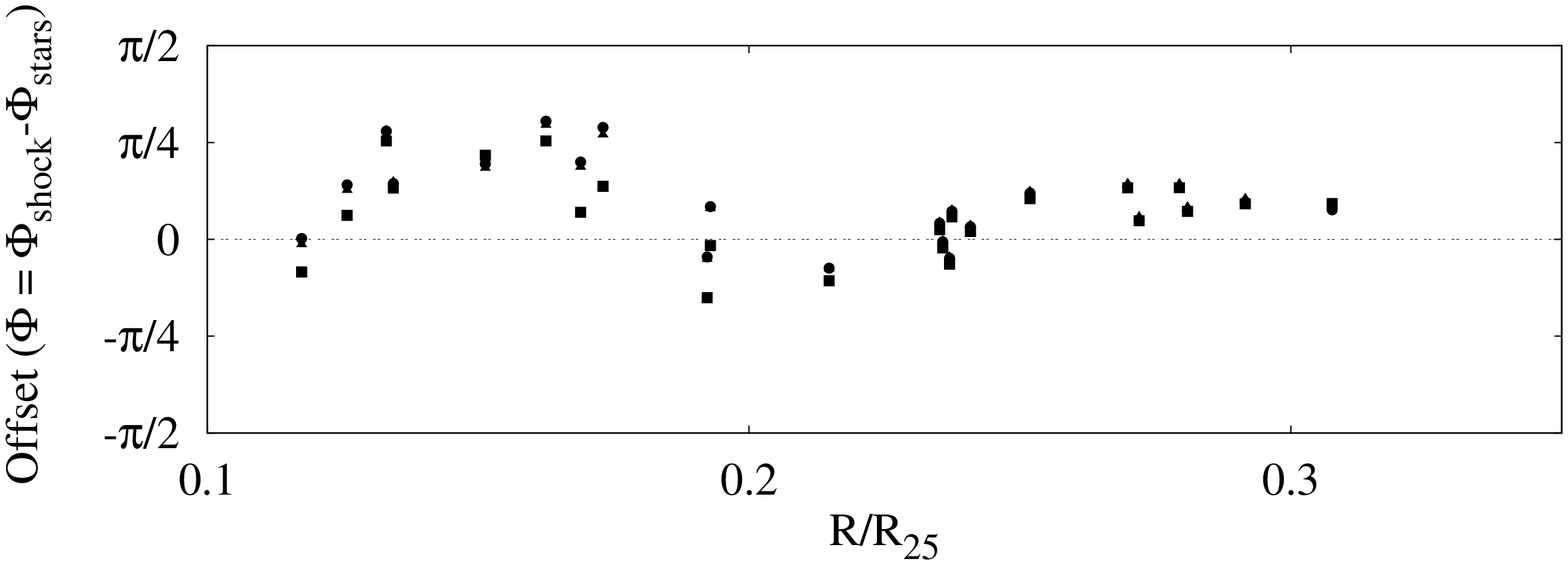}
  \caption{The offset between the gas shock and stellar mass for NGC 1566. In the case of NGC 1566 an upstream gas shock will give a positive offset.}
  \label{offset_n1566}
\end{figure}

Figure \ref{offset_n1566} shows the offset between the stellar spiral and the gas shocks. As can be seen the gas does not appear to form a shock consistently up or downstream of the stellar spiral, but instead oscillates about its position. This oscillation may be in part a product of the fact that the m=2 Fourier component is not dominant until R$>$0.2R$_{25}$.

\subsubsection{NGC 2403}\label{ch4_2403}
NGC 2403, which is an SABcd spiral, is located in the M81 group but is isolated by the catalogue criteria in \cite{1997yCat.7082....0K}. 

\begin{figure}   \centering
  \includegraphics[width=80mm]{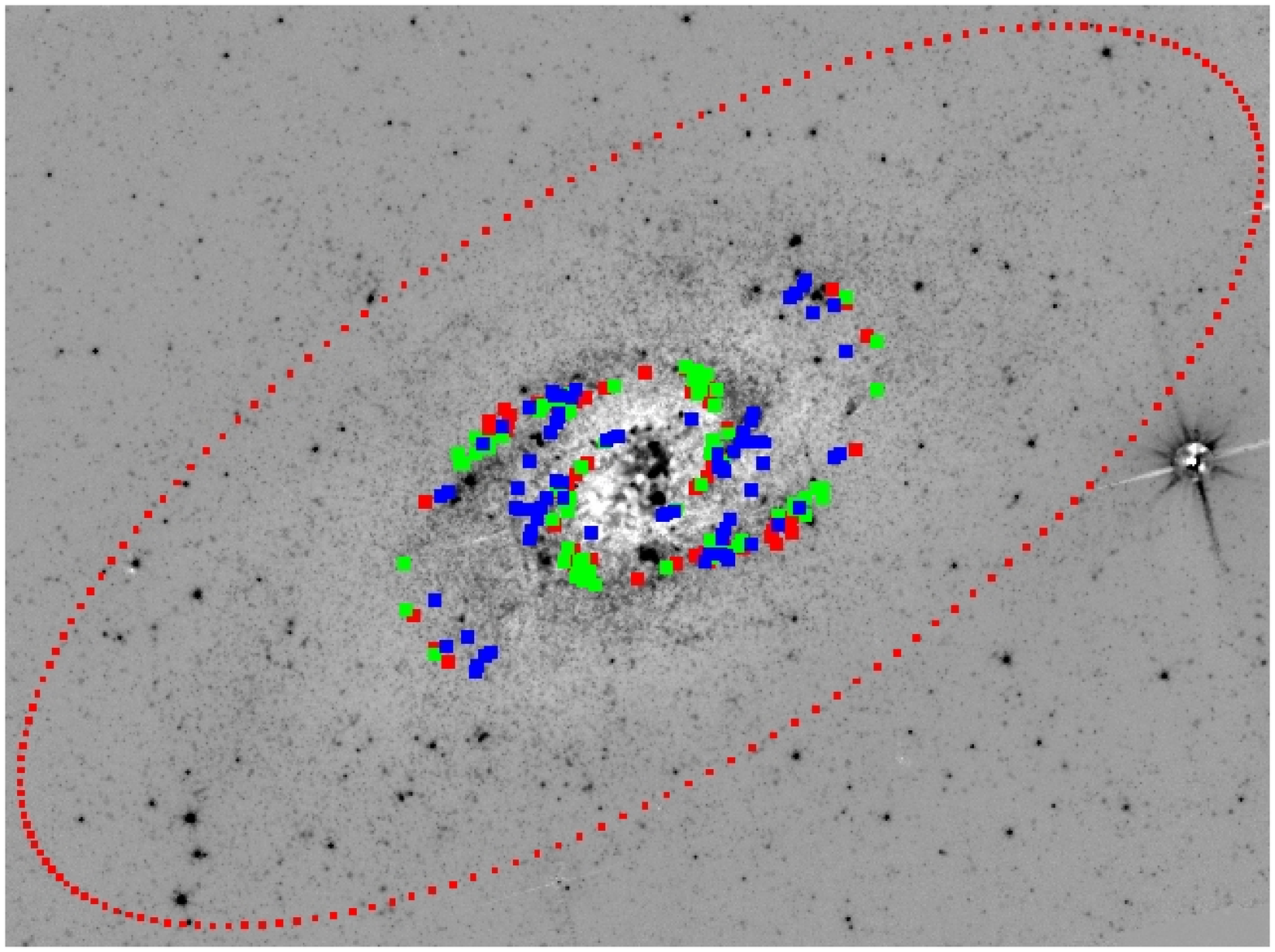}
  \includegraphics[width=80mm]{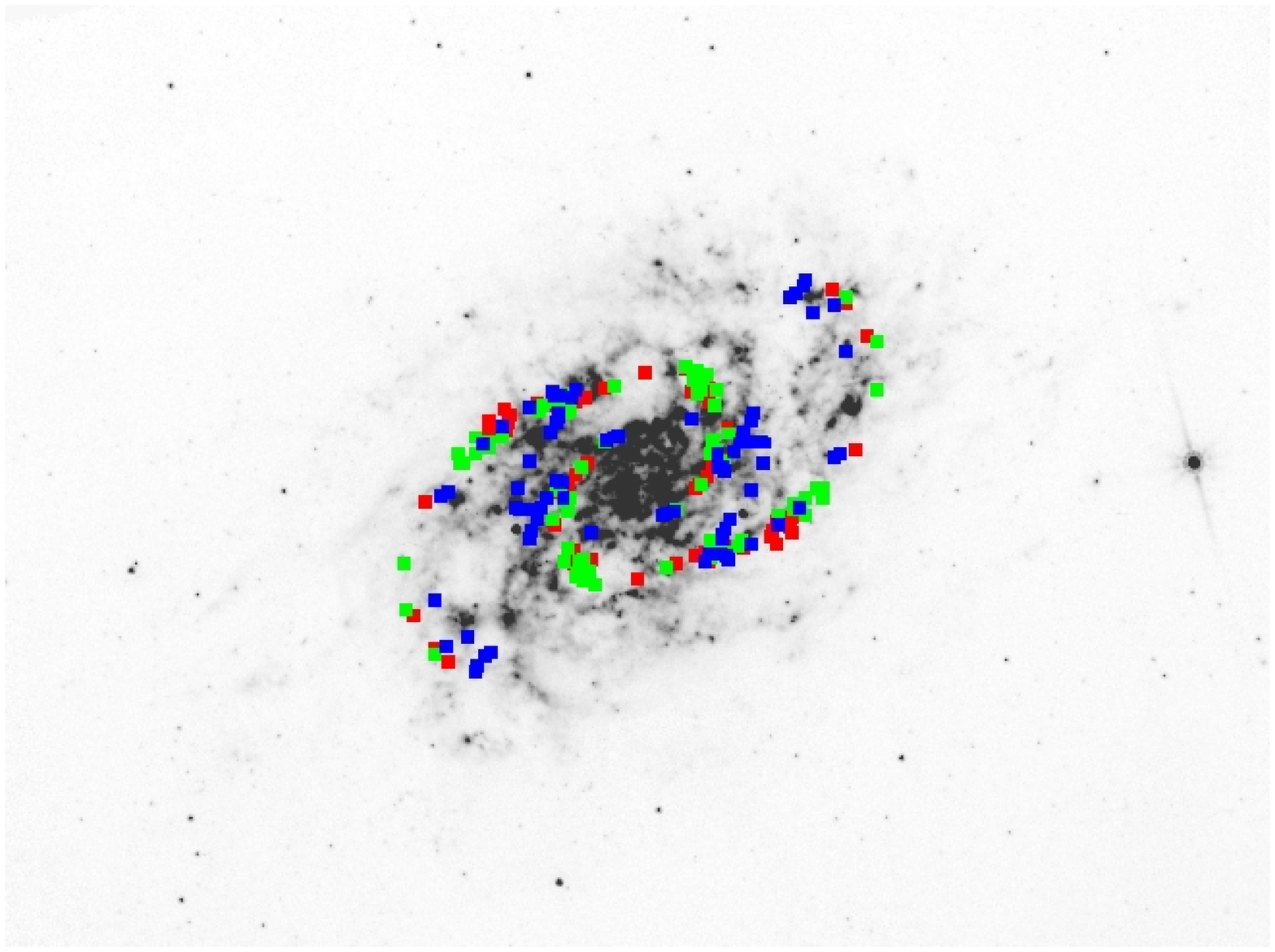}
  \caption{Images. NGC 2403 (as Figure \protect\ref{n0628_resids}). R$_{25}$ = 10.95 arc minutes.}
  \label{n2403_resids}
\end{figure}

\begin{figure}   \centering
  \includegraphics[width=83mm]{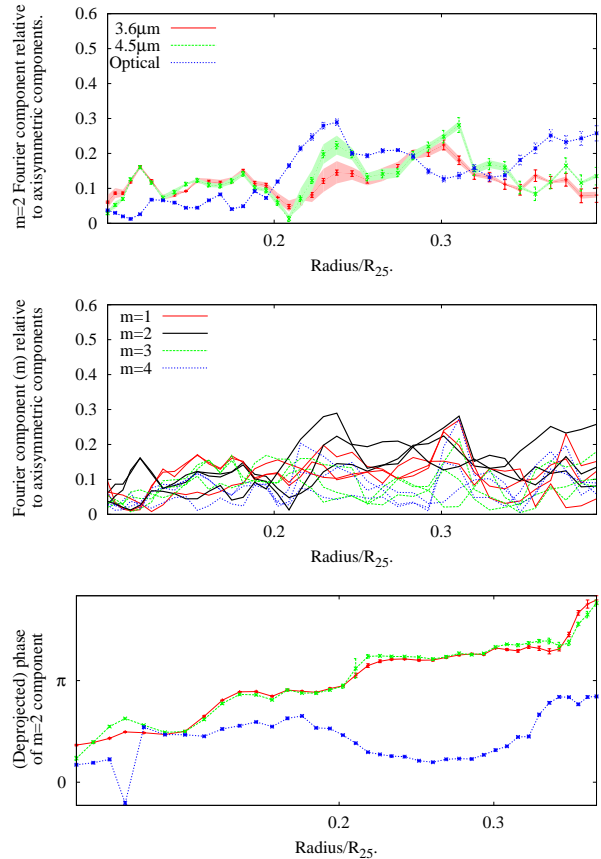}
  \caption{Azimuthal profile data for NGC 2403 (as Figure \ref{n0628_prof_dat}).}
  \label{n2403_prof_dat}
\end{figure}

This is an interesting galaxy because the optical structure is very flocculent in appearance (Elmegreen type 4 \citep{1982MNRAS.201.1021E, 1987ApJ...314....3E}), and even in the NIR the spiral arms are hard to distinguish. Nevertheless, as may be seen in Figures \ref{n2403_resids} and \ref{n2403_prof_dat}, an underlying spiral is detected. An examination of the IRAC data suggest that an (approximately) logarithmic spiral can be detected to $\sim$0.5R$_{25}$, beyond which it gets mostly lost to noise, although there are hints that very faint spiral structure might exist for at least another 180${^o}$ beyond the current radial limit of the data. In the optical the arms cannot be followed so clearly, but since the IRAC data do not have the same trouble it was decided to use the azimuthal profile data (preferable to radial profile data), but exclude the optical data from further analysis. Instead the optical data is presented here for comparison, in order to demonstrate that the amplitude and phases largely agree.

There is no offset data for NGC 2403; the gas response is too flocculent to define a shock. The average pitch angle measured from the two IRAC bands, but excluding the optical data, is 20$^o$.

\subsubsection{NGC 2841}\label{ch4_2841}

NGC 2841 is classified as an isolated \citep{1997yCat.7082....0K} regular SAb spiral without global spiral patterns \citep{1979ApJ...233..539K}. \cite{1997A&A...326..941S} find evidence for decoupled ionized gas around the nuclear region, and a possible counter-rotating stellar disc (inner few arc seconds only). They attribute these features to the accretion of a companion galaxy some time in the past. The galaxy also has a central hole in the HI distribution \citep{2008AJ....136.2648D}.

\begin{figure}   \centering
  \includegraphics[width=60mm]{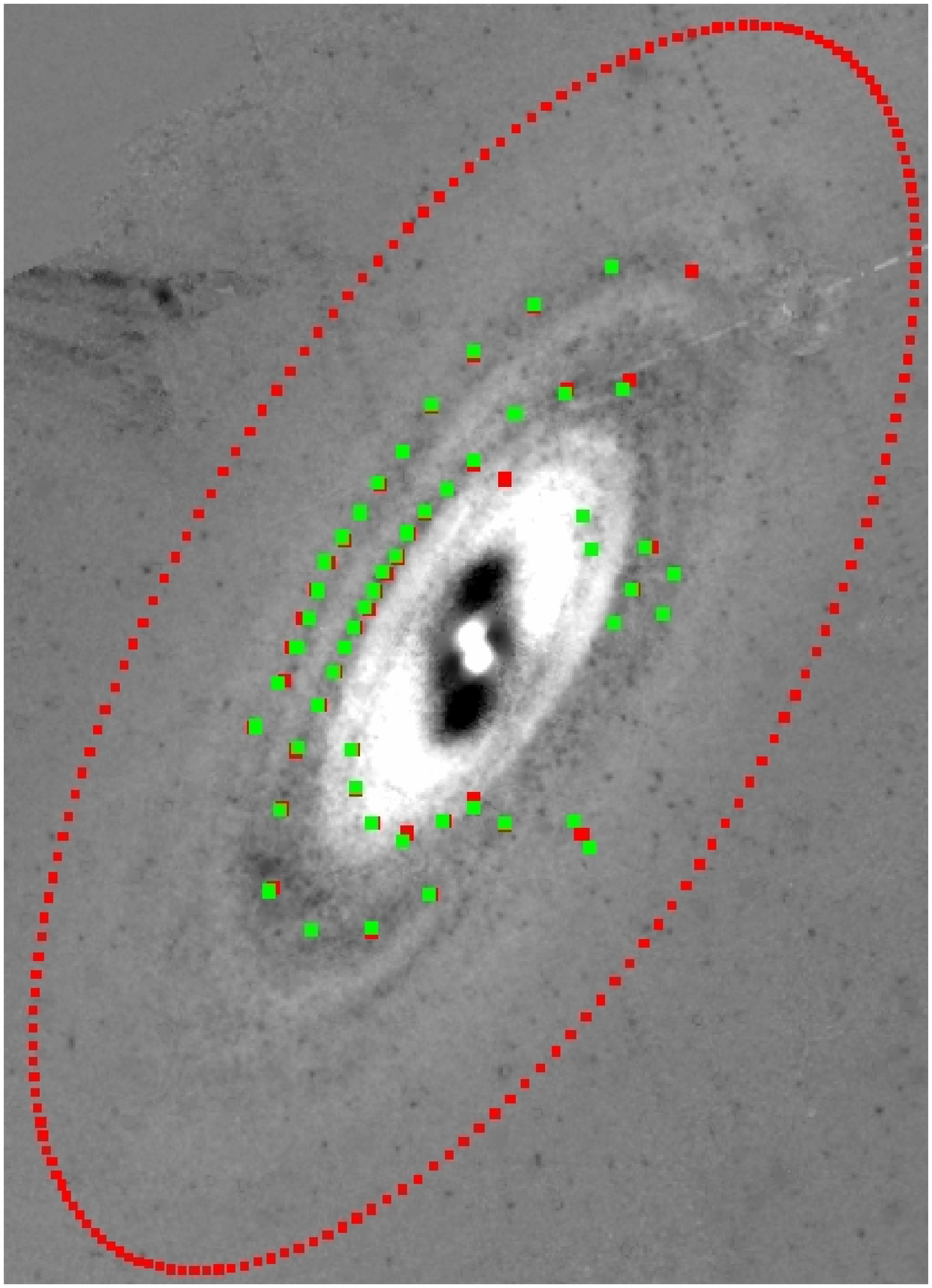}
  \includegraphics[width=60mm]{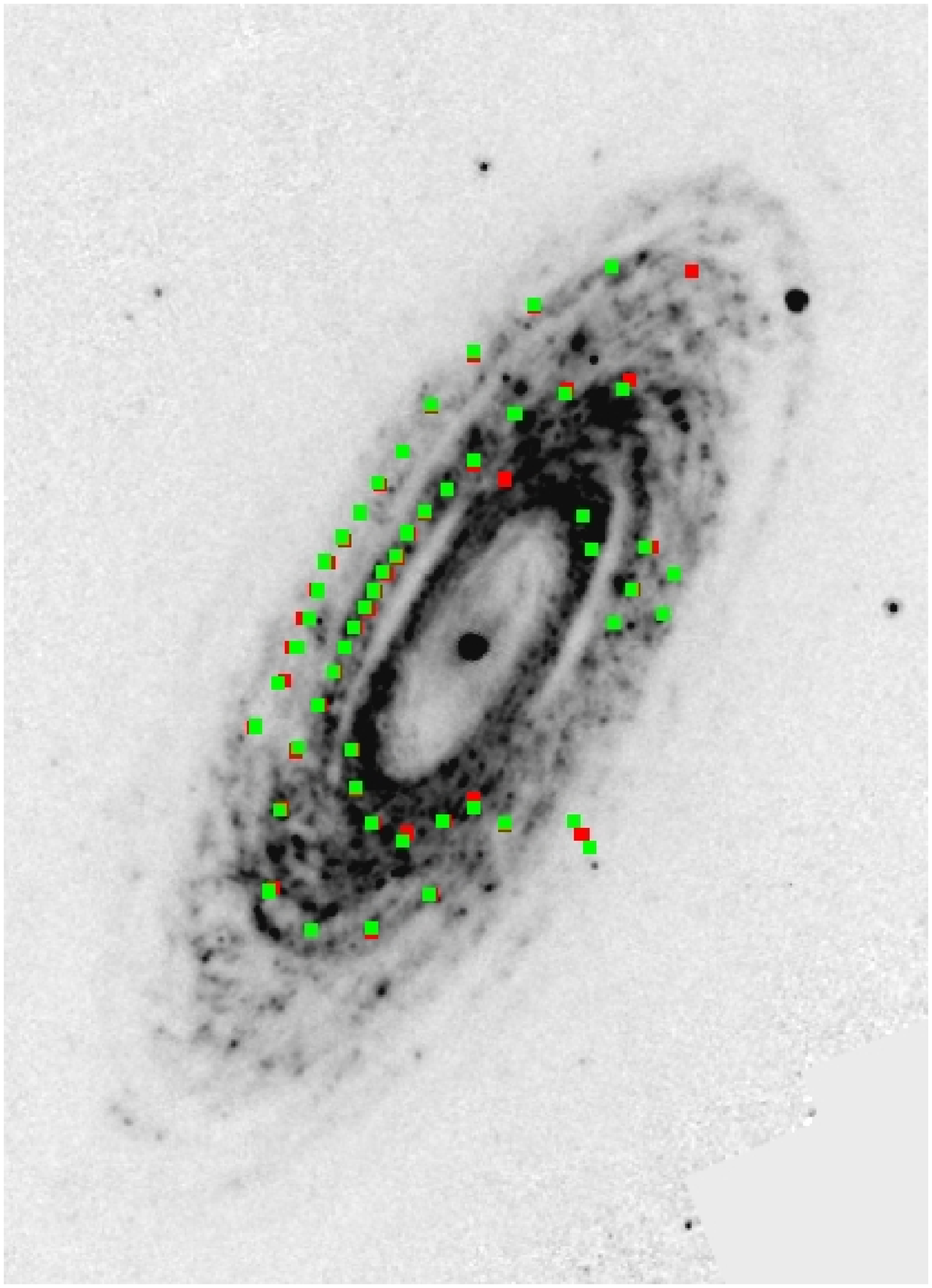}
  \caption{NGC 2841 (as Figure \protect\ref{n0628_resids2}). R$_{25}$ = 4.05 arc minutes. There are no optical results from radial profiles as it was impossible to determine peaks in the radial profiles (probably due to large dust extinction).}
  \label{n2841_resids}
\end{figure}

\begin{figure}   \centering
  \includegraphics[width=83mm]{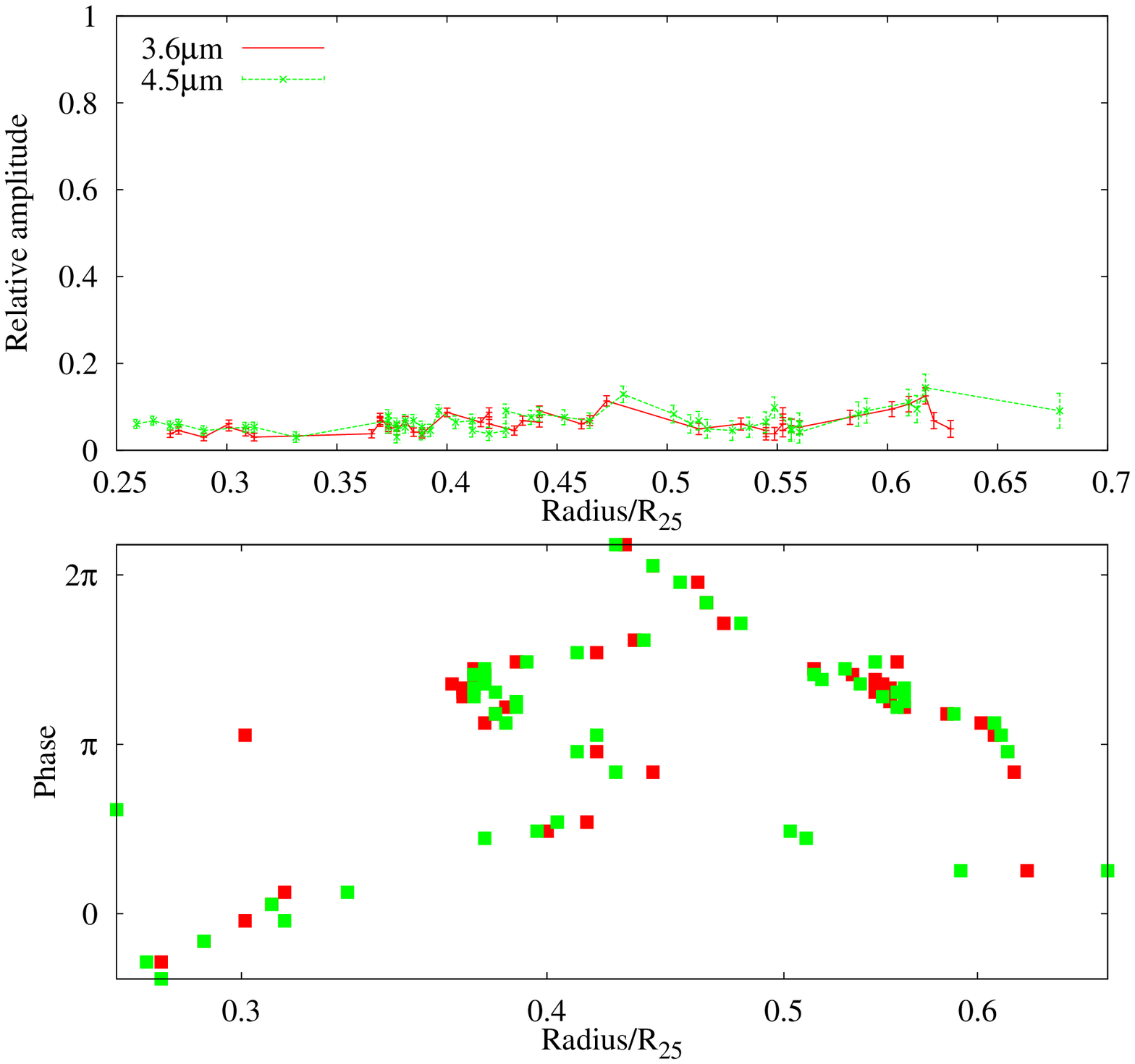}
  \caption{Radial profile data for NGC 2841 (as Figure \ref{n0628_raddat}).}
  \label{n2841_raddat}
\end{figure}

NGC 2841 is classified as Elmegreen type 3, and is very flocculent in the optical and also when viewed at 8$\mu$m. However, in the NIR there appears to be smooth, well defined spiral structure. \cite{1996Natur.381..674B} attribute the spiral features in NGC 2841 to sheared `dark clouds' - i.e. dust features rather than stellar features, and explain the lack of similar structures at optical wavelengths due to scattering of light by the dust at shorter wavelengths. The galaxy is highly inclined which is a distinct disadvantage in determining the spiral structure, especially when dealing with tightly wound spirals.

NGC 2841 was studied with the radial profile method after the azimuthal profile method proved to be unsuccessful at recovering grand design spiral structure (probably due to the complications listed above). It is difficult to draw any firm conclusions from the radial profile data, but tantalising hints emerge; a possible interpretation of the data is that there might be two different senses of spiral winding; an inner structure winding one way, with a pitch angle of $\sim$9$^{o}$, and outer spiral arms winding more tightly other way (pitch angle $\sim$7$^{o}$), with the break at a radius of $\sim$0.45R$_{25}$. This is a tenuous claim, and is difficult to verify; even with radial profiles it proved almost impossible to extract information about the spiral structure from most of the western side of the galaxy, and the galaxy is sufficiently inclined to make the image difficult to interpret. There is no offset data for NGC 2841 as the gas response is too flocculent to define a shock.

The relative amplitude of the structure that is detected is notably smaller than in optically grand design spirals (around a third of that seen for M81, and only a quarter of the relative amplitude measured for NGC 0628). This weakness of signal is also probably a contributing factor to the difficulties in picking up the spiral structure (and may therefore be a useful indicator of the limits of detectability - signals at the few percent level will be lost in the noise of these observations). However, despite the weakness of the signal, the general trend seen for the more well-defined spirals is seen, and the relative amplitude of the arms increases approximately linearly with radius.

\subsubsection{NGC 3031 (M81)}\label{ch4_3031}
NGC 3031, which is of type SAab, is the dominant member of the M81 group, and is part of an interacting triplet. M81 also has extended XUV arms, probably as a result of the tidal forces at work during the interaction.

\begin{figure}   \centering
  \includegraphics[width=60mm]{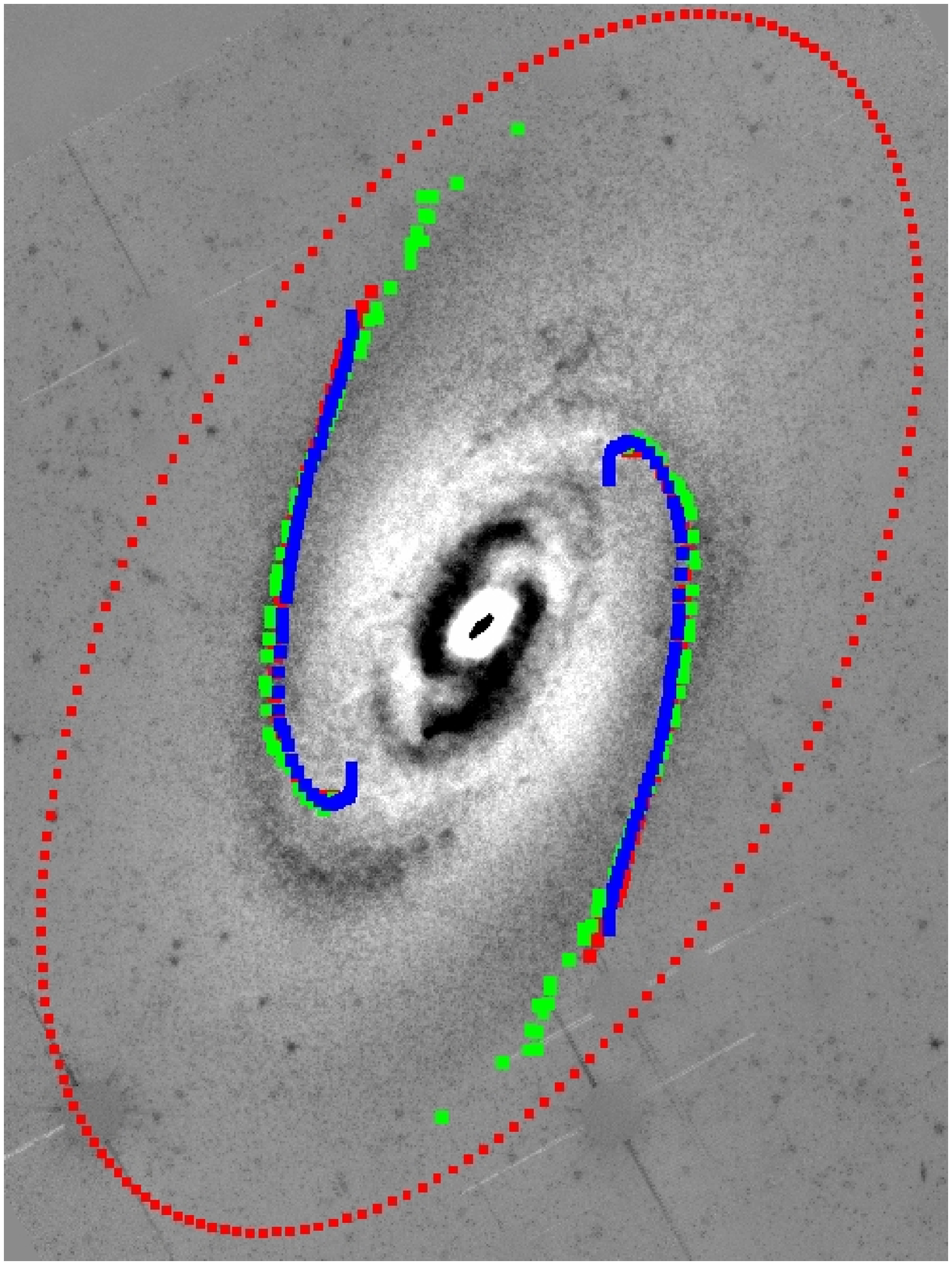}
  \includegraphics[width=60mm]{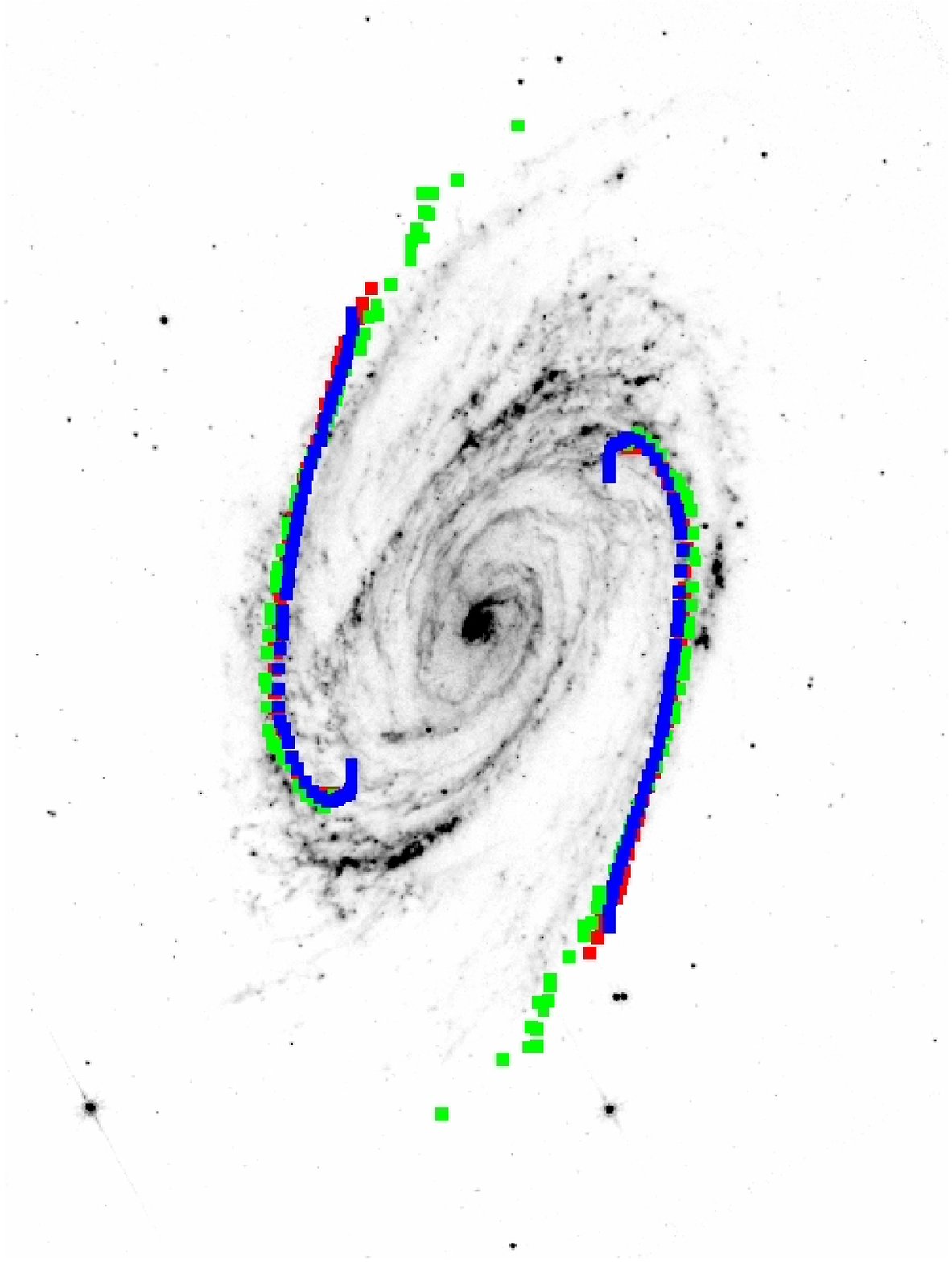}
  \caption{NGC 3031 (as Figure \protect\ref{n0628_resids}). R$_{25}$ = 13.45 arc minutes.}
   \label{n3031_resids}
\end{figure}

\begin{figure}   \centering
  \includegraphics[width=83mm]{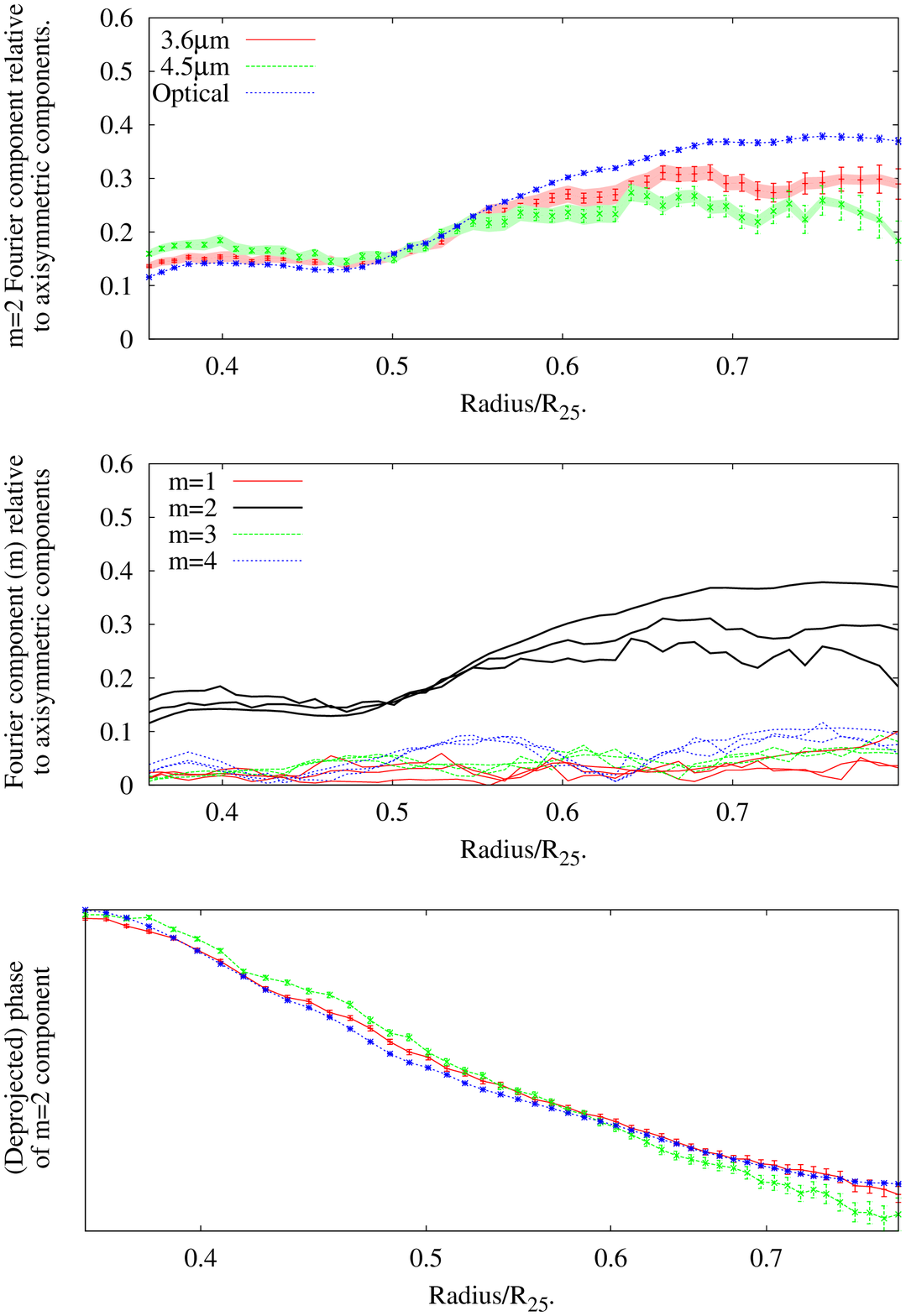}
  \caption{Azimuthal profile data for NGC 3031 (as Figure \ref{n0628_prof_dat}).}
  \label{n3031_prof_dat}
\end{figure}

Because NGC 3031 was discussed at length in KKCT08 the results are presented here simply to provide easy comparison with the other galaxies in the sample and will not be analysed further. It is also worth noting in passing that the method of calculating error bars is different for the results presented here than in KKCT08. Similarly, because the offset between the gas shock and stellar spiral was discussed in detail in KKCT08, the results will not be presented again in this section.

\subsubsection{NGC 3184}\label{ch4_3184}

NGC 3184 is classified as SABbc, and is listed as an isolated binary in \cite{2008arXiv0812.0689K}.

\begin{figure}   \centering
  \includegraphics[width=60mm]{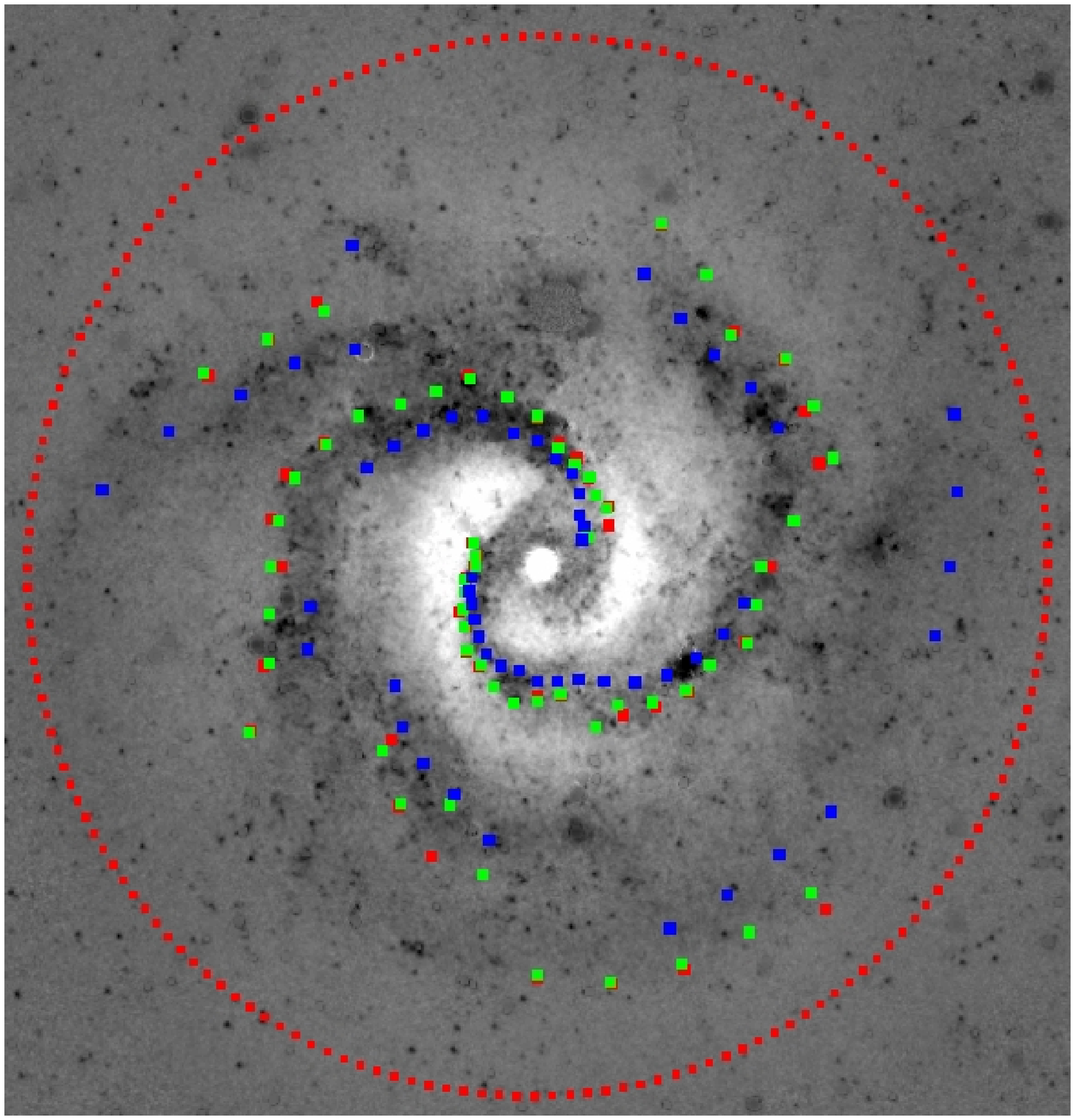}
  \includegraphics[width=60mm]{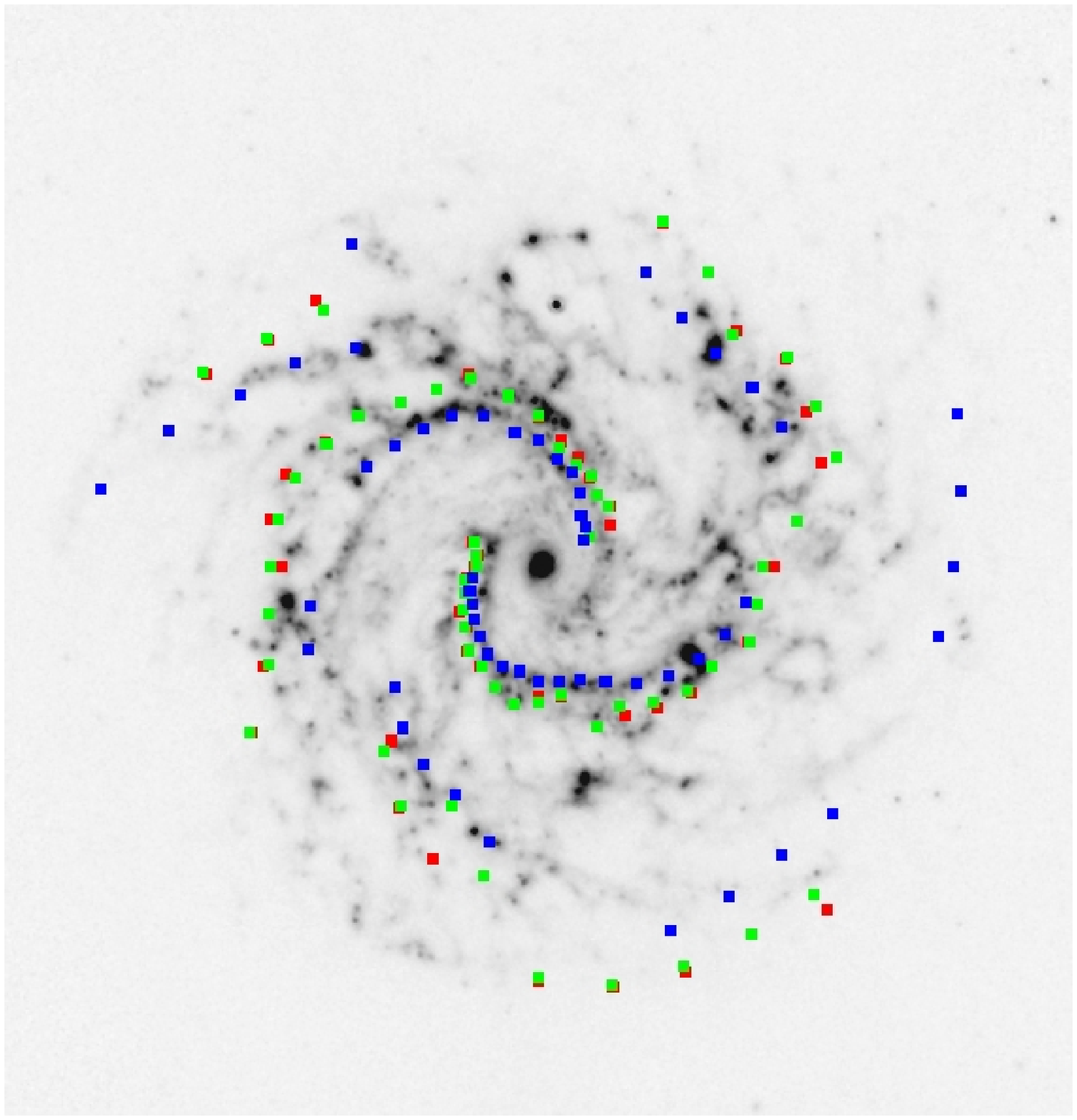}
  \caption{NGC 3184 (as Figure \protect\ref{n0628_resids2}). R$_{25}$ = 3.7 arc minutes. }
  \label{n3184_resids}
\end{figure}

\begin{figure}   \centering
  \includegraphics[width=83mm]{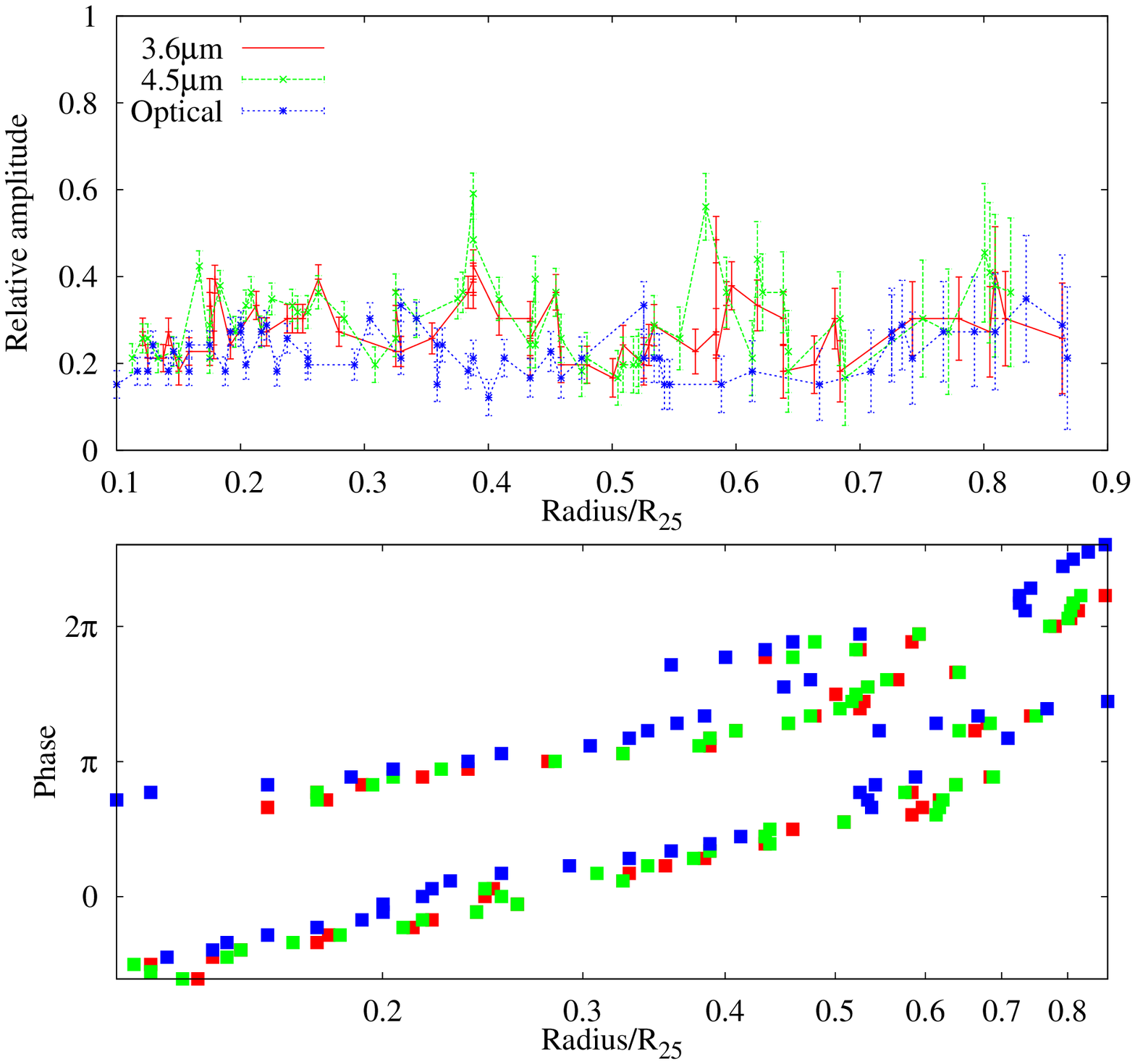}
  \caption{Radial profile data for NGC 3184 (as Figure \ref{n0628_raddat}).}
  \label{n3184_raddat}
\end{figure}

NGC 3184 had to be studied with radial profiles; similarly to NGC 5194 in Figure \ref{NGC5194_radguide}, the phase determined by the azimuthal profile method sometimes skipped sections of the arms. As can be seen from Figure \ref{n3184_resids}, one advantage of radial profiles is that subsidiary sections of spiral arm can be identified as well as the main spiral arm. These smaller spiral sections, possibly due to bifurcations in the main spirals, can be observed in Figure \ref{n3184_raddat}. Despite these extra features it is possible to trace two main spiral arms over a significant radial range, and from this determine a pitch angle which has an average value (for the three wavelengths) of 19$^o$.

\begin{figure}   \centering
  \includegraphics[width=83mm]{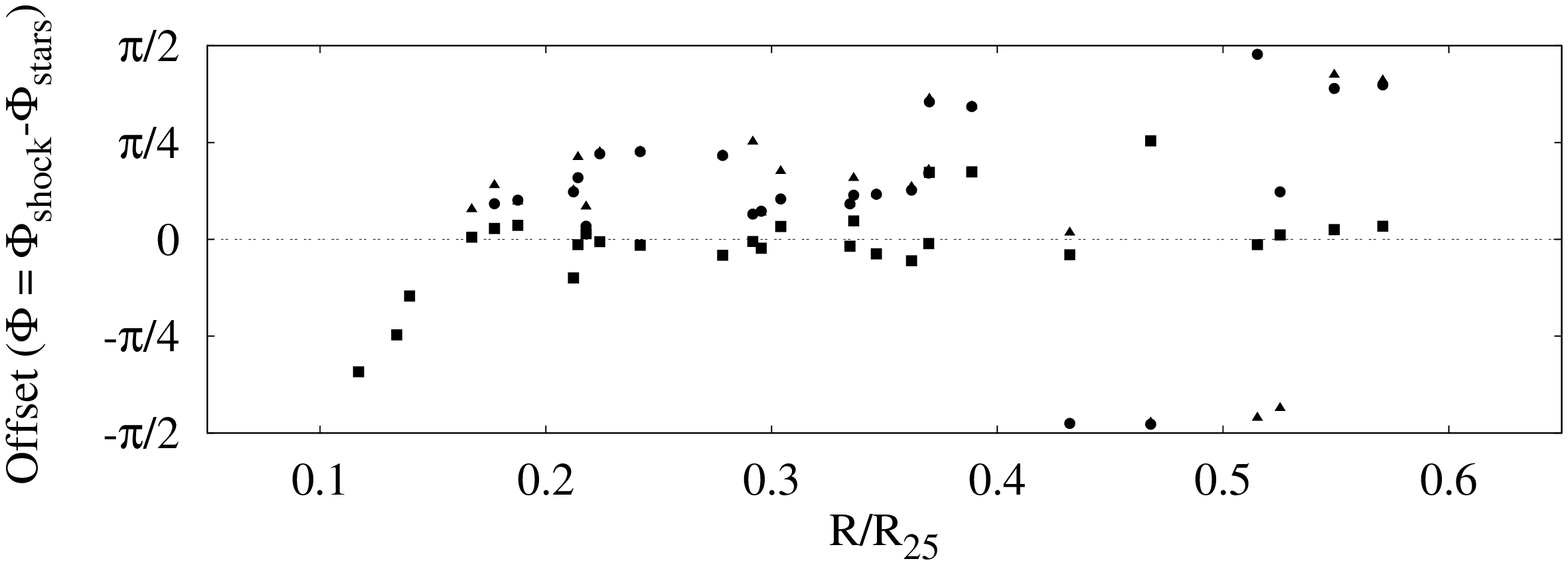}
  \caption{The offset between the gas shock and stellar mass for NGC 3184. In the case of NGC 3184 an upstream gas shock will give a positive offset.}
  \label{offset_n3184}
\end{figure}

From Figure \ref{n3184_raddat} it can be seen that the noise in the signal is significant, but the general trend is for the relative amplitude to be lower than in most cases, only reaching the 20 per cent level. Figure \ref{offset_n3184} shows the offset as a function of radius. In this galaxy there is considerable scatter in the offset, mostly due to the fact that the phase estimates from the three stellar mass maps are significantly different at some locations, but there is still a clear trend for the gas to shock upstream of the stellar spiral (i.e. inside the arm).

\subsubsection{NGC 3198}\label{ch4_3198}
NGC 3198, which is of type SBc, has a small warp in the HI disc, but shows no other signs of interaction \citep{1994MNRAS.271..427T}, and has no obvious nearby companions. As can be seen, NGC 3198 is one of the most highly inclined galaxies in the sample, with an axis ratio of 0.35. This is a distinct disadvantage when trying to determine the spiral structure, as has already been discussed. However, despite the non-ideal orientation, it is possible to follow spiral structure over a large radial range in the IRAC data. Unfortunately, the optical data cannot be traced as far out as the NIR; this is in part due to the fact that the sky noise dominates from around 0.8R$_{25}$, but even inside this radius the signal to noise is poor, and the spiral structure cannot be followed reliably beyond $\sim$0.5R$_{25}$. However, where the optical data is reliable it is in reasonable agreement with the IRAC data.

\begin{figure}   \centering
  \includegraphics[width=60mm]{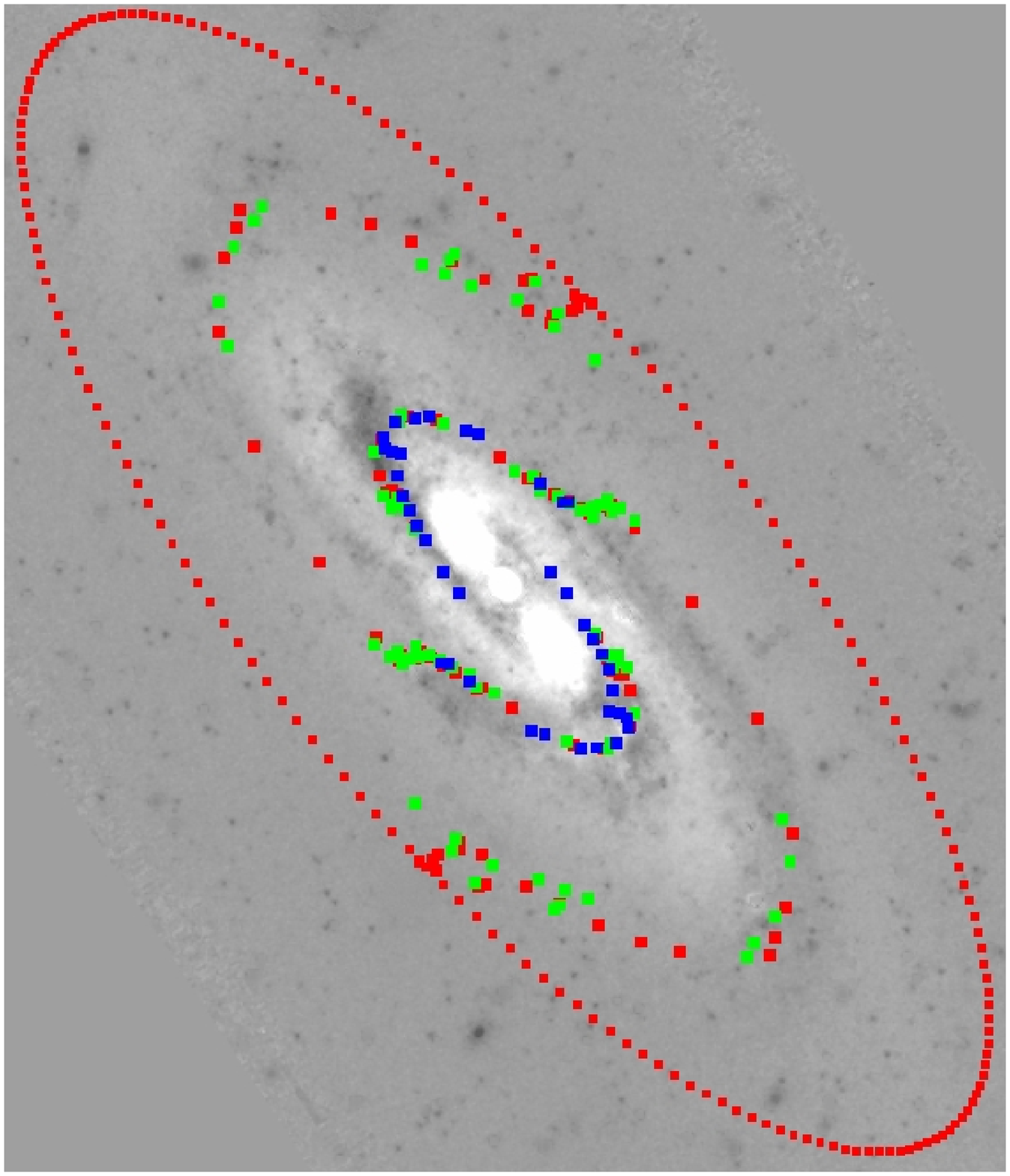}
  \includegraphics[width=60mm]{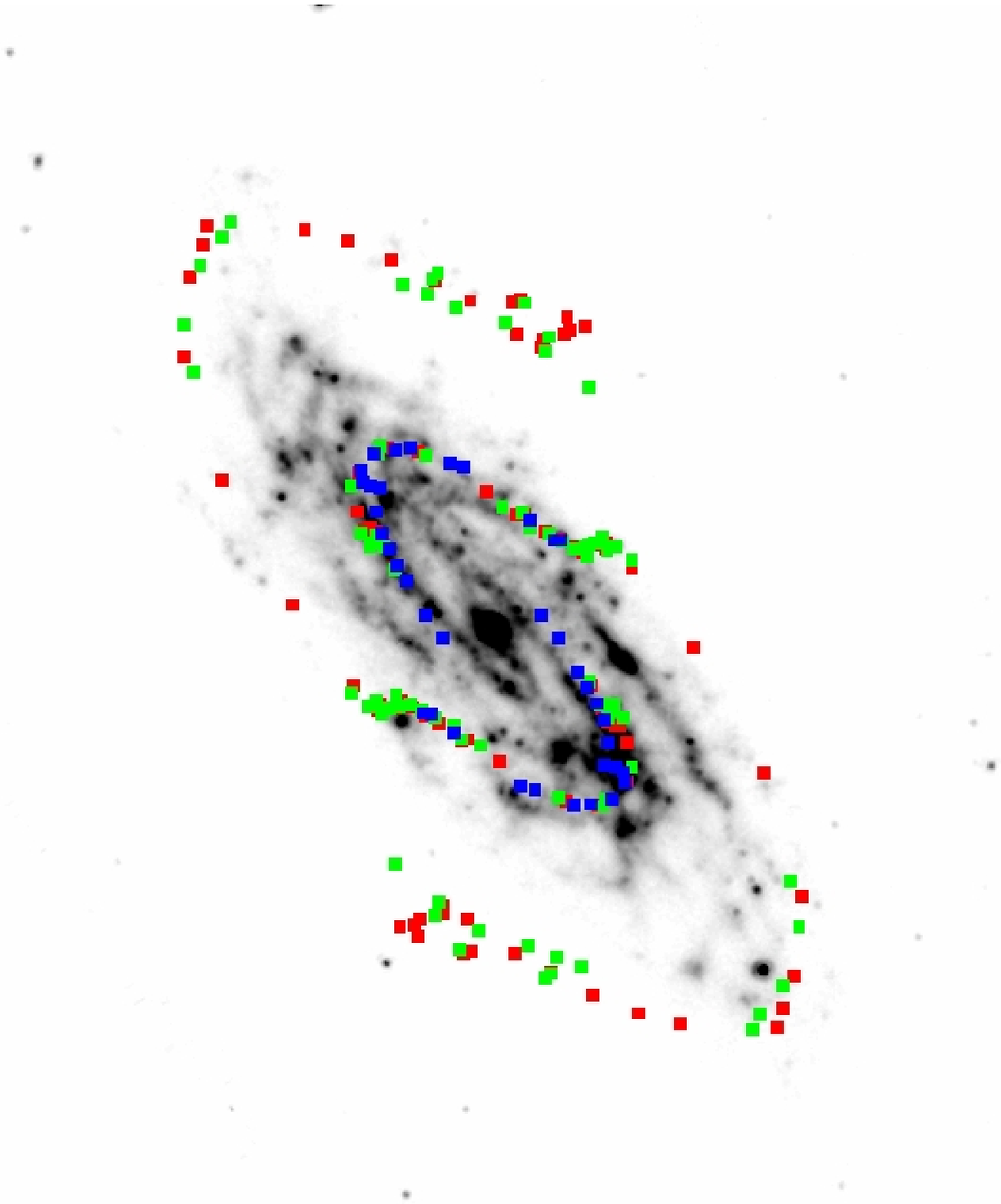}
  \caption{NGC 3198 (as Figure \protect\ref{n0628_resids}). R$_{25}$ = 4.25 arc minutes.}
  \label{n3198_resids}
\end{figure}

\begin{figure}   \centering
  \includegraphics[width=83mm]{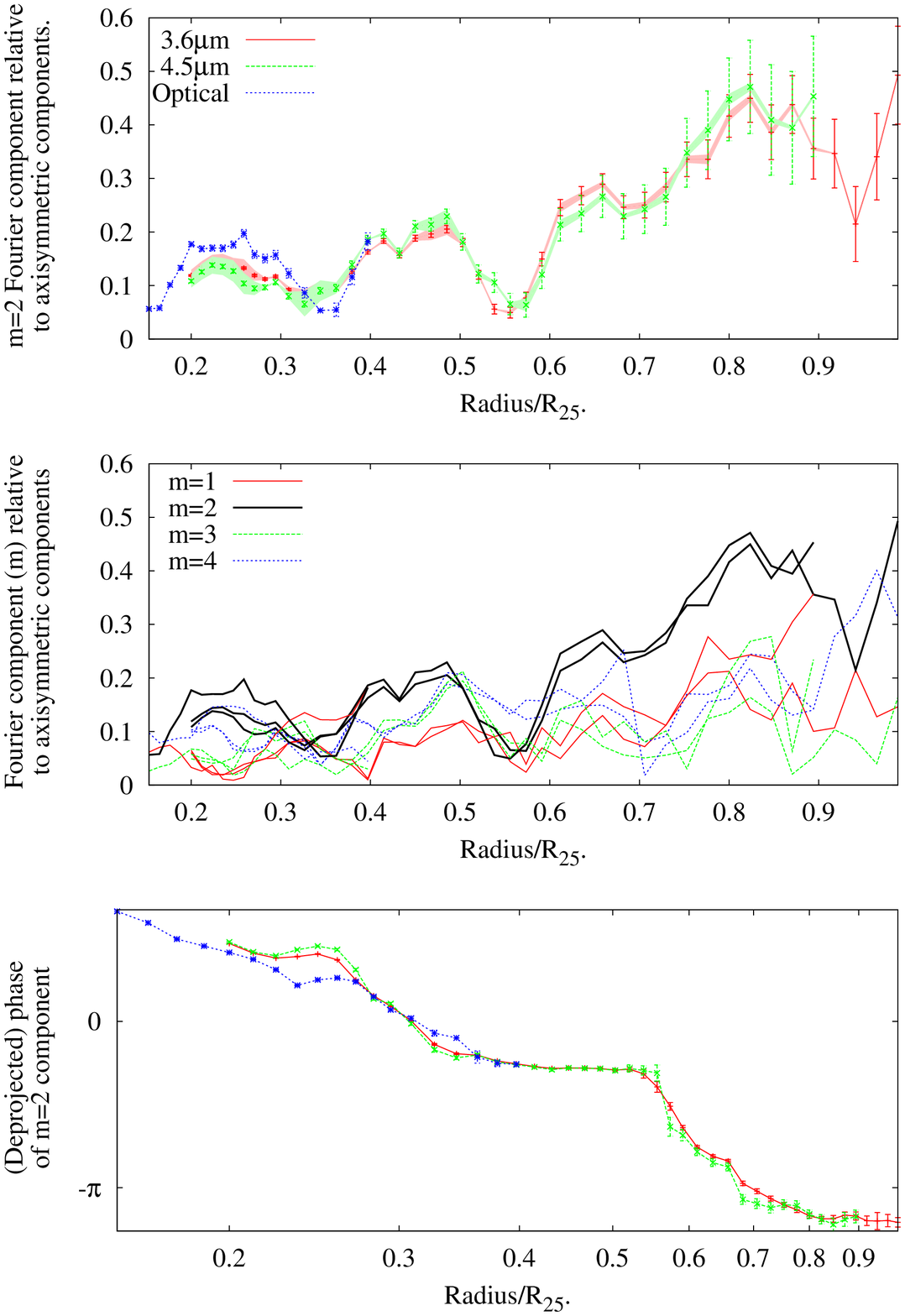}
  \caption{Azimuthal profile data for NGC 3198 (as Figure \ref{n0628_prof_dat}).}
  \label{n3198_prof_dat}
\end{figure}

Figure \ref{n3198_prof_dat} shows that the relative amplitude of the m=2 component seems to follow the standard trend of increasing with radius and it can be seen that m=2 dominates over the majority of the radial range. The local minima in the m=2 amplitude are worth considering: in previous studies, some authors have linked features such as these to resonances such as the ILR or inner 4:1 resonance (e.g \citep{1989ApJ...343..602E}). However, caution should be exercised in this interpretation; other features such as contamination from star forming regions or projection effects should also be considered. (Projection effects would probably decrease the observed arm strength near the ends of the minor axis, which is approximately where the m=2 amplitude is weakest, therefore this is a serious possibility).

\begin{figure}   \centering
  \includegraphics[width=83mm]{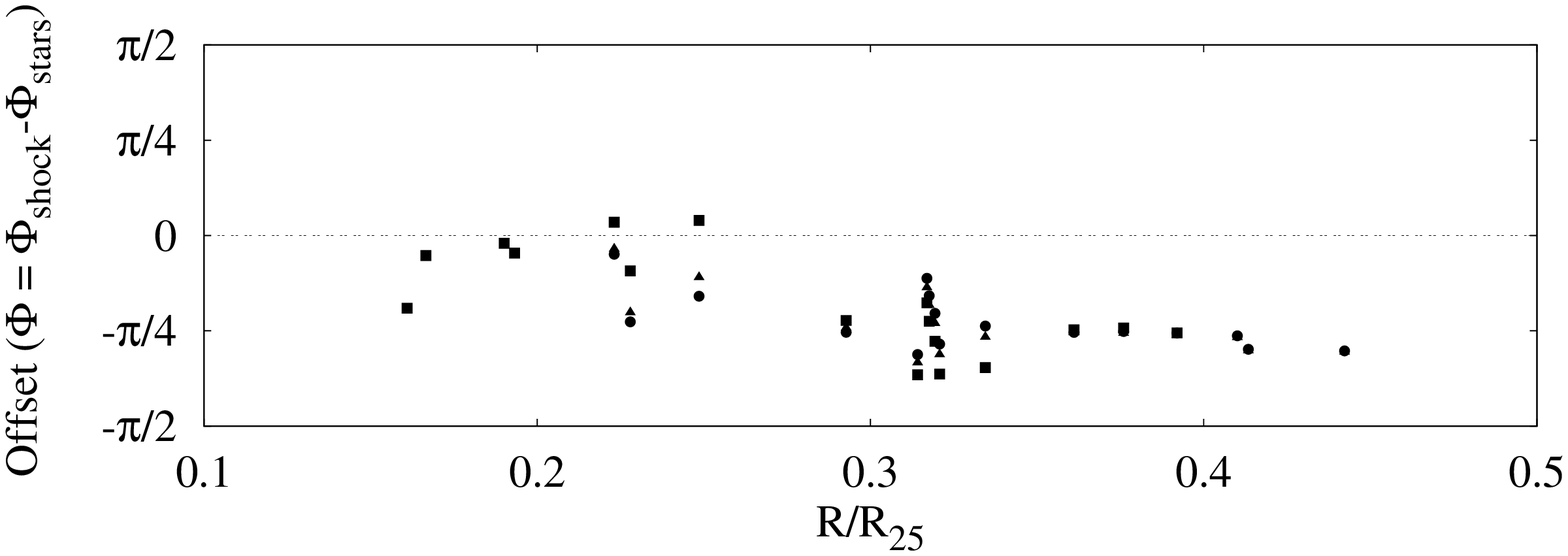}
  \caption{The offset between the gas shock and stellar mass for NGC 3198. In the case of NGC 3198 an upstream gas shock will give a negative offset.}
  \label{offset_n3198}
\end{figure}

One feature of the phase-radius plot in Figure \ref{n3198_prof_dat} is the regions such as those between 0.4-0.5R$_{25}$ where the phase does not change. From Figure \ref{n3198_resids} it can be seen that these features align with the minor axis, suggesting that this may well be an artefact due to the high inclination angle of NGC 3198, rather than a genuine feature. There is a patch of particularly strong PAH emission (and probably contributions from young stars too) on the eastern arm; when combined with the poor radial resolution along the minor axis this can probably explain the effect. If these `warps' in the phase of the arms are artefacts then the effect should smooth out over 360 degrees, and the best way to calculate the pitch angle is to trace the arms over a full revolution. If the warps are genuine features then clearly one pitch angle can not satisfactorily describe the shape of the spiral in NGC 3198. However, assuming the spirals are in fact approximately logarithmic, the average pitch angle is found to be 15$^{o}$. As with NGC 0628, the shocks could not be followed over as large a radial range as the stellar spiral, but Figure \ref{offset_n3198} shows that the shock in NGC 3198 has a clear tendency to be upstream of the stellar spiral.

\subsubsection{NGC 3938}\label{ch4_3938}
NGC 3938 is an SAc galaxy, located in the Ursa Major Cluster. It has no close ($<$ 100 kpc) companions \citep{1999A&A...342..417J}.

\begin{figure}   \centering
  \includegraphics[width=60mm]{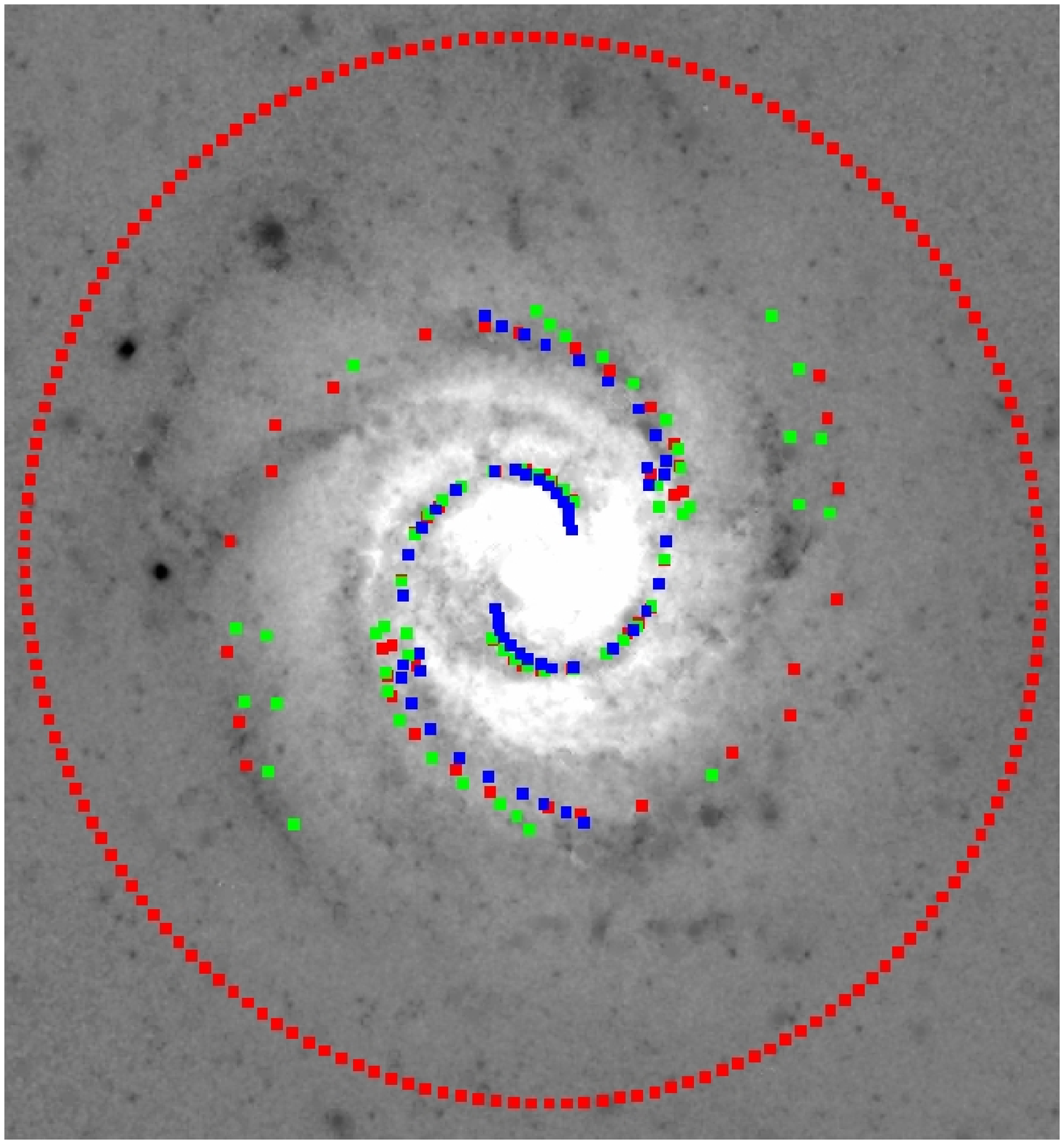}
  \includegraphics[width=60mm]{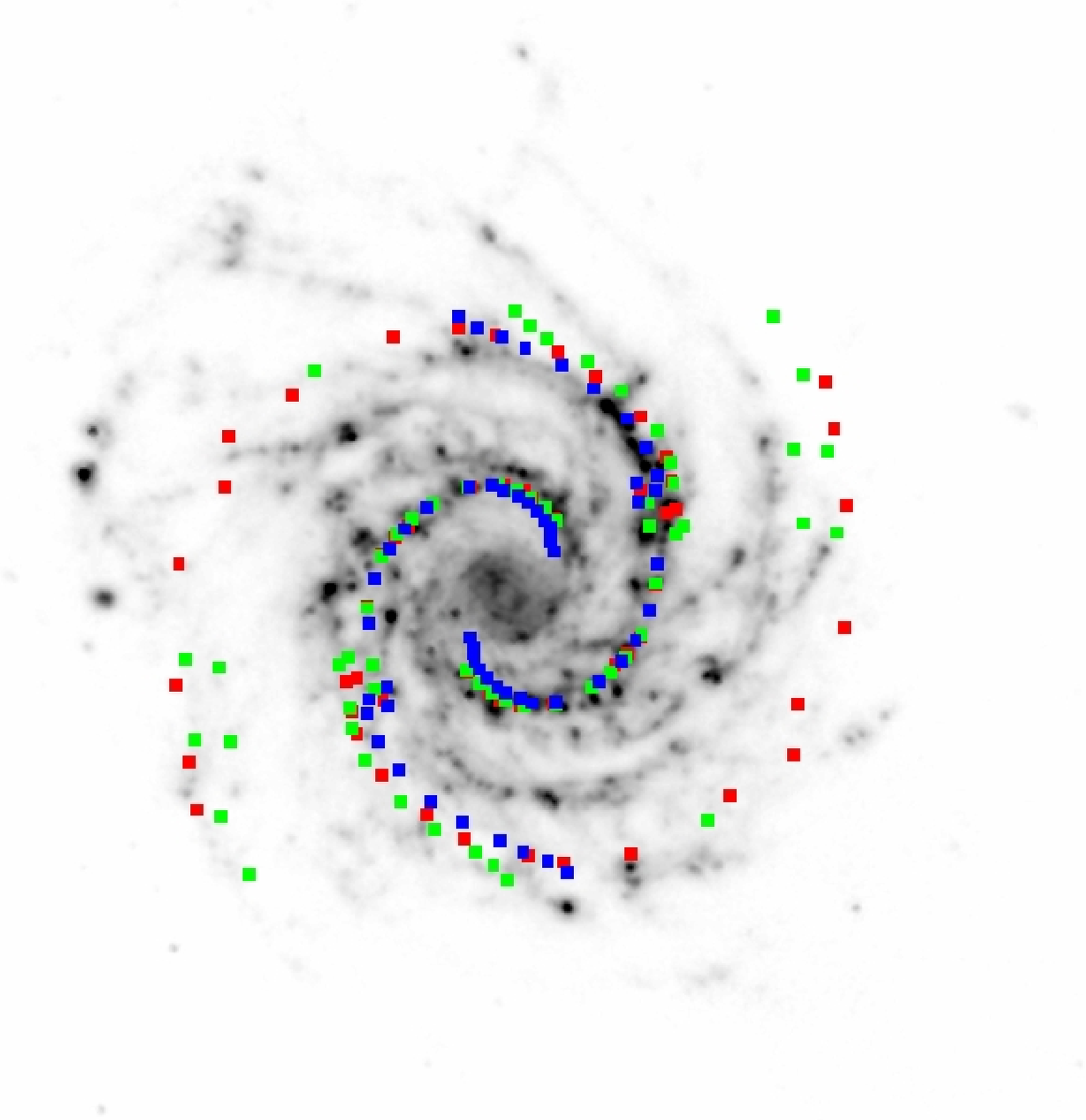}
  \caption{NGC 3938 (as Figure \protect\ref{n0628_resids}). R$_{25}$ = 2.7 arc minutes.}
  \label{n3938_resids}
\end{figure}

\begin{figure}   \centering
  \includegraphics[width=83mm]{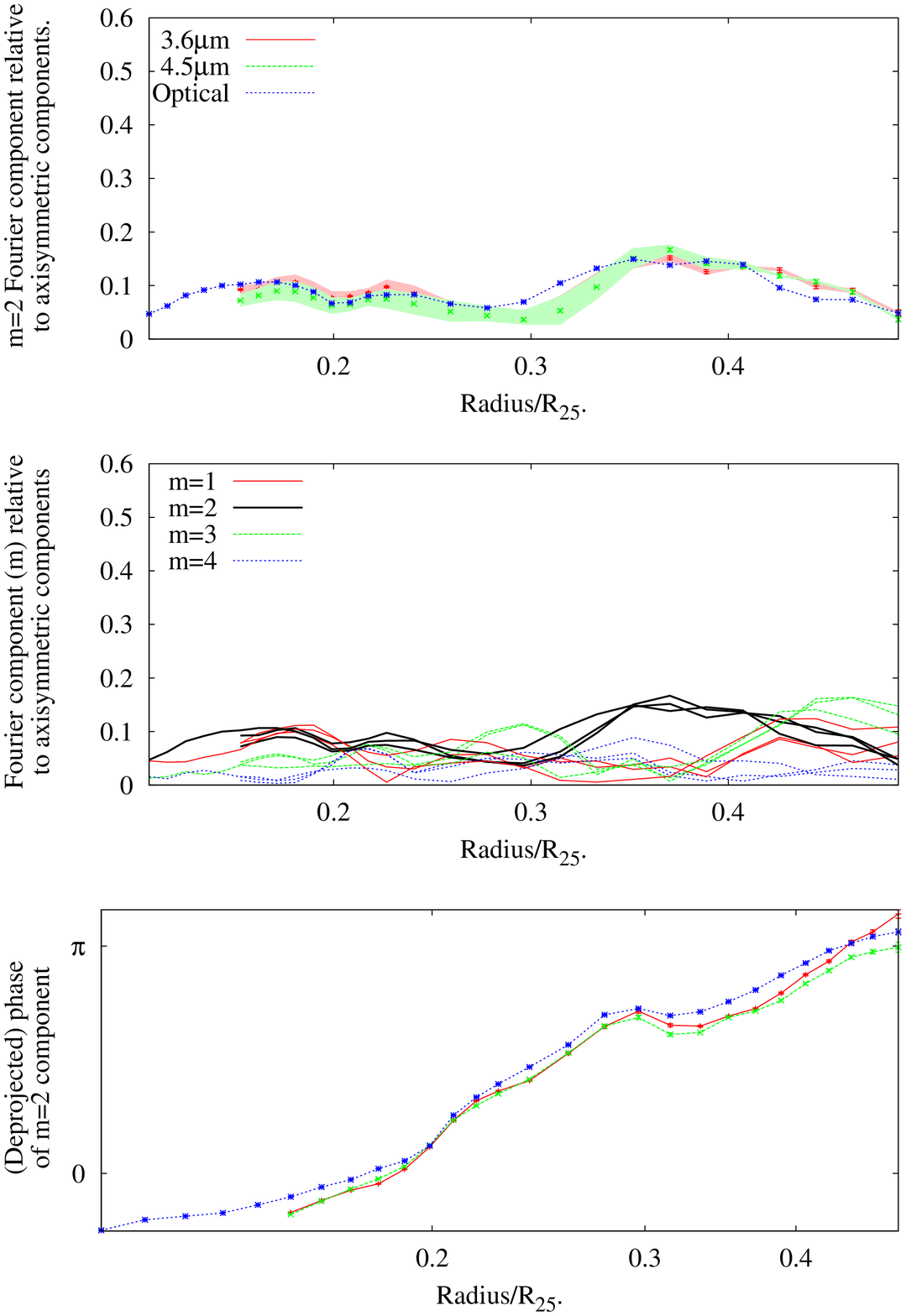}
  \caption{Azimuthal profile data for NGC 3938 (as Figure \ref{n0628_prof_dat}).}
  \label{n3938_prof_dat}
\end{figure}

NGC 3938 is a very similar spiral to NGC 0628, although interestingly the relative amplitude does not reach as large values as NGC 0628 (despite being at similar radii relative to R$_{25}$). Part of the reason for this may be that the m=1 and m=3 components appear to dominate the Fourier decomposition equally after R$\sim$ 0.45R$_{25}$, thus leaving less power for a single component (be it m=2 or anything else). The minimum in the m=2 relative amplitude between 0.5-0.6R$_{25}$ is reasonable; from examining the image it can be seen that the m=3 component should be strong around this radius as there are three approximately equally spaced regions with strong emission (at approximately 2, 6 and 10  o'clock in Figure \ref{n3938_resids}). The deviations from exact m=3 symmetry can explain the strength of the m=1 component too.

However, using the m=2 signal over the radial range indicated in Figure \ref{n3938_prof_dat} allows a pitch angle for the spiral to be calculated, which has an average value of 15$^o$. Using the m=2 component to calculate the pitch angle even when it is no longer the dominant component may seem unwise and does limit the available information, but in the case of NGC 3938 it appears to be justified; examining Figure \ref{n3938_resids} shows that the phase of the m=2 Fourier component still appears to follow the maxima in the mass distribution.

\begin{figure}   \centering
  \includegraphics[width=83mm]{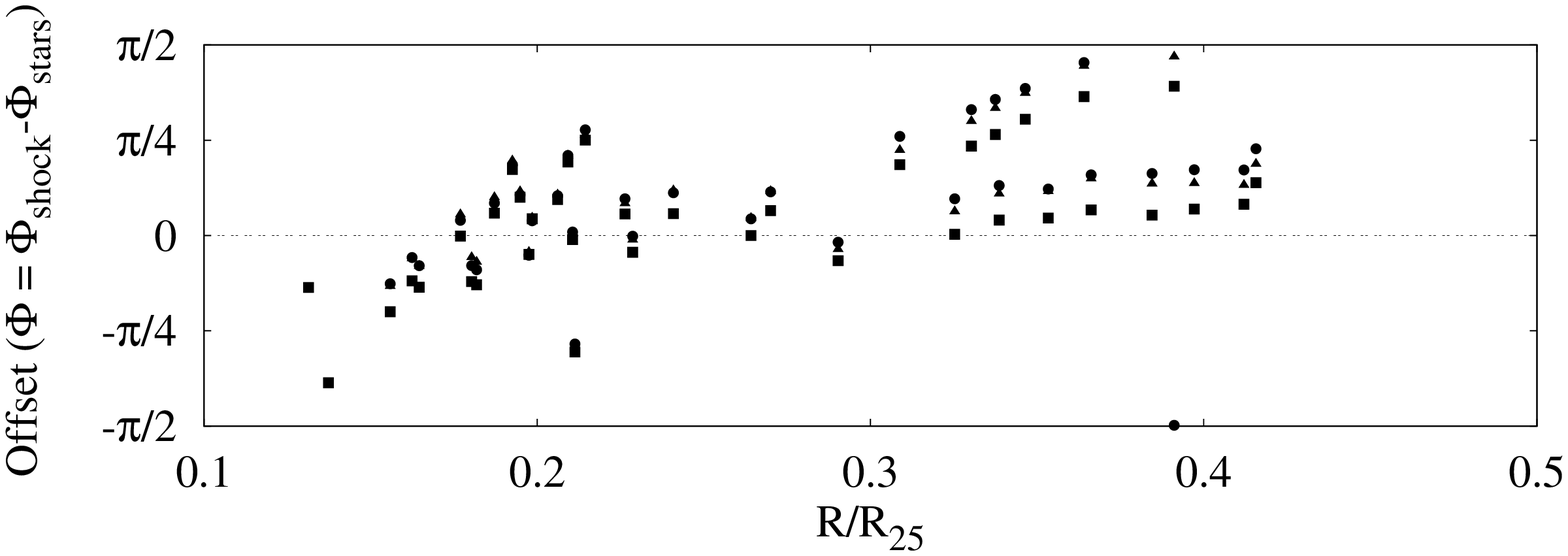}
  \caption{The offset between the gas shock and stellar mass for NGC 3938. In the case of NGC 3938 an upstream gas shock will give a positive offset.}
  \label{offset_n3938}
\end{figure}

Figure \ref{offset_n3938} is interesting, in that the offset plot clearly distinguishes the two spiral arms. The trend is for the shock to lie inside the arm (upstream) for both arms, but the offset is significantly larger for one than the other. As has already been noted this spiral is not perfectly symmetric, as evidenced by the relatively strong m=1 and m=3 components, and this will account for the different offsets measured from arm to arm.

\subsubsection{NGC 4321}\label{ch4_4321}
NGC 4321 is an SABbc type spiral in the Virgo cluster. It has two apparent dwarf companions, VCC 608 (NGC 4323) and VCC 634 (NGC 4328), which are at projected distances of only 24 and 28 kpc respectively. The HI disc is lopsided, possibly due to a past interaction, or ram pressure stripping in the Virgo cluster \citep{1993ApJ...416..563K}, and does not extend far the optical radius, again probably due to stripping of the gas.

\begin{figure}   \centering
  \includegraphics[width=60mm]{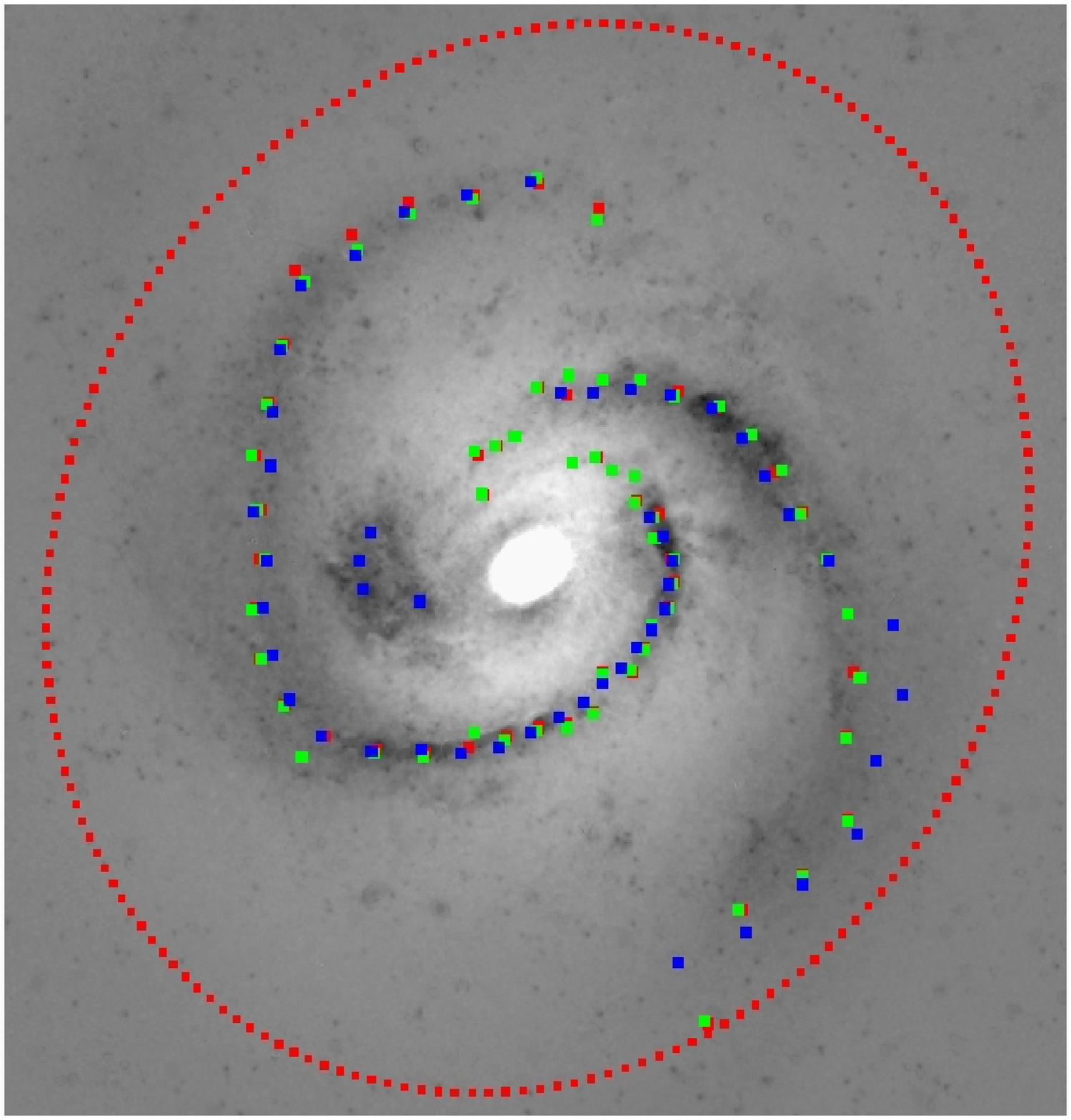}
  \includegraphics[width=60mm]{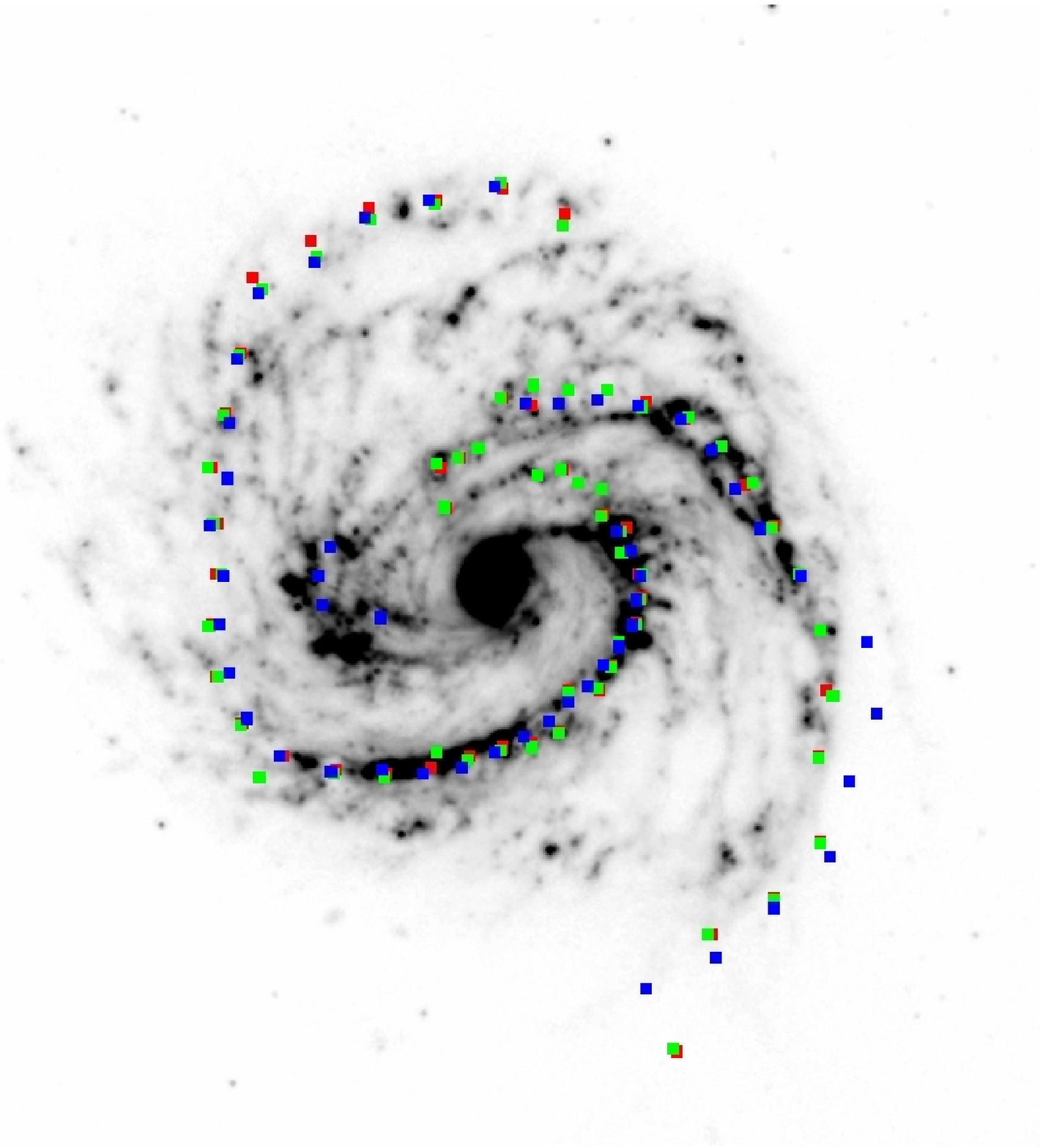}
  \caption{NGC 4321 (as Figure \protect\ref{n0628_resids2}). R$_{25}$ = 3.7 arc minutes.}
  \label{n4321_resids}
\end{figure}

\begin{figure}   \centering
  \includegraphics[width=83mm]{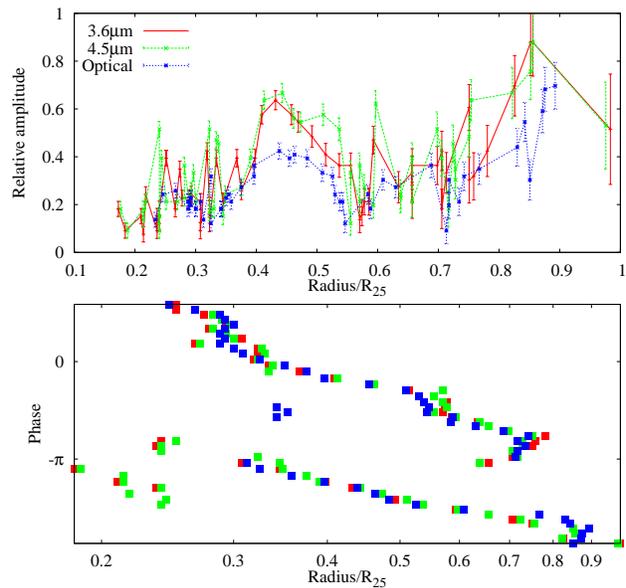}
  \caption{Radial profile data for NGC 4321 (as Figure \ref{n0628_raddat}).}
  \label{n4321_raddat}
\end{figure}

NGC 4321 had to be studied with the radial profile method, for the same reasons as NGC 3184 and NGC 5194. From the positions of the maxima in Figure \ref{n4321_resids} it can be seen that the arms appear to have a kink (particularly clear in the eastern arm), possibly due to a warp in the disc \citep{1997ApJ...479..723C}; this may well be due to the influence of one or both companion galaxies.

As can be seen from Figure \ref{n4321_raddat}, the relative amplitude has a slight peak just inside the kink, but this may well be due to the strong star formation around that region (and being radial profile data, this is particularly vulnerable to contamination). On average, the relative amplitude is stronger than that observed for most galaxies; this may be partly explained by the likelihood of contamination from localised star formation (as seen in NGC 0628 where the relative amplitude from radial profiles was observed to be higher than the corresponding azimuthal profile data), but NGC 4321 has extremely well defined spiral arms, so a large relative amplitude is not unexpected.

\begin{figure}   \centering
  \includegraphics[width=83mm]{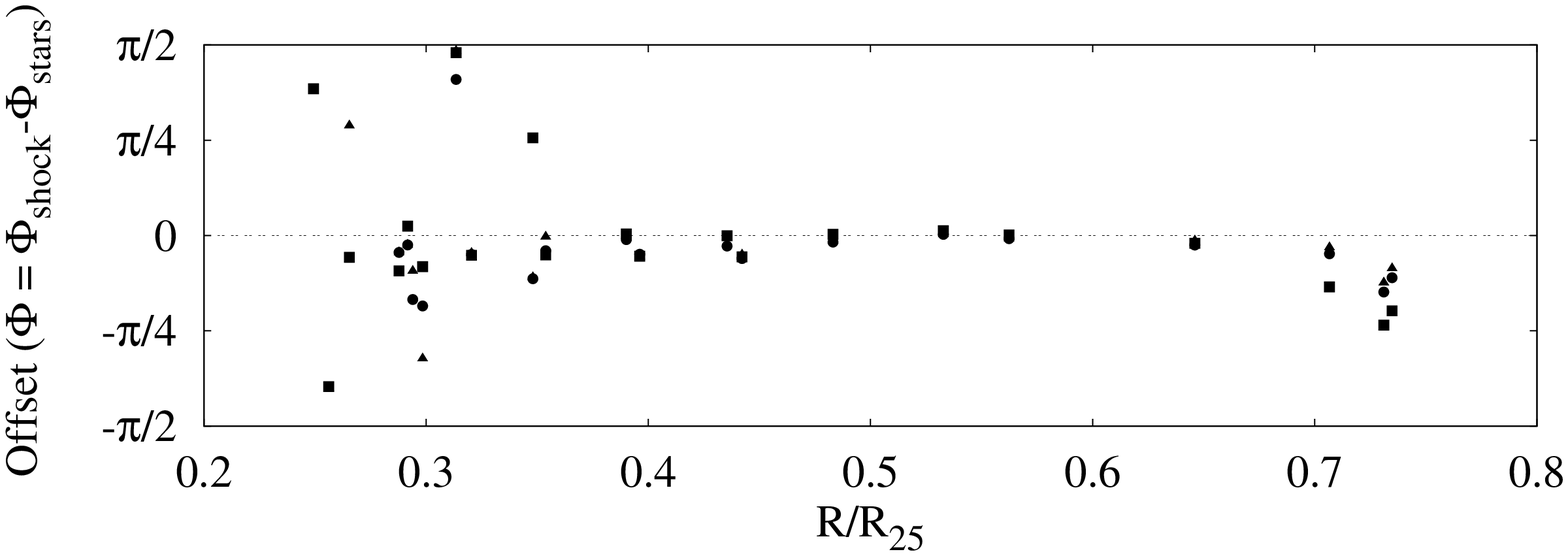}
  \caption{The offset between the gas shock and stellar mass for NGC 4321. In the case of NGC 4321 an upstream gas shock will give a negative offset.}
  \label{offset_n4321}
\end{figure}

It can be seen in Figure \ref{n4321_raddat} that, although the kink in the eastern spiral arm is clear, the trend in the azimuthal angle of the peak of
the spiral with radius is approximately logarithmic. The average pitch angle, calculated from all three wavelengths, is 20$^o$. The trend in offset with radius for NGC 4321, shown in Figure \ref{offset_n4321}, seems to be rather different to most galaxies. There is considerable scatter at the smallest radii (as can be seen in Figure \ref{n4321_resids}, the spiral arms are not well defined here, so the scatter should not be surprising). However, the shock does not appear to diverge from the stellar spiral arms except at the largest radii sampled, where the shock moves only slightly upstream of the stellar wave.

\subsubsection{NGC 4579}\label{ch4_4579}
NGC 4579 is of type SABb, and is a member of the Virgo cluster. As with NGC 2841, the radial profile method was applied to NGC 4579 after the azimuthal profile method had failed to recover evidence for an underlying spiral.

\begin{figure}   \centering
  \includegraphics[width=80mm]{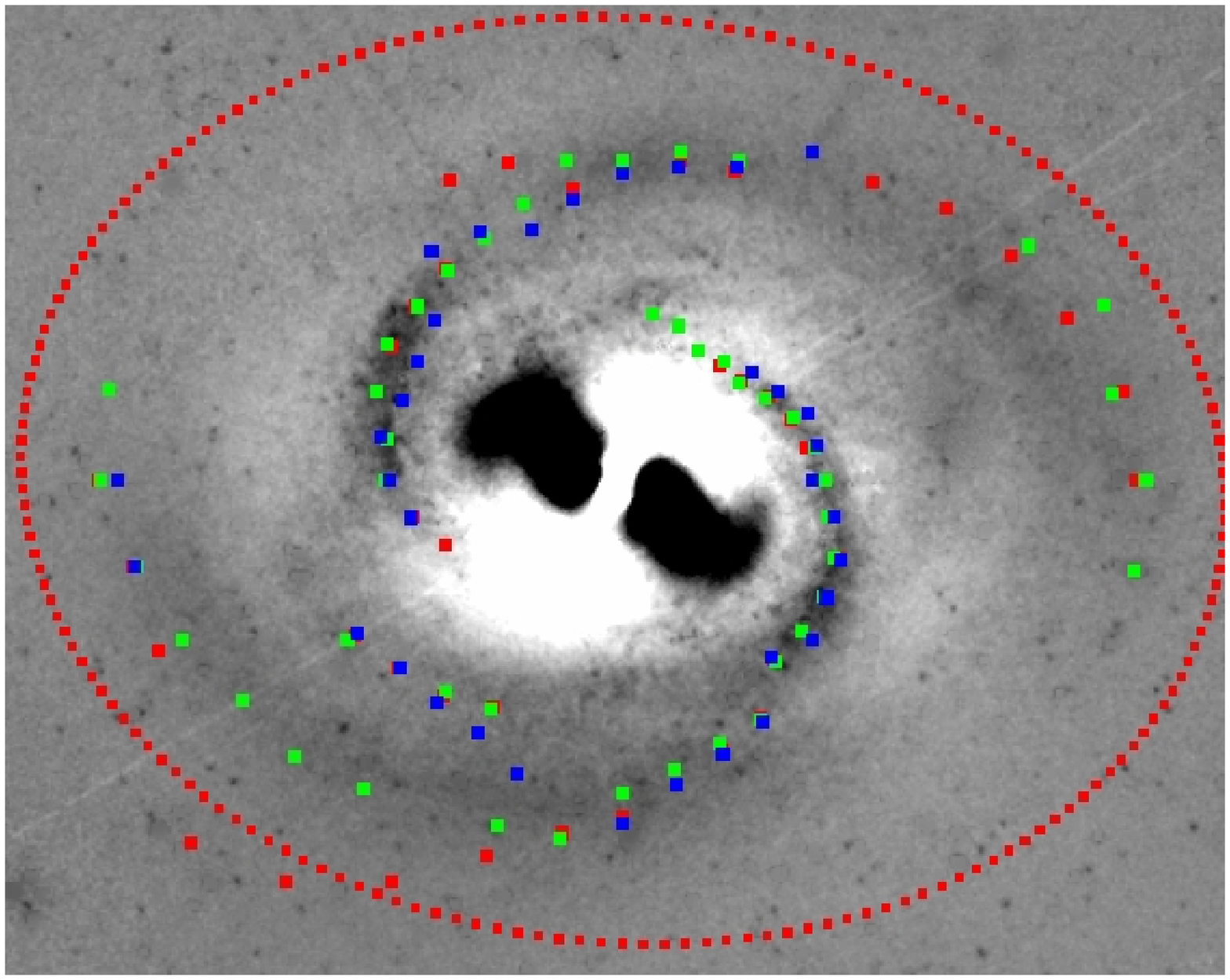}
  \includegraphics[width=80mm]{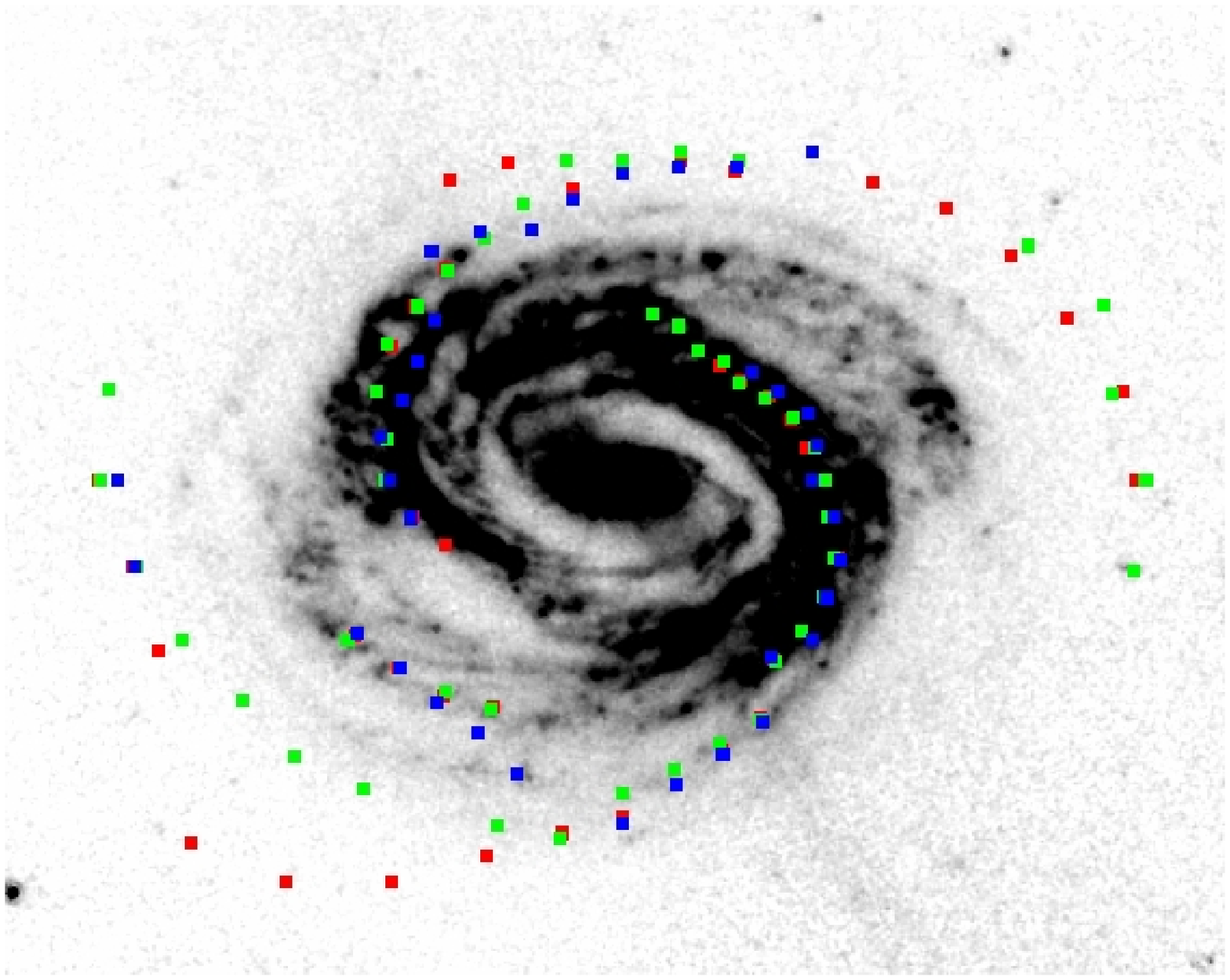}
  \caption{NGC 4579 (as Figure \protect\ref{n0628_resids2}). R$_{25}$ = 2.95 arc minutes.}
  \label{n4579_resids}
\end{figure}

\begin{figure}   \centering
  \includegraphics[width=83mm]{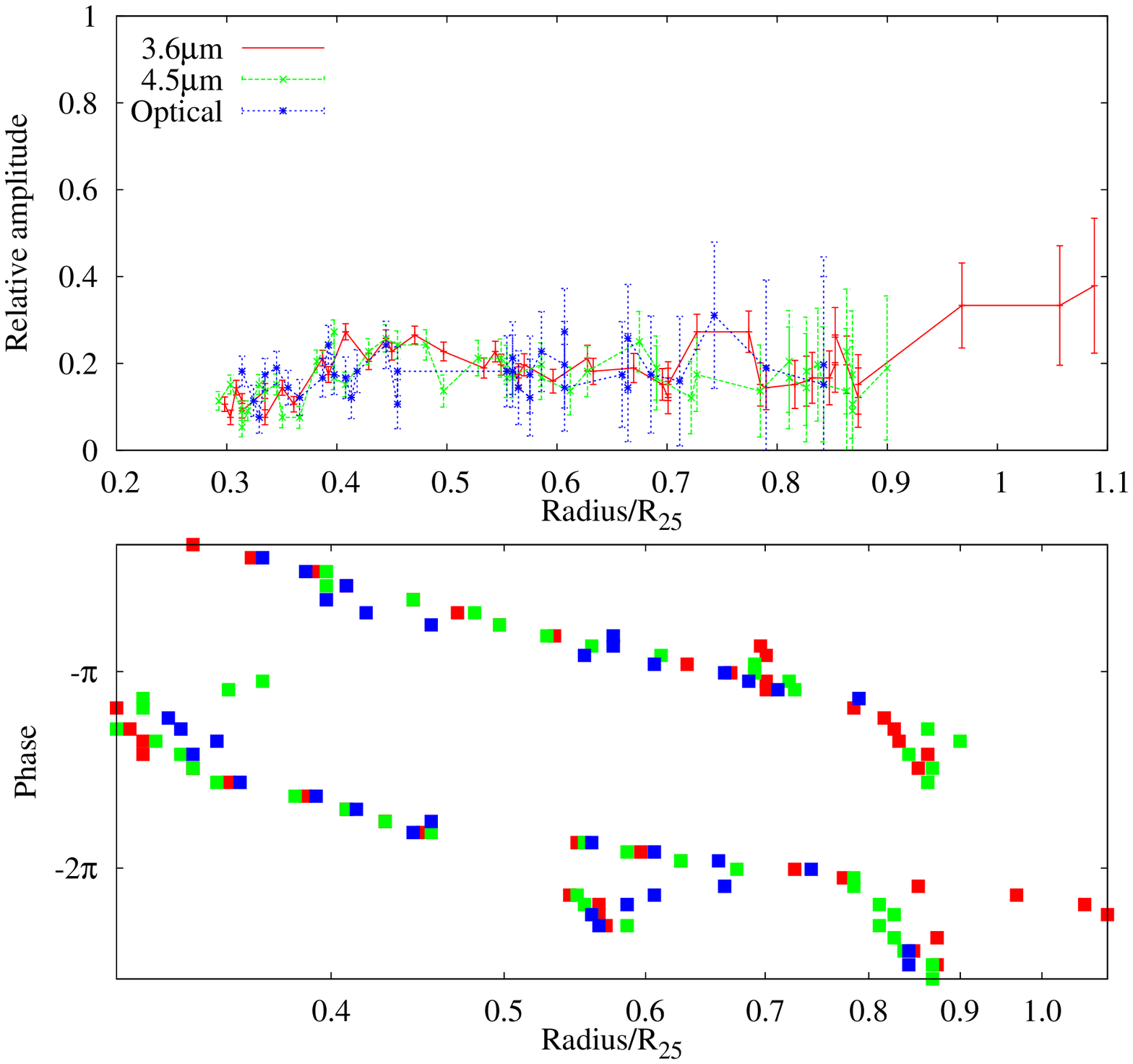}
  \caption{Radial profile data for NGC 4579 (as Figure \ref{n0628_raddat}).}
  \label{n4579_raddat}
\end{figure}

From Figure \ref{n4579_raddat} it can be seen that the relative amplitude of the spiral arms is about 20 per cent, with the expected large scatter. This amplitude is lower than for many galaxies, but not exceptionally so, and although there is a slight increase with radius the trend in relative amplitude is flatter than most. The cause is not immediately clear, but may be related to the fact that the PAH emission is much stronger inside 0.4R$_{25}$ than outside (again, the possibility of contamination from young stars/PAHs is much larger than for azimuthal profiles). The phase-radius behaviour is also fairly normal; the causes of the deviations from the logarithmic behaviour (around 0.6 and 0.8 R$_{25}$) can be seen more clearly in the 8$\mu$m image, where the points are clearly coincident with features in the gas which are mirrored in the residuals, although the contrast is less clear. Fitting a logarithmic spiral to the phase data gives an average pitch angle of 18$^o$.

\begin{figure}   \centering
  \includegraphics[width=83mm]{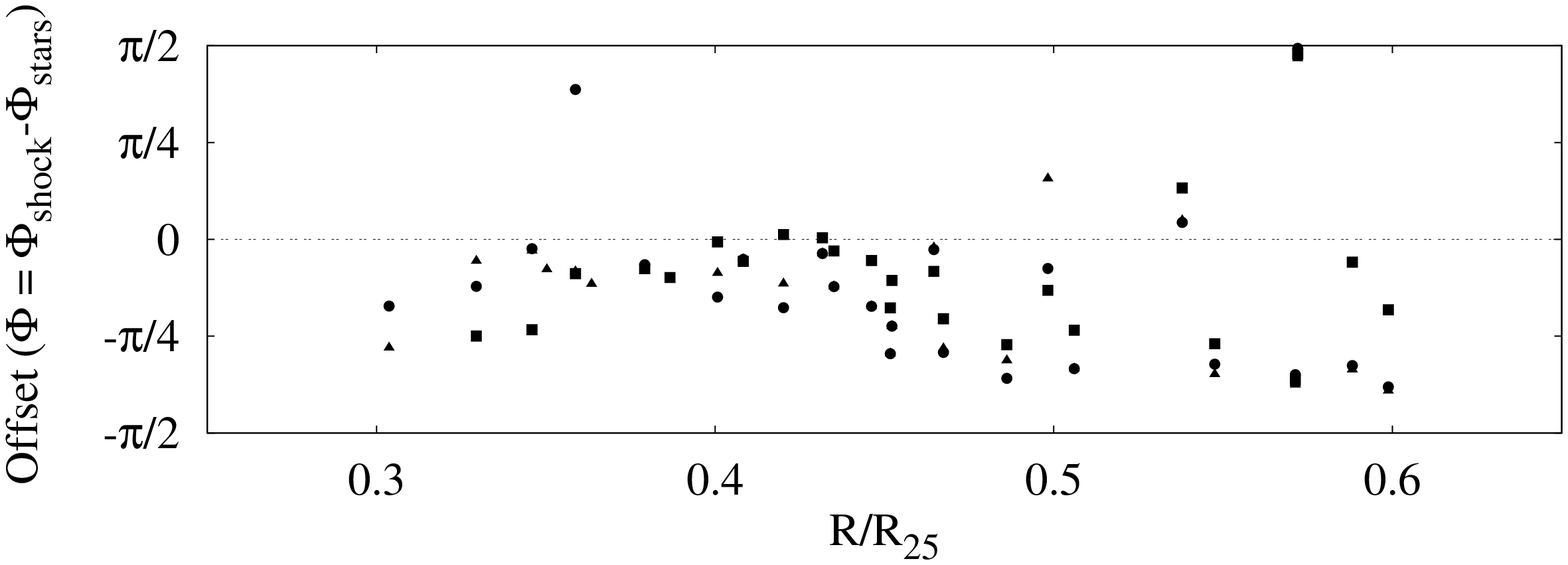}
  \caption{The offset between the gas shock and stellar mass for NGC 4579. In the case of NGC 4579 an upstream gas shock will give a negative offset.}
  \label{offset_n4579}
\end{figure}

Figure \ref{offset_n4579} shows the same general behaviour as many of the galaxies encountered so far, in that the gas shock tends to be upstream from the stellar potential minimum, albeit with a large amount of scatter in the offset at some radii (due to the deviations from logarithmic spiral behaviour around 0.6R$_{25}$). NGC 4579 is perhaps unusual in that the shock is consistently upstream, even at the smallest radii sampled.

\subsubsection{NGC 5194}\label{ch4_5194}
NGC 5194, which is classified as type SABbc, forms a strongly interacting pair with NGC 5195 (M51b). Many simulations of the interaction between M51 and M51b have been performed; recent work includes models by \cite{2000MNRAS.319..377S} and \cite{2003Ap&SS.284..495T} which indicate that the most likely scenario is for M51b to be in a bound orbit around M51.

\begin{figure}   \centering
  \includegraphics[width=60mm]{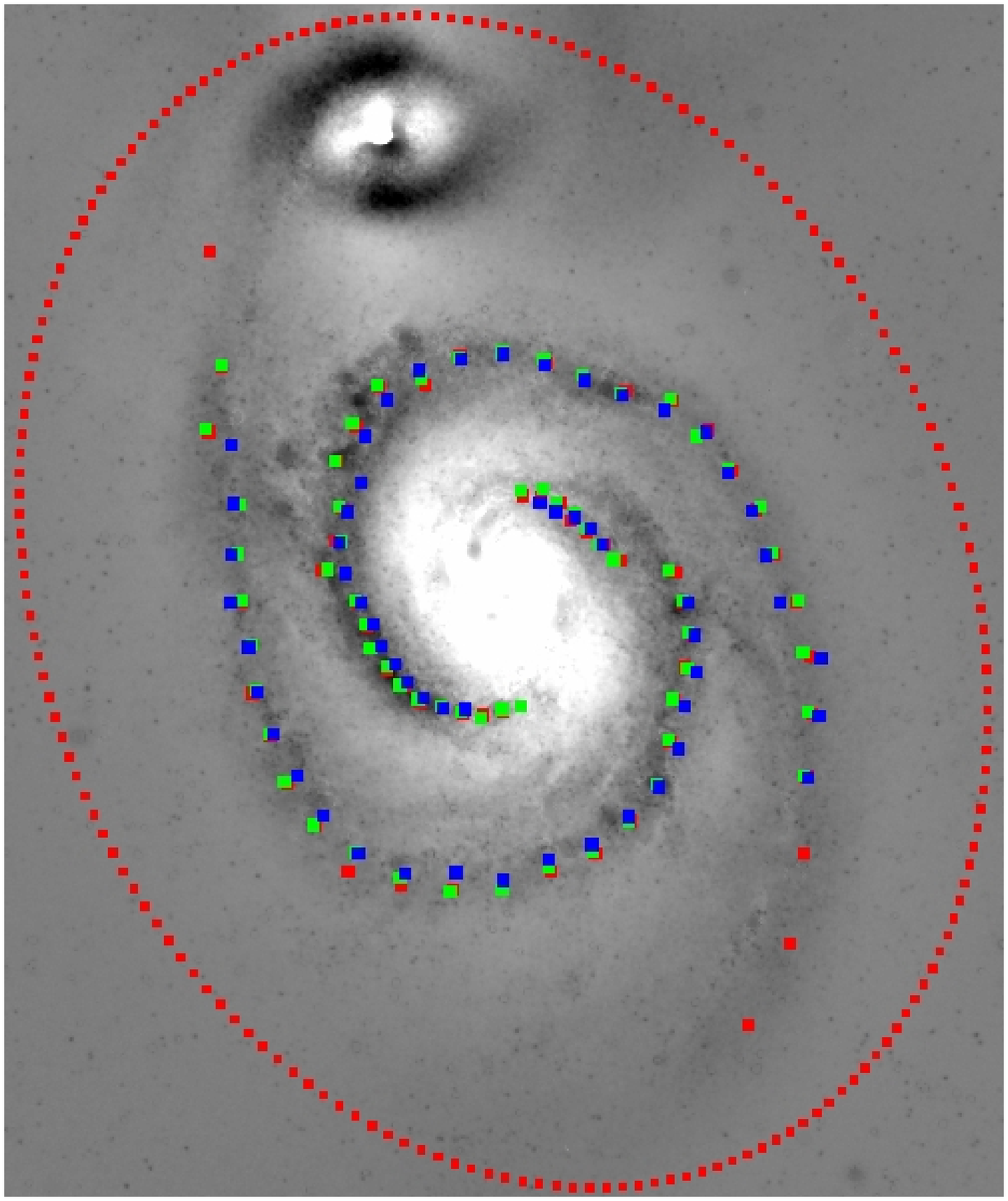}
  \includegraphics[width=60mm]{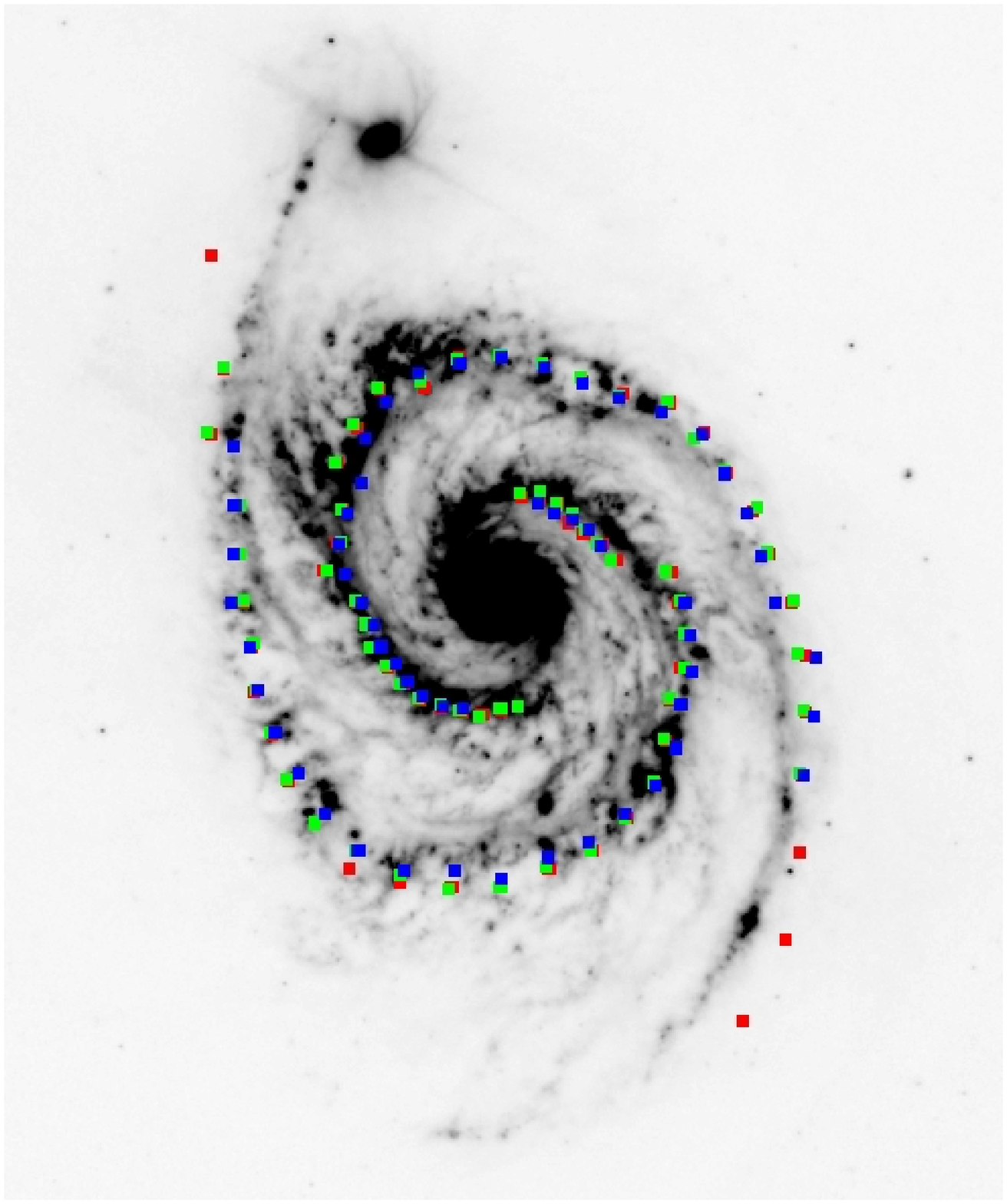}
  \caption{NGC 5194 (as Figure \protect\ref{n0628_resids2}). R$_{25}$ = 5.6 arc minutes.}
  \label{n5194_resids}
\end{figure}

\begin{figure}   \centering
  \includegraphics[width=83mm]{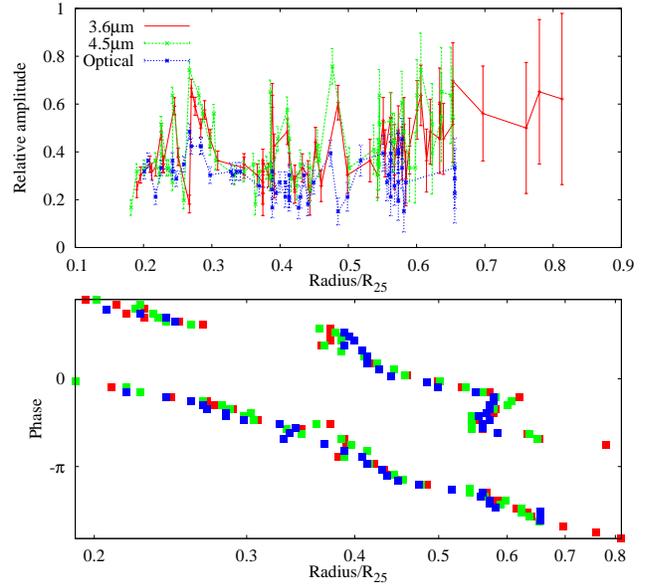}
  \caption{Radial profile data for NGC 5194 (as Figure \ref{n0628_raddat}).}
  \label{n5194_raddat}
\end{figure}

As already described, NGC 5194 was not successfully analysed with azimuthal profiles, so instead the radial profile method was used. The general trend for relative amplitude to increase with radius is not obeyed by NGC 5194, although the scatter in measurements is particularly large. As with NGC 4579, the fact that NGC 5194 does not show an increase in relative amplitude with radius may be due to the fact that the gas response appears to be stronger in the central regions than in most other galaxies in this sample (as witnessed by increased star formation). The scatter may be attributed to the high rates of star formation on the arms, and susceptibility of the radial profile method to contamination from areas of star formation. 

As with many of the galaxies for which the radial profile method was used, the spiral arms do not follow perfect logarithmic spirals. In this case, the arms are clearly kinked, possibly due to a warp in the disc of NGC 5194 caused by the close passage of the companion NGC 5195, or alternatively the kinks may be a direct result of tidal forces caused by the companion. (The latter interpretation is preferred by \cite{2010MNRAS.403..625D} who run simulations of the interaction between NGC 5194 and NGC 5195 using both stellar and gaseous components, and find a distinct switch in behaviour at $\sim$4kpc ($\sim$0.3R$_{25}$) between a more stable spiral inside and purely shearing spiral arms outside). In either case, the irregularities show up clearly on the plot of phase vs radius in Figure \ref{n5194_raddat}. It is also notable that NGC 5194 does not display 180$^o$ rotational symmetry - one of the arms is much closer to a logarithmic spiral than the other, which is also easily attributed to the effects of the companion. However, despite the irregularities it is possible to fit a straight line to the phase-radius plot and calculate an average pitch angle of 14$^o$, making NGC 5194 one of the tighter wound spirals in this sample.

\begin{figure}   \centering
  \includegraphics[width=83mm]{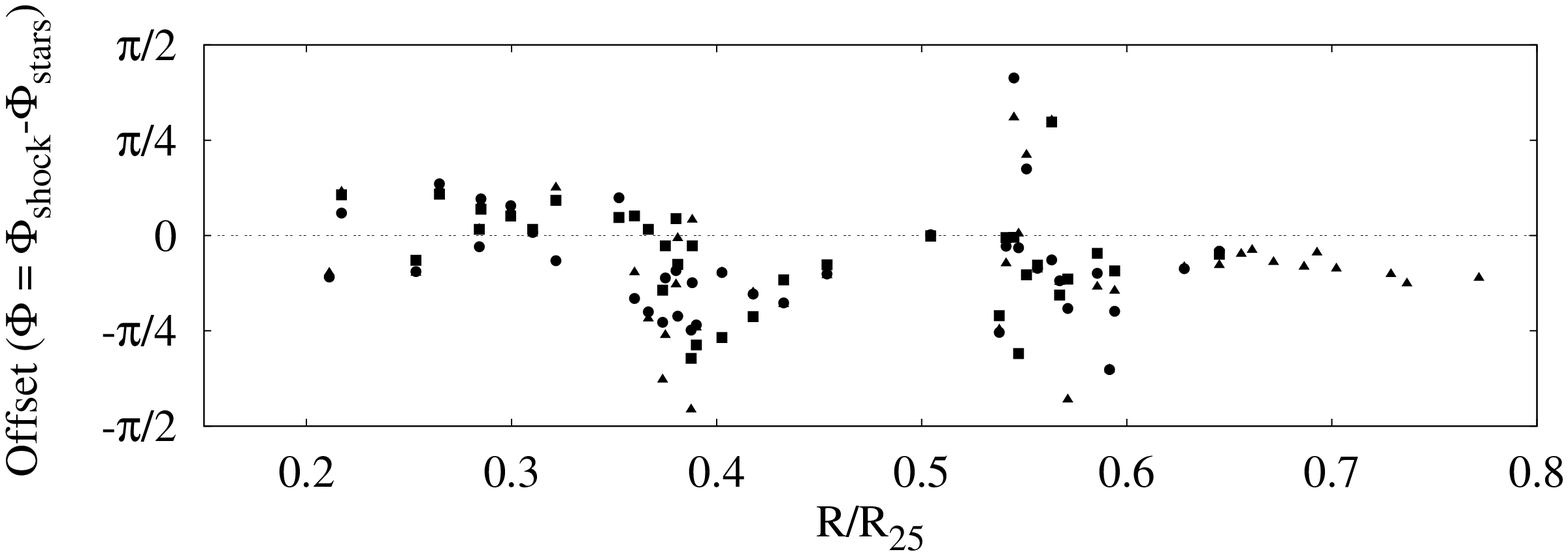}
  \caption{The offset between the gas shock and stellar mass for NGC 5194. In the case of NGC 5194 an upstream gas shock will give a negative offset.}
  \label{offset_n5194}
\end{figure}

Figure \ref{offset_n5194} shows the offset between the stellar spiral and gas shock. This plot shows that the shock is generally, but not consistently, upstream of the stellar wave, however there is a great deal of fluctuation. This variation is probably a manifestation of the strong tidal interactions that NGC 5194 and NGC 5195 are undergoing, and may well be further influenced by the fact that the disc may not be flat. \citeauthor{2010MNRAS.403..625D} look for offsets between the peak of the gas density and the stellar wave, and find that offsets tend to be small, with any variations being local and transient. The offsets found by \citeauthor{2010MNRAS.403..625D} are slightly smaller than those
that we observe but the simulations, like the data for this galaxy, find
that the sign of the offset fluctuates with radius.

\subsubsection{NGC 6946}\label{ch4_6946}
NGC 6946, which is of type SABcd, is described as `relatively isolated' \citep{2001A&A...371..433I}. Surveys suggest that there are a number of companions in bound orbits around NGC 6946, but that tidal effects are undetected, and so the system cannot be strongly interacting \citep{2000MNRAS.319..821P}.

\begin{figure}   \centering
  \includegraphics[width=80mm]{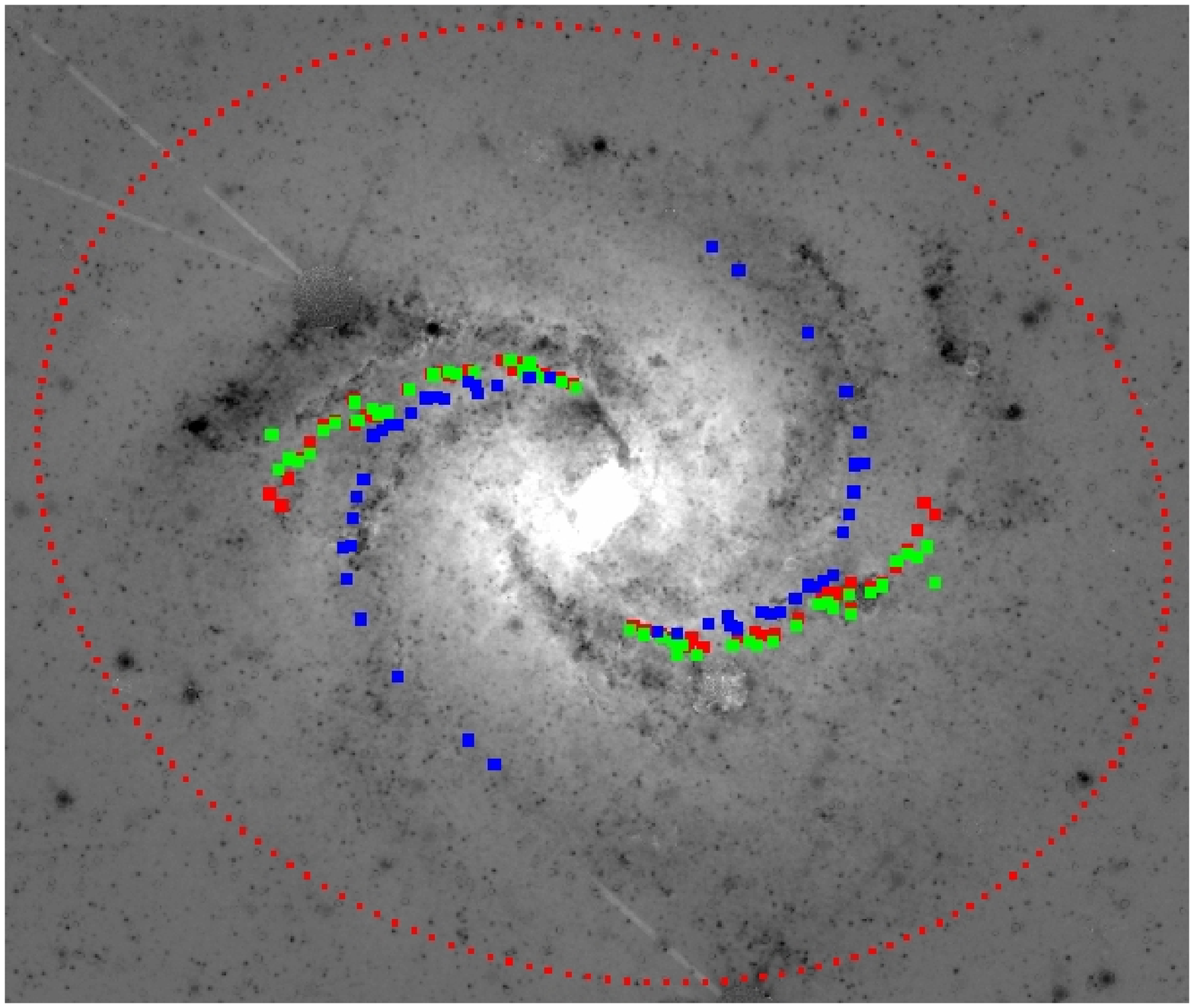}
  \includegraphics[width=80mm]{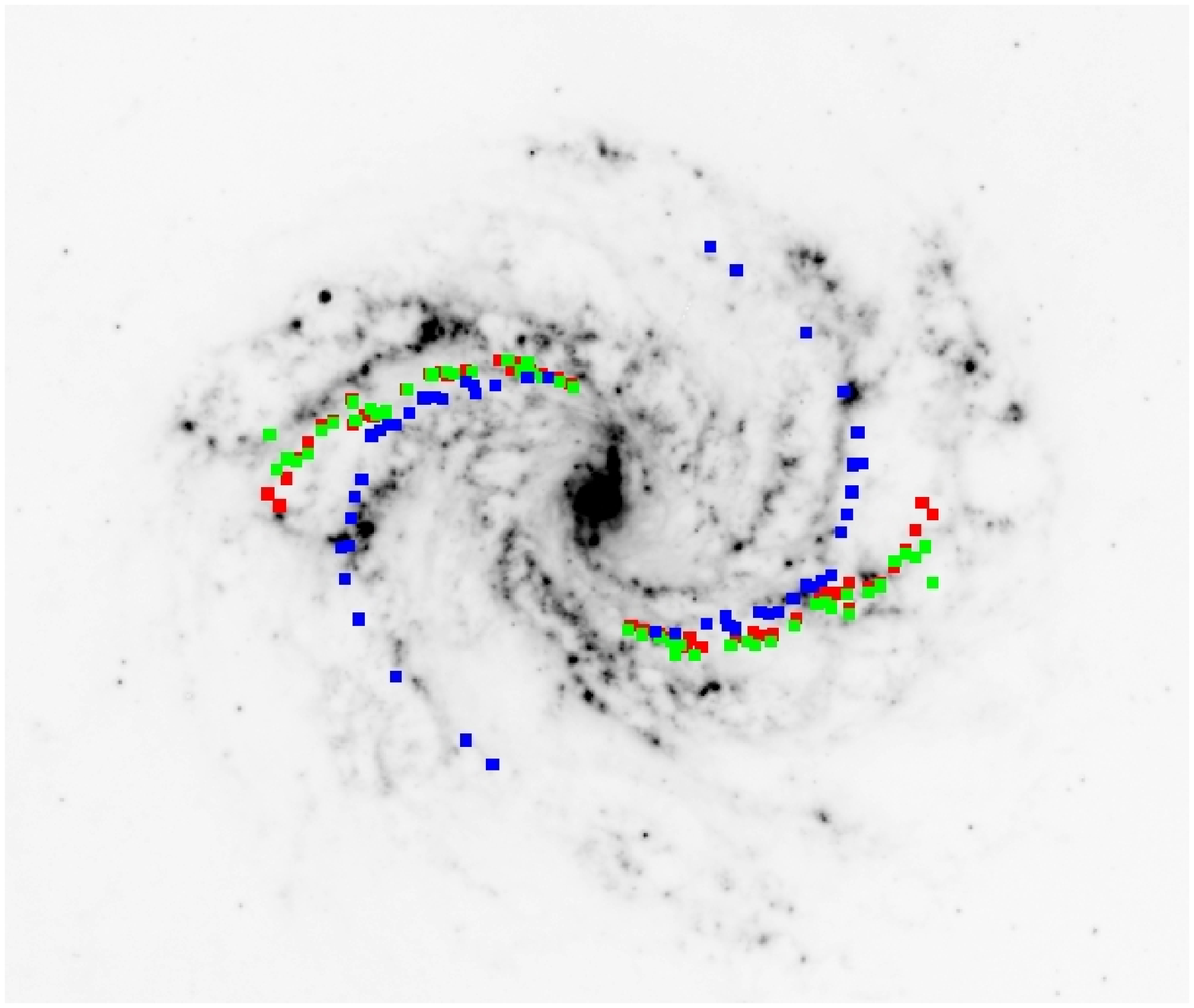}
  \caption{NGC 6946 (as Figure \protect\ref{n0628_resids}). R$_{25}$ = 5.75 arc minutes.}
  \label{n6946_resids}
\end{figure}

\begin{figure}   \centering
  \includegraphics[width=83mm]{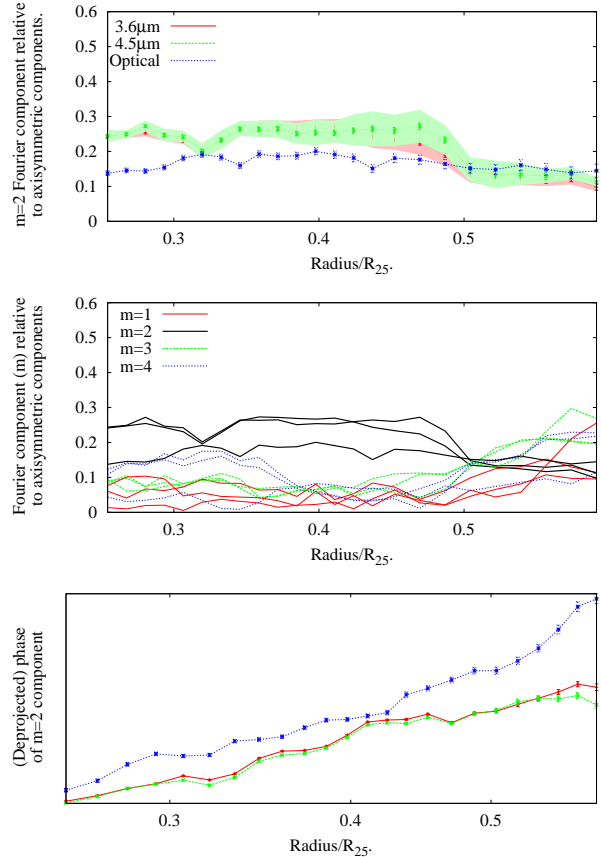}
  \caption{Azimuthal profile data for NGC 6946 (as Figure \ref{n0628_prof_dat}).}
  \label{n6946_prof_dat}
\end{figure}

This is an interesting case. Clearly if there is a single underlying two armed spiral it is very weak, because the gas response and residual image both show that from $\sim$0.3R$_{25}$ the spiral structure show indications of four armed structure, and indeed the Fourier breakdown shows that m=2 is only dominant up to 0.5R$_{25}$. Beyond 0.5R$_{25}$ the pattern is asymmetric, as witnessed by the relative strength of m=1 (optical) and m=3 (optical and IRAC data), and in the IRAC bands the m=4 component is also of a similar strength. As can be seen from Figure \ref{n6946_resids}, there is some disagreement between the IRAC and optical data over the position of the dominant mode. The \textit{I} band colour-corrected data is clearly detecting more tightly wound m=2 spiral components than the IRAC data, with a significant difference in amplitude. This is the only galaxy where the results vary to this degree; the IRAC data places the main peak of the m=2 component on the outer of the two spiral arms, and the optical data places the main m=2 peak on the inner spiral. The reason for the difference is probably due to a combination of factors; firstly the four-armed and strongly asymmetric nature of NGC 6946 means that the m=2 component is clearly not going to provide an ideal description of the pattern. In addition, the relative importance of regions with strong star formation and associated PAH emission probably explain the emphasis on one set of spiral arms or the other between the IRAC and optical data; it is clear from looking at the residual image that the fine structure is not fully removed even after the PAH correction and XZAP have been applied, and in the optical mass surface density image the effects of dust lanes are still apparent, even after the colour correction. It is not easy to determine which of these interpretations of the data is more correct. Examining the 2MASS \textit{K$_{s}$} image of NGC 6946 suggests that the optical data may in fact be more accurate in this case, but as described in KKCT08, the 2MASS data are not deep enough to make a full analysis worthwhile. In the analysis of the results the data for each of the three wavelengths are presented separately, so the IRAC bands and optical data can be considered independently.

The average pitch angle for the three wavelengths is 28$^o$, but it should be noted that this hides a large spread in measured pitch angles between the two IRAC bands (29$^o$ average) and the more tightly wound spiral detected by the optical data (24$^o$). There is no offset data for NGC 6946 because the gas response is too flocculent to be able to identify a shock.

\subsubsection{NGC 7793}\label{ch4_7793}
NGC 7793 is an SAd type spiral galaxy in the Sculptor Group, which has an unusually large HII disc. NGC 7793 is also listed in the \cite{2008AstBu..63..299K} catalogue of isolated binary galaxies, with a companion at a projected distance of 165kpc.

\begin{figure}   \centering
  \includegraphics[width=80mm]{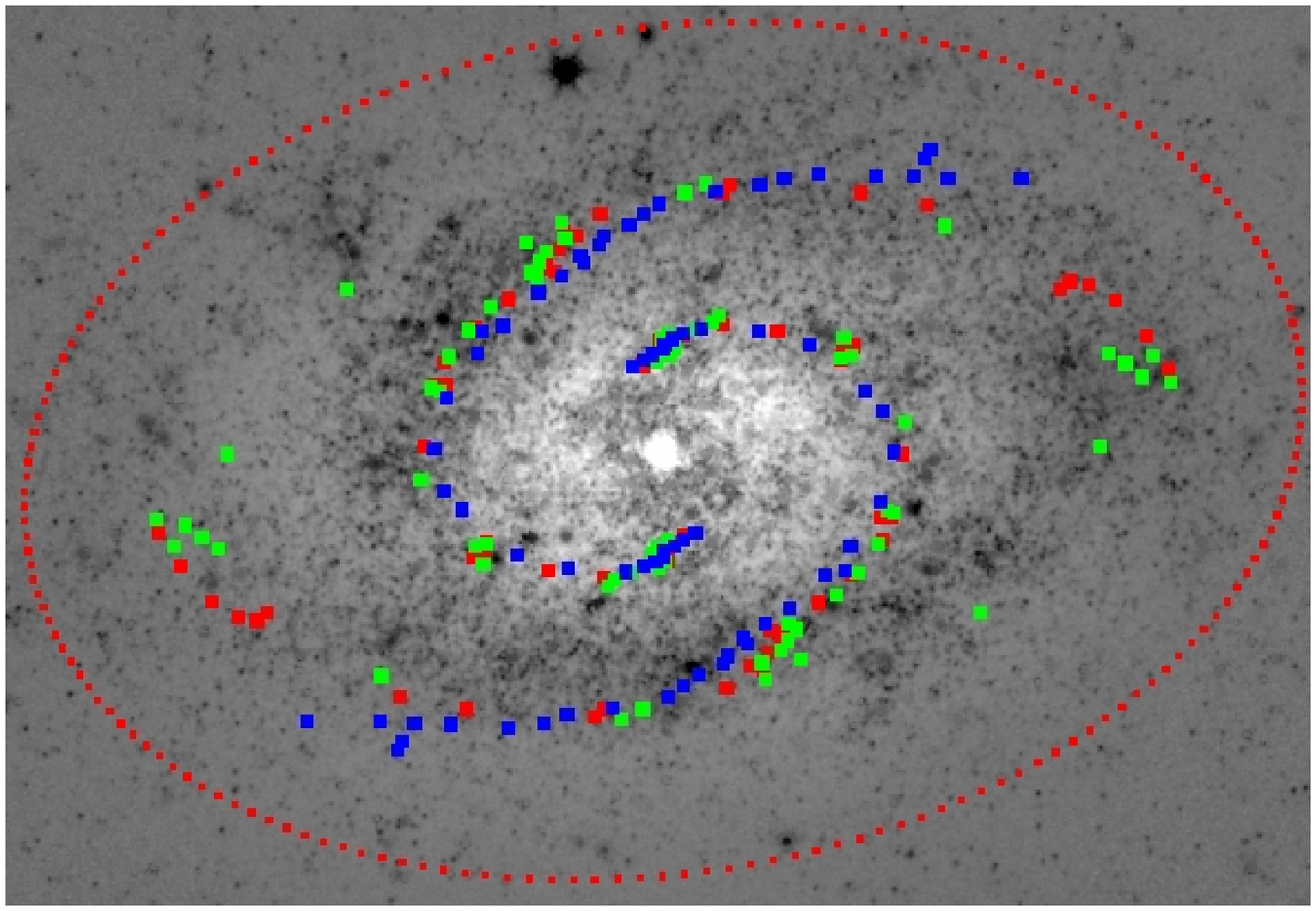}
  \includegraphics[width=80mm]{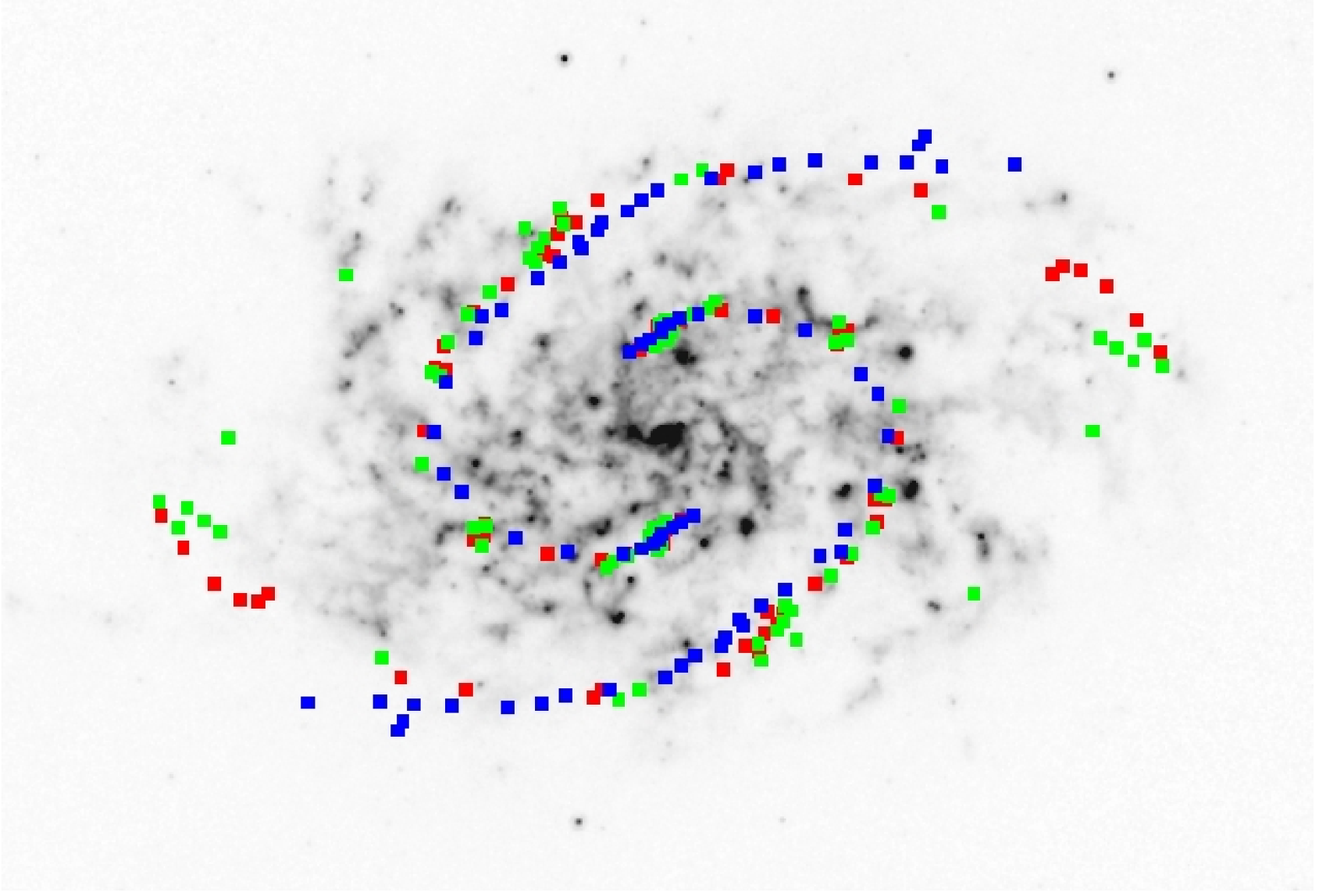}
  \caption{NGC 7793 (as Figure \protect\ref{n0628_resids}). R$_{25}$ = 4.65 arc minutes.}
  \label{n7793_resids}
\end{figure}

\begin{figure}   \centering
  \includegraphics[width=83mm]{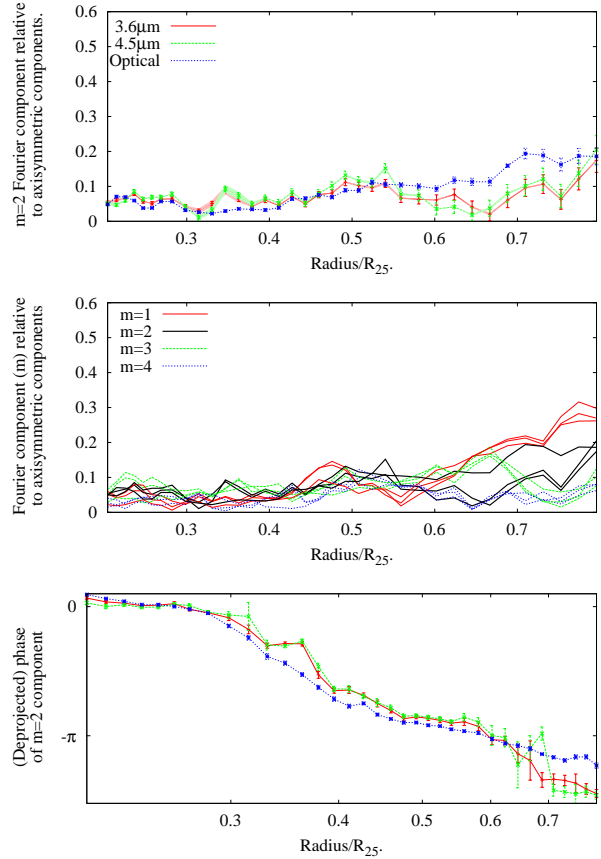}
  \caption{Azimuthal profile data for NGC 7793 (as Figure \ref{n0628_prof_dat}).}
  \label{n7793_prof_dat}
\end{figure}

NGC 7793 is extremely flocculent in the optical (Elmegreen class 2), but as with NGC 2403 and NGC 2841, there appears to be underlying spiral structure in the stellar mass, and although this is not obvious from the residual image (Figure \ref{n7793_resids}), with the identified m=2 phase to guide the eye the fit looks plausible. It should be noted, however, that the m=2 mode is not dominant over any of the other Fourier components, and indeed m=1 is in fact the strongest component beyond 0.6R$_{25}$, suggesting that the galaxy is slightly lopsided. The lack of dominance of a single mode may in part explain the fairly low values of the relative amplitude (either m=2 or any other), but this is consistent with expectations for a flocculent galaxy, where more power is predicted to be in the higher order Fourier components (m $\gg$ 4) rather than low-m values.

Examining the m=2 phase as a function of radius (Figure \ref{n7793_prof_dat}), it can be seen that the spiral arms are not perfect logarithmic spirals, but the general trend is approximately linear, and can be used to calculate a pitch angle for the m=2 spiral arms of 16$^o$ (averaged over the three wavelengths used). The gas response is too flocculent to be able to identify a shock, so no offset data is available for NGC 7793

\subsection{Other spirals.}\label{ch4_results_other}

The spirals which are not included in the detailed sample are listed in Table \ref{ch4_tab4} (hereafter the `non-detailed' sample). The reasons for excluding these galaxies vary; in some cases it is simply because there does not appear to be grand design spiral structure present. 

 However, as noted at the beginning of Section 3, there are five galaxies
(i.e. NGC 4254, NGC 3627, NGC 1097, NGC 4536, NGC 4450, which together
constitute the `additional sample' ) which we consider
to be `grand design' in the NIR but which we find to be unsuitable
for detailed analysis. Briefly considering each of these in turn:
NGC 4254 is a highly asymmetric galaxy, with one strong arm with
high pitch angle on one side but with a disordered set of two or three
stubby arms on the opposite side. NGC 3267 is likewise asymmetric with
bifurcation of the arm on one side. NGC 1097 is a textbook grand design spiral
but whose analysis is complicated by its strong bar, from whose ends
emanate  two very
narrow spiral arms. These  wrap around  tightly through 180$^o$ and then form 
more open spiral arms after the first half-turn. The strength of the bar and associated morphology present a unique challenge in studying its spiral structure. In addition, correcting for the resulting PAH emission is hindered by the fact that the 8$\mu$m image is saturated in the nucleus, leading to charge bleeding effects in a direction approximately parallel to the bar. In an attempt to account for
the presence of the bar, an extra component was added to the GALFIT model
of this galaxy. However, the morphology around the bar is not well described
by a combination of bar plus logarithmic spiral and we were finally only able to
characterise the structure in a narrow radial range, well beyond
the bar. For this reason, NGC 1097 is omitted from the detailed sample. 
Turning to NGC 4536, analysis is again complicated by a bar and
strong circumnuclear star formation. Finally, NGC 4450 is an early type
galaxy with essentially no star formation in the disc and where  the spiral
arms (although discernible by eye) are too faint for our analysis.

\begin{table}
  \centering
  \begin{tabular}{|l|l|l|l|}
    \hline
    Galaxy & R$_{25}$ (arc min) & axis ratio (b/a) & PA (degrees)\\
    \hline
    NGC 0925 & 5.25 & 0.51 & --76 \\
    NGC 1097 & 4.65 & 0.75 & --56 \\
    NGC 1291 & 4.9 & 0.94 & --53  \\
    NGC 1512 & 4.45 & 0.60 & 47  \\
    NGC 2976 & 2.95 & 0.47 & --38 \\
    NGC 3351 & 3.7 & 0.87 & 4 \\
    NGC 3521 & 5.5 & 0.44 & --16 \\
    NGC 3621 & 6.15 & 0.46 & --21 \\
    NGC 3627 (M66) & 4.55 & 0.49 & 1 \\
    NGC 4254 (M99) & 2.7 & 0.75 & 46 \\
    NGC 4450 & 2.6 & 0.62 & --6 \\
    NGC 4536 & 3.8 & 0.42 & --68 \\
    NGC 4559 & 5.35 & 0.40 & --40 \\
    NGC 4569 (M69) & 4.75 & 0.39 & 21 \\
    NGC 4725 & 5.35 & 0.53 & 42 \\
    NGC 4736 & 5.6 & 0.71 & --88 \\
    NGC 4826 & 5.0 & 0.54 & --67 \\
    NGC 5055 & 6.1 & 0.46 & --81 \\
    \hline
  \end {tabular}
  \caption{Galaxies that could not be included in the sample with grand design spiral structure. R$_{25}$ is from \protect\cite{2003PASP..115..928K}. The axis ratios and position angles are the results of our fitting procedures (see text for details).}
  \label{ch4_tab4}
\end{table}

\section{Discussion}\label{discuss}

\begin{figure}   \centering
  \includegraphics[width=83mm]{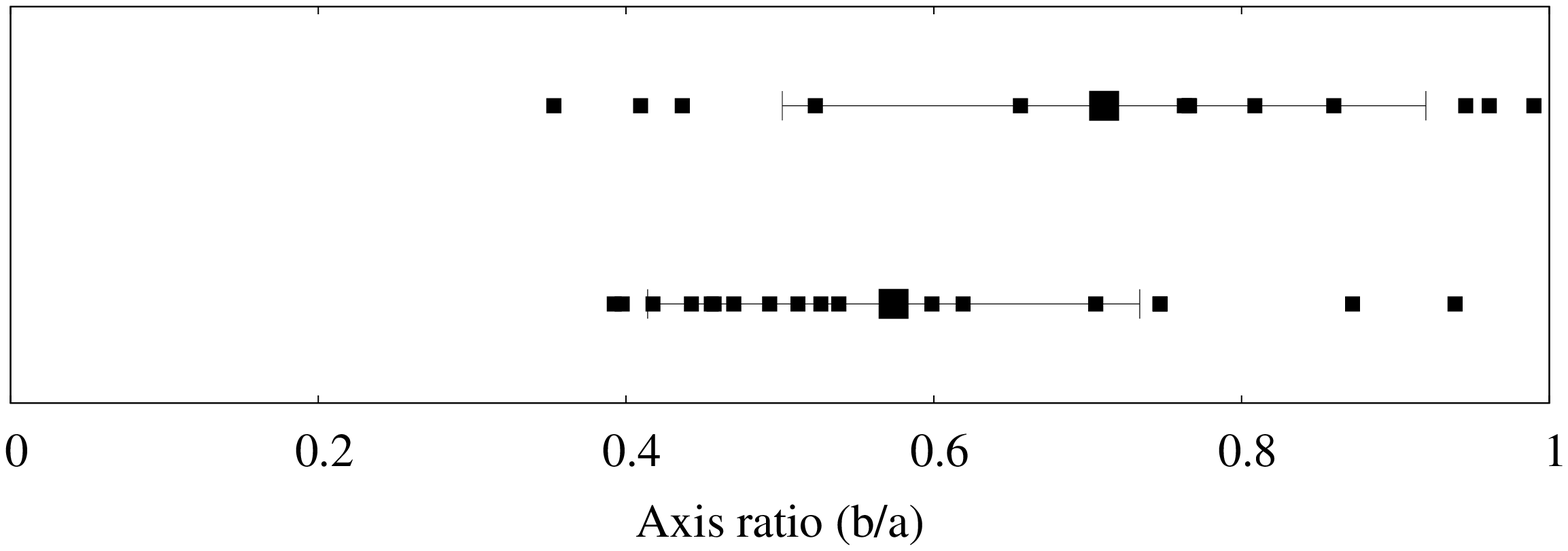}
  \caption{Plot of axis ratio vs the category; those in the detailed sample (top line) v.s. those in the non-detailed sample (bottom line). As can be seen, both types cover nearly the whole of the available range of values of b/a (the 70$^o$ cut off in inclination corresponds to an axis ratio of 0.34). The larger points with error bars give the average values and standard deviation of the two categories. As can be seen, the detailed sample are slightly more face on on average, but the two categories have the same average value to within the error bars.}
  \label{axis_ratio_plot}
\end{figure}

One important question raised by this work is whether there are any systematic differences between the galaxies in the detailed sample and those in the non-detailed sample (beyond the obvious presence or lack of detectable coherent or grand spiral structure); specifically whether the b/a and R$_{25}$ selection criteria affect the category in to which a galaxy is assigned. As discussed earlier, the more inclined a galaxy, the harder it is to extract a signature of spiral structure. However, the presence of galaxies in the detailed sample with axis ratios below 0.4 (see Figure \ref{axis_ratio_plot}) suggests that the cutoff in inclination was applied at a suitable position, and galaxies with relatively small axis ratios can still be analysed successfully (although there may be some for which the inclination does cause problems).

As well as the inclination effects, the galaxy size is also a potential influence
on the detectability of spiral structure; all other things being equal, a larger galaxy will be better resolved and thus easier to analyse. However, from Figure \ref{axis_ratio_plot}, it can be seen that the size of R$_{25}$ is not a contributing factor to the detection of spiral structure in this sample. A similar conclusion can be drawn from Figure \ref{gal_dist_plot}, where there is no systematic difference between the distances to the galaxies with grand design spiral structure and those without.

\begin{figure}   \centering
  \includegraphics[width=83mm]{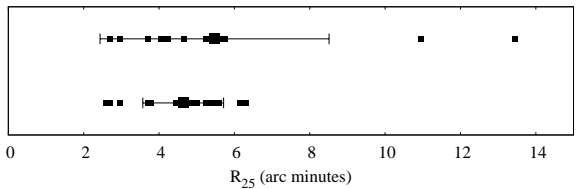}
  \caption{Plot of galaxy size vs the category; either those in the detailed sample (top line) v.s. those in the non-detailed sample (bottom line). As can be seen, the detailed sample are slightly larger on average, but this is almost entirely due to the two biggest galaxies in the detailed sample, NGC 2403 and NGC 3031.}.
  \label{r25_size_plot}
\end{figure}

\begin{figure}   \centering
  \includegraphics[width=83mm]{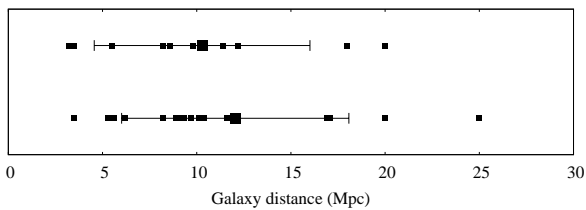}
  \caption{Plot of the average galaxy distance for the detailed sample (top line) and non-detailed sample (bottom line). The detailed sample are, on average, closer, but by far less than the scatter in the values.}
  \label{gal_dist_plot}
\end{figure}

From these results it can be concluded that the limits imposed on axis ratio and R$_{25}$ have not unduly biased the results: although one might suspect, for
example, that tightly wound arms would be harder to detect in high inclination
systems, our distribution of derived pitch angles shows no deficit of
small pitch angles for larger axis ratios.  

  Instead, the major limitation that affects this work is simply  one of sample size; of the 75 SINGS galaxies, only 31 satisfy all the selection criteria and of these only 13 have quantifiable spiral structure. Thus  any conclusions drawn from a comparison of galaxy properties with spiral structure is going to have to be tempered by the consideration of the statistical noise.

\subsection{Relationship between grand design status and non-axisymmetric structure}

\begin{figure}   \centering
  \includegraphics[width=83mm]{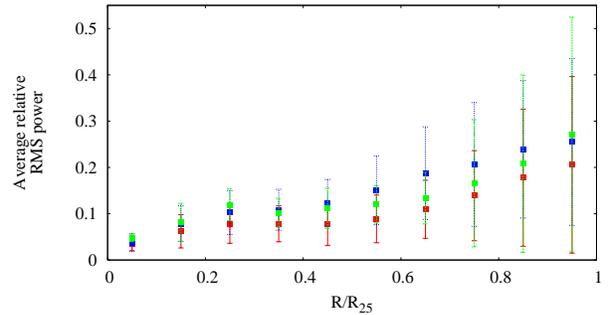}
  \caption{Plot showing amplitude of normalised 
non-axisymmetric power as a function of radius, averaged over the detailed sample (green symbols), the
additional sample (blue symbols) and the non-grand design sample (red symbols).}
  \label{nonaxi}
\end{figure}

  Grand design spirals are defined as those with coherent and large
amplitude structure in the m=2 mode. We can therefore use our classification
to enquire whether  such galaxies necessarily exhibit larger amplitude 
non-axisymmetric structure than those not deemed to be grand design. In Figure 
\ref{nonaxi} we plot the rms power in the
residual images (normalised to the local axi-symmetric value) as a function of 
radius (normalised to $R_{25}$). Each point represents a mean for all
the galaxies that are not grand design in the NIR (red symbols) and those
in the detailed  and additional samples (green and blue  
respectively) with the error bars representing $1 \sigma$ limits. 
 
 Figure \ref{nonaxi} demonstrates that the grand design galaxies have higher non-
axisymmetric power levels (as expected given their definition) but that
the difference is relatively modest. Thus non-grand design galaxies are not
significantly smoother than their grand design counterparts because they
in many cases possess a wealth of structure that is either concentrated
in higher order modes or is not coherent over a significant radial range. 
One might then not expect any strong correlation between grand design status
and phenomena (such as star formation rate) that are correlated with
regions of locally enhanced density. 

\begin{figure}   \centering
  \includegraphics[width=83mm]{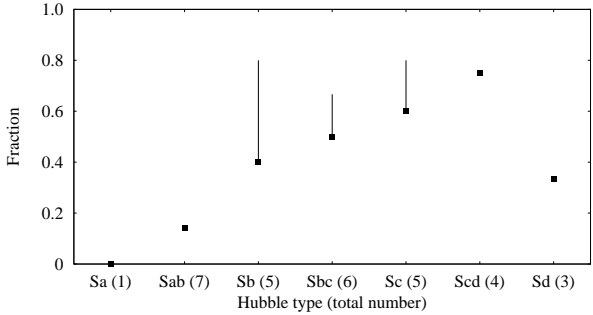}
  \caption{Plot showing fraction of galaxies that were included in the detailed sample by Hubble type. The vertical bars indicate how the fractions would change if we instead included also the  galaxies in the additional sample (see Section 3.2).}
  \label{hubb_frac}
\end{figure}

\subsection{Relationship between grand design status and Hubble type} 
Figure \ref{hubb_frac} shows the fraction of galaxies with strong spiral structure relative to those without, plotted by Hubble type. It seems that around half of the mid-to-late type spirals have detectable spiral structure in the NIR, except the very late type spirals, where the optical discs are extremely flocculent and a smaller fraction ($\frac{1}{3}$) have underlying spirals. A possible explanation
for the low fraction of early type  (Sab and probably Sa) galaxies 
showing  grand design spiral structure is that such galaxies have a low
gas fraction. It is often argued that gas plays an important role in
amplifying and sustaining spiral structure since in many simulations,
the rising velocity dispersion in the stars otherwise washes out spiral structure
over a few galactic rotation periods  (e.g. \citep{1984ApJ...282...61S, 2002A&A...388..826S}). Gas  (together with star formation, which
supplies new stars on circular, i.e. dynamically cold orbits) provides an effective dynamical cooling which helps to maintain the disc in a gravitationally
unstable state. According to this line of argument, one might not expect to
see long lived spiral structure in isolated, gas poor galaxies. However,
this statement needs to be carefully quantified, especially as the
requirement on the amount of gas required is almost certainly linked to
numerical factors (discreteness effects) that over-estimate the heating of
the stellar disc in most simulations.  
 Of the six Sab galaxies in our  sample which do not make it into the detailed sample, two are anaemic or have low star formation rates outside the centre (NGC 4569 and NGC 4826. The latter has the added complication of a counter-rotating gas disc as well). 

\subsection{Relationship between grand design status and bar strength}

In addition to Hubble type, the prevalence of spiral structure as a function of bar strength was examined. There is no statistical link (within the errors) between a galaxy's optical bar classification (SA, SAB or SB) and whether or not measurable spiral structure was detected. Clearly a larger sample of galaxies is needed to help investigate this link further, but this result is in agreement with \cite{2003MNRAS.342....1S} who find no evidence that strong bars (classified in the optical) drive spiral structure.

\begin{figure}   \centering
  \includegraphics[width=83mm]{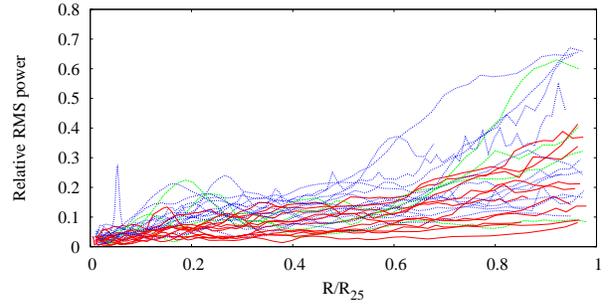}
  \caption{Relative r.m.s. power in the residual images colour coded by bar type; red=SA, blue=SAB, green=SB.}
  \label{asympower_bar}
\end{figure}

\begin{figure}   \centering
  \includegraphics[width=83mm]{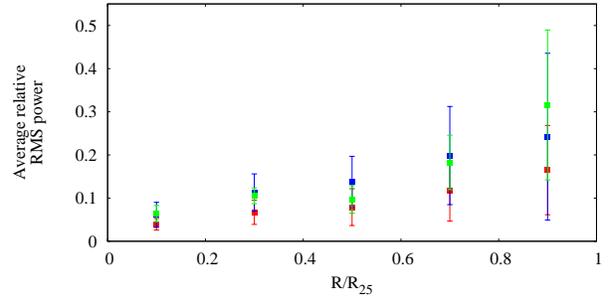}
  \caption{As Figure \ref{asympower_bar} but, in order to show the trends more clearly, the data are averaged by bar type and over sections 0.2$R_{25}$ in radius. The error bars give the standard deviation for each data point. Bar types are as follows: red=SA, blue=SAB, green=SB.}
  \label{bar_frac_av}
\end{figure}

Figure \ref{asympower_bar} shows the non-axisymmetric power in the residual images plotted against bar class; SA, SAB or SB. Note that all 31 spiral galaxies
in SINGS with suitable size and axis ratios have been plotted in this
figure, since evidently it is not necessary for a galaxy to display
a well defined logarithmic spiral in order to be able to assess its
level of non-axisymmetric power. It can be seen that barred (SAB and SB) galaxies tend to have more power, a factor of 1.6 on average, in their non-axisymmetric components than non-barred (SA) galaxies. This trend is statistical rather than absolute: some barred galaxies have relatively low non-axisymmetric power, but the overall behaviour is unmistakable, and is illustrated further in Figure \ref{bar_frac_av}. As can be seen in this figure, the distinction between SAB and SB galaxies is much less clear than with SA galaxies; this may well be because many galaxies are more strongly barred than their optical appearances suggest \citep{2004A&A...423..849G}. Of course, the bar itself will contribute to the non-axisymmetric power at small radii but the enhancement of non-axisymmetric power
at radii well beyond the bar's extent is more surprising. When combined with the fact that no link was found between bar type and m=2 spiral structure, this result suggests that  bars do drive non-axisymmetric structure, but not
necessarily in the main m=2 mode.

\subsection{Relationship between grand design status and the influence of companion galaxies}

A similar investigation may be carried out into the effects of companion galaxies on spiral structure. A search for companion or satellite galaxies to those in this sample was made using the NASA/IPAC Extragalactic Database (NED). 
We make an `inclusive' definition of a companion galaxy as being one
that satisfies a number of criteria; velocities must fall within $\pm$400kms$^{-1}$ of the target galaxy, (projected) distances  satisfy the criteria $\frac{R_{proj}}{R_{25}}$ $<$ 10.0, and a \textit{B} band apparent magnitude cutoff of +15 is imposed (most galaxies in this survey have magnitudes of around $+$10, so the companions are no less than $\sim$100$^{th}$ of the mass). The projected distances are of course a lower limit to the separation in three dimensions.

 We also make a `restricted' definition of a companion as being one where
the tidal pull, P, exceeds 0.01 (\citep{1992AJ....103.1089B}; P = $\frac{M_{c}/M_{g}}{(r/R)^{3}}$; M$_{g}$ is the galaxy mass, M$_{c}$ the companion mass, R the galaxy radius, and r is the projected separation). This selects out only
the pairs where there is most likely to be a strong dynamical influence and
represents a limited subset of the companions identified in the
`inclusive' companion sample. 

Table \ref{ch4_tab5} details the fractions of galaxies that have
companions in each category according to whether or not they are classified
as grand design galaxies in the infrared. We see a positive correlation between the presence of companions and grand design structure (particularly if we
only consider the strongest interactions) so long as we include
all the galaxies in the additional sample in our definition
of those with a well defined spiral structure. The figures in square
brackets show the results if we instead classify the additional galaxies as
lacking spiral structure: this reclassification changes the correlation
significantly because $4$ out of the $5$ of these galaxies have a companion,
and in two of these cases the companions exert a strong tidal pull ($P > 0.01$).
We can also see from Figure \ref{nonaxi} that these galaxies in the additional
sample have high values of non-axisymmetric power. We thus see that
companions are  associated with  large amplitude structures 
in the disc but that these structures are  not necessarily ones 
that can be analysed
as a coherent m=2 structure. 

  Table \ref{ch4_tab5} demonstrates that the presence of a close companion
is (almost) a sufficient  condition for associated structure: only
one galaxy with a close companion in projection (NGC 1512) ) is in our
`non grand design' category and inspection of Figure 12 demonstrates that
this galaxy, although not classically grand design, shows significant structure
in its outer regions.

 However, possession of a close companion  is evidently {\it not} a necessary
condition for galaxies to be placed in our `NIR grand design'
category, since there are  $7$ galaxies in this category that are `isolated'
according to our `inclusive' definition. (Note that some of
these galaxies may have bound companions but that they do not
fulfil the requirements in terms of projected separation and
magnitude difference set out above). Inspection of these galaxies
on a case by case basis however reveals that such galaxies are
either barred or else, more frequently, exhibit grand design
structure in the NIR that is rather weak (and in some cases, these same galaxies have
been classified as flocculent in the optical).
Thus weak spiral structure in the NIR does not apparently require the presence
of a companion.

 These findings are broadly in line with results on the correlations
between companions and grand design structure found in previous optical
studies.
For example,  \cite{1998MNRAS.299..685S}
find an increased number of m=2 spirals amongst those galaxies with companions within 6 galaxy diameters. \cite{1979ApJ...233..539K} found that galaxies with grand design structure either rotate with solid-body rotation over the radial range of the spiral pattern, or have a companion or bar which can drive the pattern if the rotation is differential. Similarly, \cite{1982MNRAS.201.1021E} found that the majority (68 per cent) of isolated SA spirals have flocculent structure, and that galaxies without companions or bars are more likely to be flocculent.

\begin{table}
  \centering
  \begin{tabular}{|l|l|l|}
    \hline
     & Well defined spiral & No well defined spiral \\
     \hline
     fraction with  companion     &  10/18  [6/13]      &  5/13 [9/18]   \\
     (inclusive definition)   &     &      \\
     \hline
     fraction with companion   & 4/18  [2/13]      & 1/13   [3/13]    \\
     (restrictive definition)   &     &       \\
    \hline
  \end {tabular}
  \caption{Table showing the fraction of galaxies with grand design spiral structure (left column) and without such structure (right column) that have companions,  defined `inclusively' in the text (upper line ) and defined restrictively
in the text (in terms of their tidal pull) on the lower line. In each category,
the first number is the fraction if the well defined spirals are taken
to be the $13$ objects in the detailed sample plus the $5$ objects in the additional sample,  whereas the following
figure in brackets is the fraction in the case that the well defined
objects are taken to be the $13$ objects in the detailed sample only.}
 \label{ch4_tab5}
\end{table}

\section{Conclusions}.\label{conclusions}

Underlying two-armed spiral waves in the old stellar population are not ubiquitous in disc galaxies; $\frac{13}{31}$ of the galaxies studied in this work have detectable two armed spiral structure that follows approximately logarithmic behaviour with radius, and which could be analysed in detail. We judged a further 5 galaxies to be grand design even though we were unable to quantify their structure in detail (see discussion in Section 3.2).   Thus around half of the disc galaxies analysed have grand design spiral structure in the old stellar population. The galaxies in which grand design spiral structure could not be detected exhibit a variety of structures but significant non-axisymmetric power is detected in almost all cases, with the range of values overlapping those found in grand design galaxies (see Figure \ref{nonaxi}). This simply means that in many non grand design galaxies there is significant power in higher order modes or else
that an m=2 mode is strong only over a restricted radial range.

  In general there is a good correlation between galaxies being grand
design in the infrared and in the optical (see Table \ref{armclass_tab}). There are no
galaxies that exhibit grand design structure in the optical that do not
also exhibit such structure in the near infrared, as is consistent with
the expectation that the spiral structure involves the dominant stellar
mass component in the galaxies, rather than merely the products of
recent star formation. On the other hand, as noted by \cite{1991Natur.353...48B, 1996ApJ...469L..45T, 1999AJ....118.2618E}, we have several examples of objects that are optically flocculent but which do exhibit near infrared grand design
structure.

Bars have been linked to spiral structure in some previous research \citep{1982MNRAS.201.1021E}, but in this work, no link is found between a galaxy being barred and having spiral structure, suggesting that bars do not trigger spirals in the majority of cases. This result is in agreement with \cite{1998MNRAS.299..685S}. When the non-axisymmetric power in the whole sample is analysed it is found that bars do seem to increase the degree of substructure in the disc, even beyond the radial extent of the bar; SAB and SB galaxies are found to have $\sim$1.5 time more power in their non-axisymmetric components than SA galaxies (see Figure
\ref{bar_frac_av}). 

 The sample was also examined for evidence of spirals being triggered by interactions with companion galaxies. Here, we find that those galaxies with
strong spiral structure have a higher incidence of companions than those
without, and that this correlation increases if one restricts the definition
of a companion to those exerting the strongest tidal
influence (see Table \ref{ch4_tab5}). Interestingly enough, we find that
$4$ out of $6$ of the systems undergoing the strongest interactions
are placed in our `additional sample', which comprises a handful of galaxies
that - though being clearly `grand design' and with a large amplitude
of non-axisymmetric structure - are nevertheless not possible to analyse
as a simple m=2 logarithmic spiral (see Section 3.2 for a discussion on 
a case by case basis). This result may thus relate to the temporal sequence
in which modes develop during a tidal interaction (see \citep{1992ApJS...79...37E}
and simulations of spiral generation during encounters by \cite{2008ApJ...683...94O, 2010MNRAS.403..625D}). 
We do also find a number of galaxies with grand design spiral structure and with no evidence of a suitable companion. However, these are generally either barred galaxies or else are objects where
the level of grand design structure in the NIR is relatively weak (and
which may even be classified as flocculent in the optical). 
We therefore conclude that
whereas strong prominent grand design structure is generally associated
with either a bar or a companion, weaker structure in the NIR may also
occur in isolated, unbarred systems.

\section{Acknowledgments.}

Many thanks to Bob Carswell for much useful advice regarding the use of IDL, in particular the CURVEFIT program. We are indebted to Hans-Walter Rix for the idea of using colour-correction to make mass maps from optical data. Many thanks also go to Jim Pringle, Clare Dobbs and Stephanie Bush for useful discussions over the course of this work. We would like to thank the referee, Phil James, for his thorough reading of the paper and helpful suggestions.

This work makes use of IRAF. IRAF is distributed by the National Optical Astronomy Observatories, which are operated by the Association of Universities for Research in Astronomy, Inc., under cooperative agreement with the National Science Foundation.

\bibliography{references}

\label{lastpage}

\end{document}